\documentclass{aa}
\usepackage{graphicx}
\begin{document}
\title{VLBA images of High Frequency Peakers}
\subtitle{   }

\author{M. Orienti\inst{1,2} \and
D. Dallacasa\inst{1,2} \and
S. Tinti\inst{3}
 \and C. Stanghellini\inst{2} 
}
\offprints{M. Orienti}
\institute{
Dipartimento di Astronomia, Universit\`a di Bologna, via Ranzani 1,
I-40127, Bologna, Italy \and 
Istituto di Radioastronomia -- CNR, via Gobetti 101, I-40129, Bologna,
Italy \and
SISSA/ISAS, Via Beirut 4, 34014, Trieste, Italy
}
\date{Received \today; accepted ?}
\abstract{
%Context heading
{}
We propose a morphological classification based on the parsec scale
  structure of fifty-one High Frequency Peakers (HFPs) from the
  ``bright'' HFP sample.
VLBA images at two adjacent frequencies (chosen among 8.4, 15.3, 22.2
and 43.2 GHz) have been used to investigate the morphological
properties of the HFPs in the optically thin part of their spectrum.
We confirm that there is quite a clear distinction between the 
pc-scale radio structure of galaxies and quasars:
the 78\% of the galaxies show a ``Double/Triple''
morphology, typical of Compact Symmetric Objects (CSOs), while
the 87\% of the quasars are characterised by Core--Jet or
unresolved structure. This suggests that most HFP candidates
identified with quasars are likely blazar objects in which a flaring
self-absorbed component at the jet base was outshining the remainder
of the source at the time of the selection based on the spectral shape.\\ 
Among the sources classified as CSOs or candidates it is possible to
find extremely young radio sources with ages of about 100 years or even
less.
}
\titlerunning{VLBA images of High Frequency Peakers}
\maketitle
\section{Introduction}
Young radio sources, despite being very rare, are important tools in
the study of the onset of radio emission. The radio plasma interacts
with a presumably inhomogeneous and quite dense ambient medium in the
vicinity of the active nucleus.\\ 
In the radio domain they have outstanding characteristics, being
scaled down versions of the extended (up to a few Mpc) radio
galaxies. Based on their radio morphology they are classified as
``Compact Symmetric Objects'' (CSOs), implying that they are
intrinsically compact ($<1$ kpc) and bright extragalactic
radio sources (P$_{\rm 1.4~GHz}$ $>$ 10$^{25}$ W/Hz) with a rather
symmetric radio structure, and characterised by convex synchrotron
radio spectra peaking at frequencies ranging from 100 MHz to few GHz
(Wilkinson et al. \cite{wil94}). 
Both kinematic and spectral studies show that CSOs with sizes of
50-100 pc have typical ages
of about 10$^{3}$ years (Polatidis \& Conway
\cite{PC03}; Murgia \cite{Mu03}), still developing their radio
emission within the host galaxy (Fanti et al. \cite{cf95};
Readhead et al. \cite{rh96}; Snellen et al. \cite{sn00}).\\
The convex spectral shape is a distinctive property of the
intrinsically small radio sources: there is an anti-correlation
between the intrinsic peak frequency and the source size
(e.g. O'Dea \cite{O98}) for the Compact Steep Spectrum sources of the
Fanti et al. (\cite{rf90}) and the GHz-Peaked Spectrum of the
Stanghellini et al. (\cite{cstan98}) samples. It is explained in
terms of synchrotron self-absorption occurring in a small radio emitting
region. As the region (a mini radio lobe) expands as the result of the
source growth, the turnover frequency moves to lower frequencies. It is
also worth to mention that free-free absorption could play a role
(e.g. Bicknell et al. \cite{bdo97}), more relevant in the smallest
objects.\\ 
In this scenario the anti-correlation implies that the youngest
objects have the highest turnover frequency, and
therefore, sources with spectral peaks at frequencies higher than few
GHz are good candidates to be {\it newly born}, with ages of about
10$^{2 - 3}$ years. \\  
A complete sample of ``bright'' radio sources with turnover frequency
above $\sim$5 GHz, termed ``High Frequency Peakers'' (HFPs), has been
selected by Dallacasa et al. (\cite{hfp0}). \\ 
Since the selection of these sources is based on their simultaneous
radio spectra at a single epoch, there is still some contamination
from beamed radio sources, whose emission is temporarily dominated by
a self-absorbed jet component. \\
Further simultaneous radio spectra from VLA observations (Tinti et
al. \cite{st05}) showed that most of the sources identified with radio
quasars modified the spectral shape due to flux density variability,
and did not fulfill anymore the selection criteria. On the other hand
most of the objects identified with galaxies (and empty field as well)
preserved their convex radio spectra. \\
Another key aspect to understand these sources is related to their
pc-scale morphology: it is well known that in young radio galaxies the
radio emission is dominated by mini-hot-spots and mini-lobes. On the
other hand, contaminating sources with temporarily peaked spectra should have
core-jet structures. 
This paper reports on the results of new VLBA observations at two
frequencies in the optically thin part of the spectrum of each source,
chosen among 8.4, 15, 22 and 43 GHz depending on the spectral peak
frequency as found in Dallacasa et al. (\cite{hfp0}).\\
Section 2 briefly recalls the sample selection criteria, section 3
describes the observations and the production of the images, section 4
presents the results, including the morphological classification, the
spectral index distribution of the source components, and, finally,
section 5 summarises the main results from this work.\\

Throughout this paper, we assume H$_{0}$= 71 km s$^{-1}$ Mpc$^{-1}$,
$\Omega_{\rm M}$ = 0.27 and $\Omega_{\rm \lambda}$ = 0.73. When
redshift is unknown, we adopt z=1.00.\\

\section{The sample}
The bright HFP sample has been assembled by cross correlating the
Green Bank survey (87GB) at 4.85 GHz (Gregory et al. \cite{87gb}) and
the NRAO VLA Sky Survey (NVSS) at 1.4 GHz (Condon et al. \cite{nvss}):
only the sources brighter than 300 mJy at 4.85 GHz with 
inverted synchrotron spectra, and in particular those with a slope
steeper than -- 0.5 (S $\propto$ $\nu^{- \alpha}$) have been
selected. Sources with $|$ b$_{II}$$ |$ $<$ 10$^\circ$ have been
excluded to avoid the galactic plane. 
Simultaneous VLA observations at eight different frequencies
ranging from 1.365 to 22.46 GHz were used to derive the radio spectrum
and to remove variable sources from the sample. \\
The final sample of
HFP candidates consists in 55 objects ($\sim 3\%$ of the total number
of sources) with peaked radio spectra (Dallacasa et al. \cite{hfp0}),
whose optical counterpart comprises: 11 galaxies, 33 quasars, 5 BL Lac 
and 6 unidentified 
sources (Dallacasa et al. \cite{dfs}, \cite{dfs6}). 
The redshift distribution is
different between galaxies and quasars: typically galaxies redshift
ranges between $\leq$ 0.1 and 0.67, while quasars, generally at higher
redshift, between 0.9 and 3.6 (see Dallacasa (\cite{dd03}) for a recent review
on the properties of this new class of objects).\\

\begin{table}
\begin{center}
\begin{tabular}{clcc}
\hline
\hline
&&&\\
Exp. Code & Date & Obs. time & Notes\\
\hline
\hline
BD077A&11 Jan 2002& 9h & Erratic T$_{sys}$ values at LA \\
BD077B&10 Feb 2002& 9h & - \\
BD077C&16 Feb 2002& 9h & - \\
BD077D&02 May 2002& 9h & - \\
&&&\\
\hline
\end{tabular} 
\vspace{0.5cm}
\end{center}   
\caption{Log of the VLBA observations} 
\label{taboss}
\end{table}  

\section{VLBA observations and data reduction}
Pc-scale resolution observations were carried out with the VLBA in
2002 (see Table \ref{taboss}) at 8.4, 15.3, 22.2 and 43.2 GHz, in
  full polarisation mode with a
recording band-width of 32 MHz at 128 Mbps.  The correlation was
performed at the VLBA correlator in Socorro. \\ 
The target sources, accommodated into four 9-hr  observing runs, were
observed at a pair of frequencies in the optically thin region of
their synchrotron spectra when this was possible: for example, the
objects with turnover frequency above 15 GHz in Dallacasa et
al. (\cite{hfp0}) were observed at 22.2 and 43.2 GHz. \\
Each target was observed typically for 20/30 min at each frequency,
spread into 4 to 7 short scans at various HA in order to improve the
{\it uv} coverage. \\ 
The strong sources \object{3C~454.3}, \object{3C345} and
\object{J2136+0041} (a ``bright'' HFP
quasar) were used as fringe finders, and a few more calibration
sources (including the ``bright'' HFP \object{J0927+3902} and 
\object{J1407+2827} alias \object{OQ208}) were
observed to verify the system performance during each experiment. \\ 
All the data handling has been carried out by means of the NRAO AIPS
package. 
A-priori amplitude calibration was derived using measurements of the
system temperatures and the antenna gains. We estimate that the error
on the absolute flux density scale is generally within 5\%, perhaps
with a slightly larger value at 43.2 GHz. \\
Fringe fitting was carried out on each source, with a short (down to 1
min) solution interval at 22.2 and 43.2 GHz in order to preserve phase
coherence. In general all sources were detected on all baselines, and
the data turned out to have good quality at all frequencies, with the
only exception of \object{J1526+6650}, which has not been detected at 43.2
GHz. \\
During the BD077A observations, the VLBA antenna located in Los Alamos
was affected by technical problems that caused the
system temperature have erratic values at 15.3 and 22.2 GHz, while
the data were good at the other two frequencies. \\
The final radio images were obtained after a number of phase
self-calibration iterations, and 
source parameters (total flux density and deconvolved sizes) have been
generally measured by means of the task JMFIT for marginally resolved
components, and with TVSTAT and IMSTAT when a component could not be
fitted with a Gaussian profile. In cases
  where multiple components are present, the VLBA total flux density
  has been derived on the image plane by means of TVSTAT.
All the values are reported in
Table 2, Table 3 and Table 4.\\

\begin{table*}[h]
\small{ 
\begin{center}
\begin{tabular}{|c|c|c|cc|cc|cc|c|c|c|c|c|c|}
\hline
Source&z&Id&S$_{\rm 8.4 }^{VLA}$&S$_{\rm 8.4 }^{VLBA}$&S$_{\rm
  15.3 }^{VLA}$&S$_{\rm 15.3 }^{VLBA}$&S$_{\rm 22.2
  }^{VLA}$&S$_{\rm 22.2 }^{VLBA}$&S$_{\rm 43.2
  }^{VLBA}$&$\alpha$&H$_{\rm eq}$&Morph\\
 & & &mJy&mJy&mJy&mJy&mJy&mJy&mJy& &mG&\\
\hline
&&&&&&&&&&&&\\
J0003+2129&0.455&G&227&228&140&131&86&-&-&1.0&17&CSO\\
J0005+0524&1.887&Q&166&163&111&105&82&-&-&0.8&36&CSO\\
J0037+0808& &G&262&256&190&173&143&-&-&0.7&31&CSO\\
{\bf J0111+3906}&0.668&G& & & & & & & & & &\\
J0116+2422& &EF&248&237&128&173&149&-&-&0.5&68&Un\\
J0217+0144&1.715&Q&810&-&856&-&838&705&521&0.4&340&Un\\
J0329+3510&0.50&Q&578&472&659&525&702&-&-&-0.2&48&CJ\\
J0357+2319& &Q&144&-&154&-&168&155&110&0.5&113&Un\\
J0428+3259&0.479&G&514& &375&357&263&220& &1.3&26&CSO\\
J0519+0848& &EF&430&-&420&-&401&387&349&0.2&128&Un\\
J0625+4440& &BL&238&-&219&-&210&143&60&1.3&167&Un\\
J0638+5933& &EF&667&-&620&-&567&505&369&0.5&72&CSO\\
J0642+6758&3.180&Q&321&320&203&189&149&-&-&0.9&116&Un\\
J0646+4451&3.396&Q&3757&-&3691&-&3318&2953&1938&0.6&462&MR\\
J0650+6001&0.455&Q&964&-&671&650&495&406&-&1.3&27&CSO\\
J0655+4100&0.02156&G&335&-&313&272&271&214&-&0.6&94&Un\\
J0722+3722&1.63&Q&198&-&138&170&96&115&-&1.1&55&MR\\
J0927+3902&0.6948&Q&11859&-&10060&-&8660&-&-&-&-&CJ\\
J1016+0513& &Q&522&-&449&-&379&253&227&0.2&106&Un\\
J1045+0624&1.507&Q&284&266&238&211&196&-&-&0.3&91&Un\\
J1148+5254&1.632&Q&512&-&501&355&458&238&-&1.0&106&CSO\\
J1335+4542&2.449&Q&592&526&359&307&234&-&-&0.9&187&CSO\\
J1335+5844& &EF&671&604&531&449&253&-&-&0.5&17&CSO\\
J1407+2827&0.0769&G&2050&-&1139&-&604&-&-& &-&CSO\\
J1412+1334& &EF&267&251&185&140&130&-&-&1.0&39&Un\\
J1424+2256&3.626&Q&460&419&251&-&145&-&-& &-&Un\\
J1430+1043&1.710&Q&752&709&546&519&385&-&-&0.6&112&MR\\
J1457+0749& &BL&188&225&141&157&109&-&-&1.1&110&Un\\
J1505+0326&0.411&Q&710&573&665&509&567&-&-&0.2&420&Un\\
J1511+0518&0.084&G&861& &843&693&617&457& &1.2&40&CSO\\
J1526+6650&3.02&Q&341&-&193&-&107&48&-& &-&MR\\
&&&&&&&&&&&&\\
\hline
\end{tabular}
\end{center}  }
\caption{The VLBA flux density of candidates HFP. Col. 1: source name
(J2000); 
Col. 2: redshift; Col. 3: optical identification from Dallacasa
  et al. (\cite{hfp0}), (\cite{dfs}), (\cite{dfs6}); 
Col. 4, 5: VLA and VLBA 8.4 GHz flux density
  respectively; Col. 6, 7: VLA and VLBA 15.3 GHz flux density;
  Col. 8, 9: VLA and VLBA 22.2 GHz flux density; Col. 10: VLBA 43.2
  GHz flux density; Col. 11: spectral index between the two
  frequencies where VLBA images are available;  Col. 12: Equipartition magnetic field; we
  assume $\alpha$ = 0.7;
Col. 13: Morphological classification from VLBA images. 
  The two
  sources J0111+3906 and J1751+0939 were not observed, since they are
  already plenty of information. The sources J0927+3902 and OQ 208
  have been observed only for an extremely short period of time, to
  verify the system performance. }
\label{Tab 1}
\end{table*}

\addtocounter{table}{-1}
\begin{table*}[h]
\small{ 
\begin{center}
\begin{tabular}{|c|c|c|cc|cc|cc|c|c|c|c|c|c|}
\hline
Source&z&Id&S$_{\rm 8.4 }^{VLA}$&S$_{\rm 8.4 }^{VLBA}$&S$_{\rm
  15.3 }^{VLA}$&S$_{\rm 15.3 }^{VLBA}$&S$_{\rm 22.2
  }^{VLA}$&S$_{\rm 22.2 }^{VLBA}$&S$_{\rm 43.2
  }^{VLBA}$&$\alpha$&H$_{\rm eq}$&Morph\\
 & & &mJy&mJy&mJy&mJy&mJy&mJy&mJy& &mG&\\
\hline
&&&&&&&&&&&&\\
J1603+1105& &BL&225&176&234&176&217&-&-&0.0&94&Un\\
J1616+0459&3.197&Q&559&496&333&268&212&-&-&1.0&123&CSO\\
J1623+6624&0.203&G&283&296&224&216&175&-&-&0.5&49&Un\\
J1645+6330&2.379&Q&618& &596& &493&493&363&0.5&272&Un\\
J1717+1917&1.81&Q&229&- &227&167&215&97&- &1.5&145&Un  \\
J1735+5049& &G&920&875&740&672&587&-&-&0.4&27&CSO\\
{\bf J1751+0939}&0.322&BL& & & & & & & & & &\\
J1800+3848&2.092&Q&1063&-&1174&-&1076&768&537&0.5&277&Un\\
J1811+1704& &BL&509&518&499&486&418&-&-&0.1&119&CJ\\
J1840+3900&3.095&Q&158&151&134&128&114&-&-&0.3&189&Un\\
J1850+2825&2.560&Q&1541&-&1318&1208&1045&867&-&0.8&150&MR\\
J1855+3742& &G&212&189&124&87&91&-&-&1.3&19&CSO\\
J2021+0515& &Q&368&359&267&242&191&-&-&0.7&43&CJ\\
J2024+1718&1.05&Q&800& &697& &569&495&294&0.8&74&Un\\
J2101+0341&1.013&Q&583& -  &499&-&478&410&341&0.3&97&Un\\
J2114+2832& &EF&792&-&749&466&685&373&-&0.6&40&CJ\\
J2123+0535&1.878&Q&2482&-  &2755& -&2560&1719&1408&0.3&219&CJ\\
J2136+0041&1.932&Q&8940&-  &7443& - &6169&3565&2218&0.8&71&CJ\\
J2203+1007& &G&234&223&129&110&77&-&-&1.2&10&CSO\\
J2207+1652&1.64&Q&223&-&188&193&163&147&-&0.7&76 &Un\\
J2212+2355&1.125&Q&1028&1242&975&1059&915&-&-&0.3&112&Un\\
J2257+0243&2.081&Q&450&-&558&-&528&427&272&0.7&212&Un\\
J2320+0513&0.622&Q&725&608&806&561&843&-&-&0.1&78&Un\\
J2330+3348&1.809&Q&463&408&532&431&548&-&-&0.1&141&MR\\
&&&&&&&&&&&&\\
\hline
\end{tabular}
\end{center}  }
\caption{Continued.}
%\label{Tab 1}
\end{table*}

\section{Results}

Low dynamic range images are not the ideal tool to perform a very
accurate morphological classification, and this is particularly true
for complex sources. However, from our short observations it is
possible to derive indications on the nature using also the spectral
information allowed by the frequency pair data available for each
source. \\
In our VLBA images fourteen out of the 51 sources we observed, show a
CSO-like morphology, six present a Core-Jet structure and one has
multiple images, being a well known gravitational lens (\object{J1424+2256}).
These sources will be described in detail in section 4.2.\\
Despite the high resolution achieved by our VLBA images, about 30
sources ($\sim$ 60\%) are still unresolved or marginally resolved. 
We consider ``Marginally Resolved'' those sources whose largest
angular size (LAS) we detect is between 0.5 and 1 beam size at the
highest frequency, and we term ``Unresolved'' all the sources whose
LAS  is smaller than half the beam size at both frequency. We then
classify as Unresolved 25 objects (2 galaxies, 17 quasars, 3 BL Lacs, 
and 3 empty
field), and ``Marginally Resolved'' 6 sources (all quasars). 
The total spectral index of unresolved and marginally resolved sources
are usually quite flat or inverted, despite we are observing the
optically thin emission of each source as from the VLA total spectra
in Dallacasa et al. (2000). \\
If we compare the flux densities in our VLBA data, substantial
variability has been found in two ``Unresolved''
and one
``Marginally Resolved'' objects. These sources, the quasars
\object{J0722+3722} and
\object{J2212+2355}, and the BL Lac object \object{J1457+0749}, show an
increment of about 20\% and 10\% in their flux density at the lower
and higher frequencies, respectively.
Therefore, they are
likely flat spectrum sources seen in different phase of their
variability, and happened to be classified as HFP radio sources since
at the time of the simultaneous VLA multi-frequency observation their
emission was dominated by a bright self-absorbed component peaking at
cm-wavelengths. As in Tinti et al. (\cite{st05}), we point out these
objects are not to be considered HFPs anymore.\\

\subsection{Source images}

Full resolution images of the fourteen sources with a CSO-like
morphology are presented in
Figs. \ref{CSO} and \ref{CSO-1f}. Figs. \ref{jet} and \ref{jet-1f} instead
show images of the objects with a Core-Jet structure. 
An example of a ``Marginally
Resolved'' source is shown in Fig. \ref{MR}.
Finally, Fig. \ref{lente} show the multiple images of the
gravitational lens. In general the sources are presented in increasing
RA order and images at the two frequencies are next to each other, the
lowest frequency always on the left panel (Fig. \ref{CSO} and
\ref{jet}). 
For the
objects resolved at one frequency only, such image is
shown (Fig. \ref{CSO-1f} and \ref{jet-1f}). Images for the objects found to be
unresolved at both frequencies are not presented.\\
Noises in the image plane are generally between 0.5 and 1 mJy, except
for few cases where bad weather conditions affected the observations
implying substantially higher noise levels, particularly at high frequency.\\
For each image we provide the following information on the plot itself:\\
a) the source name and the observing frequency on the top left corner; \\
b) the peak flux density in mJy/beam;\\
c) the first contour intensity ({\it f.c.} in mJy/beam), which is
generally 3 times the off-source r.m.s. noise level measured on
the image; contour levels increase by a factor 2;\\
d) the optical identification;\\
e) The restoring beam, plotted on the bottom left corner of each
image.\\

\begin{figure*} 
\begin{center}
\includegraphics{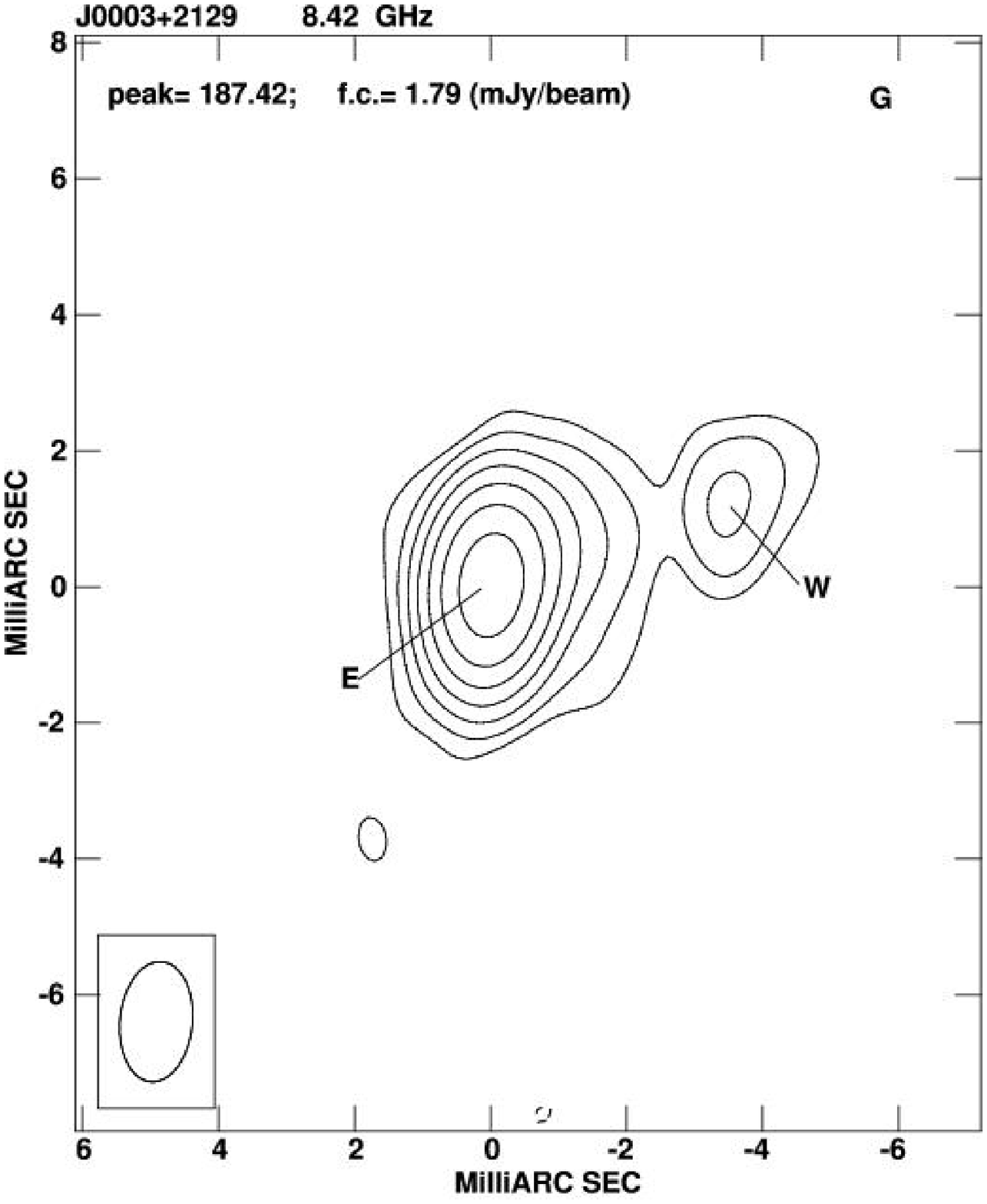}
\includegraphics{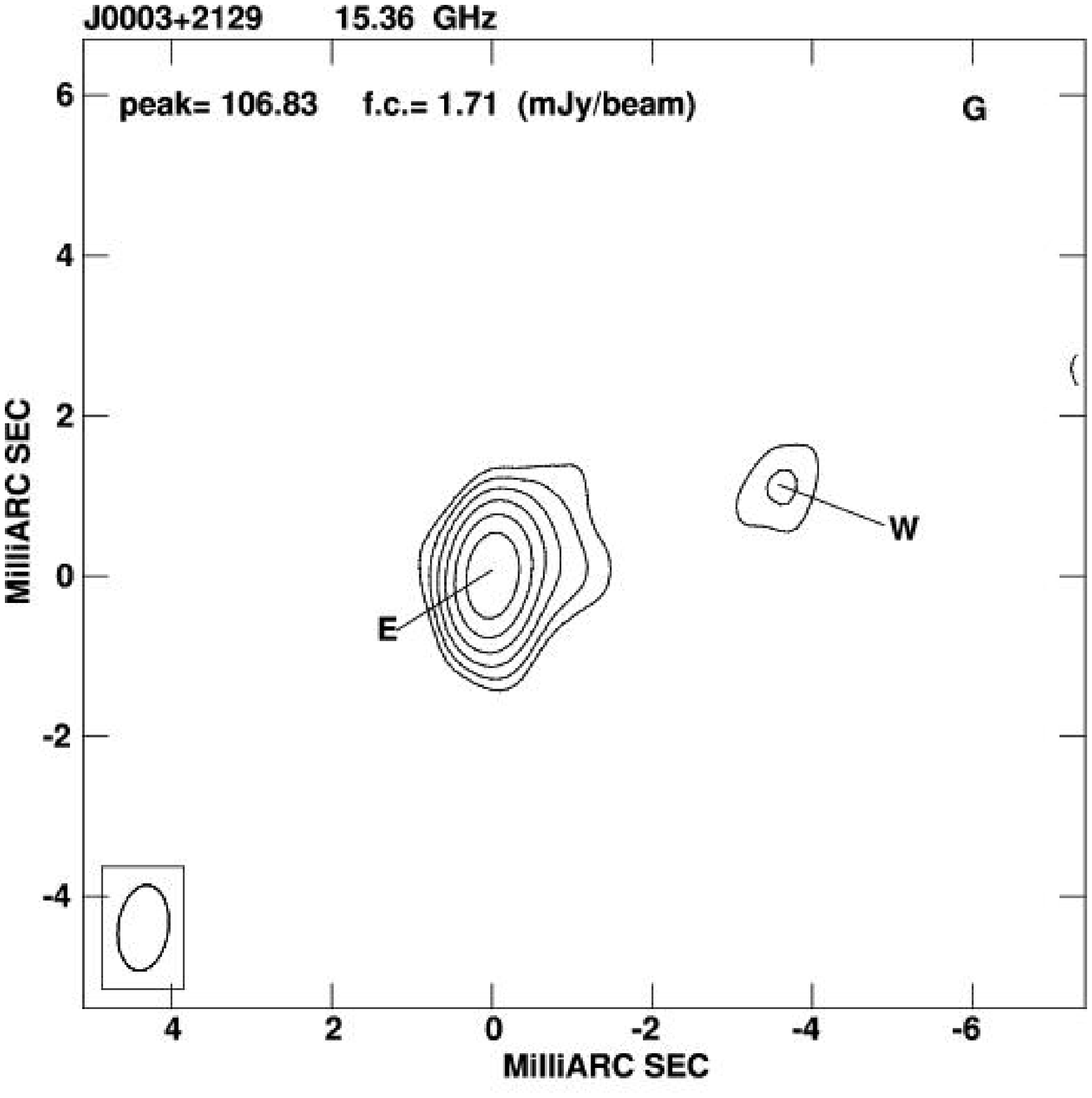}
\includegraphics{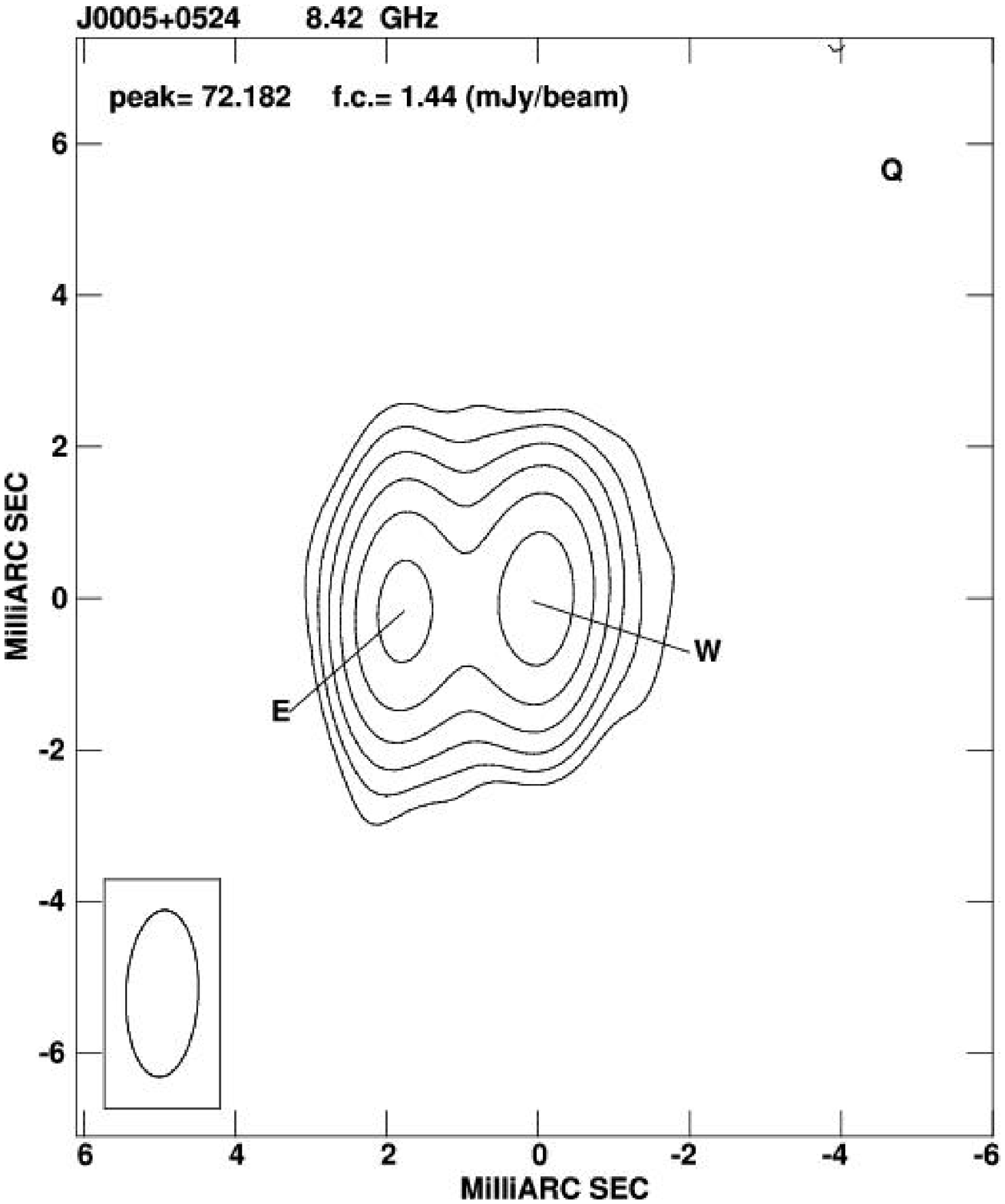}
\includegraphics{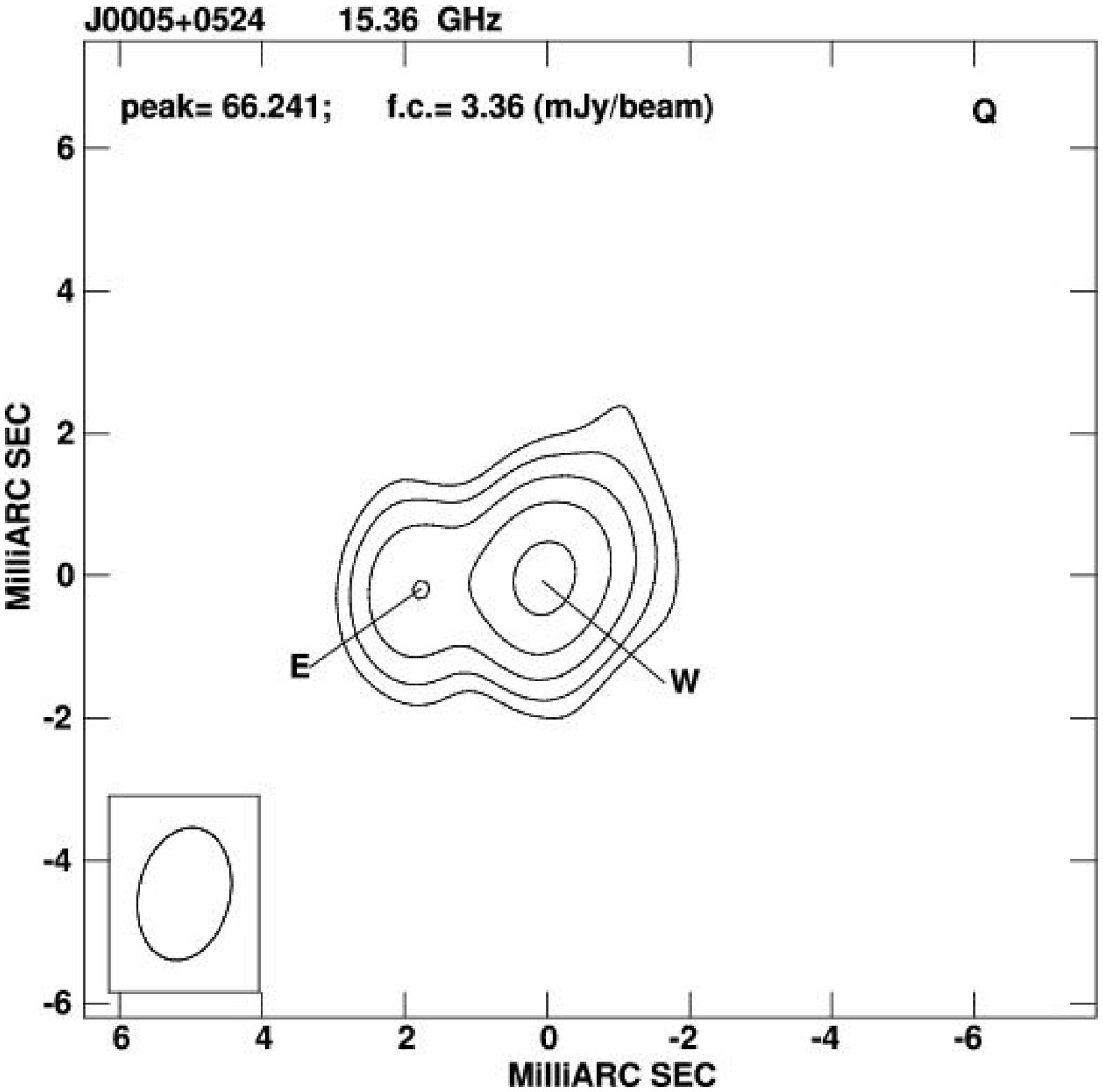}
\includegraphics{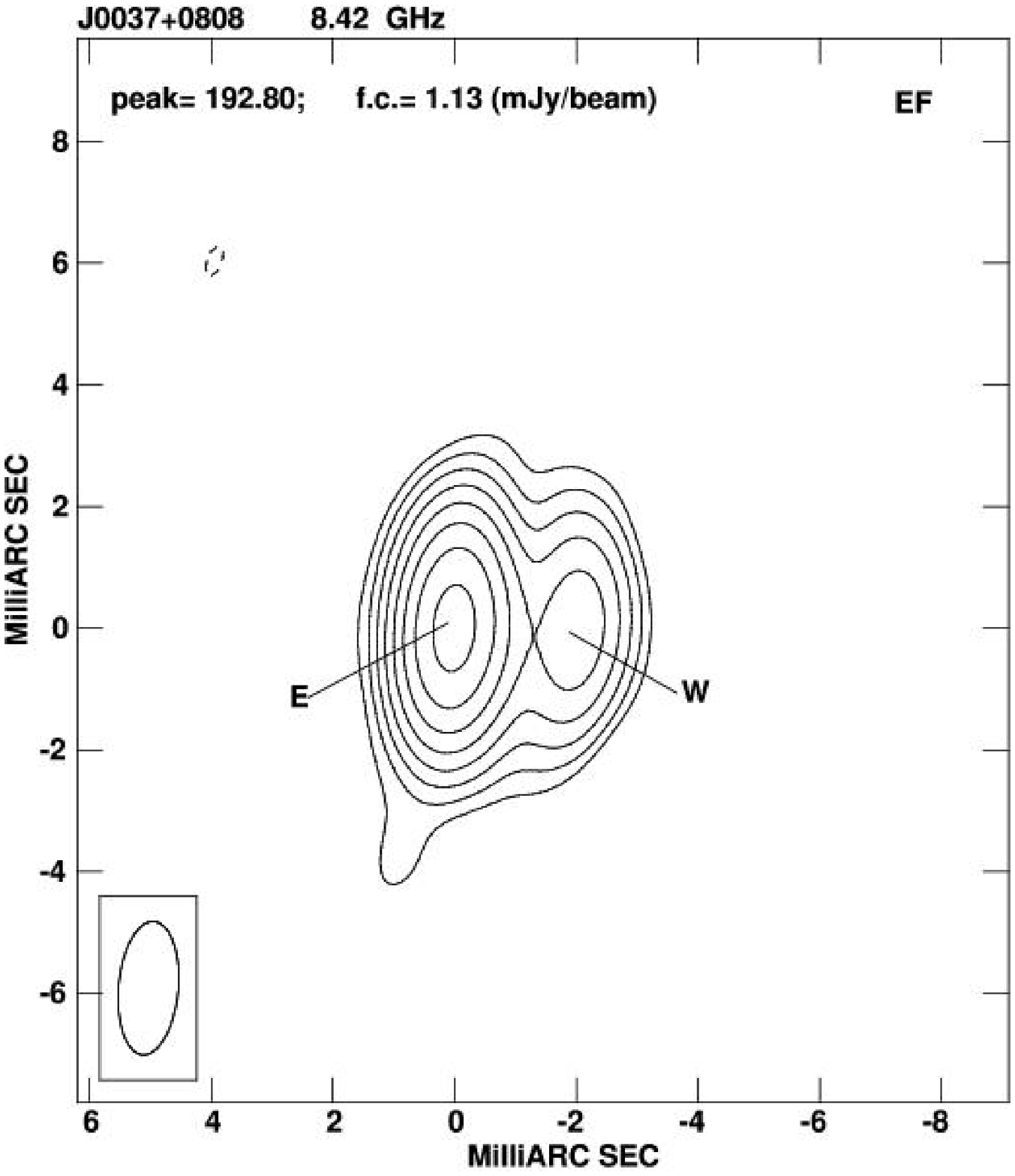}
\includegraphics{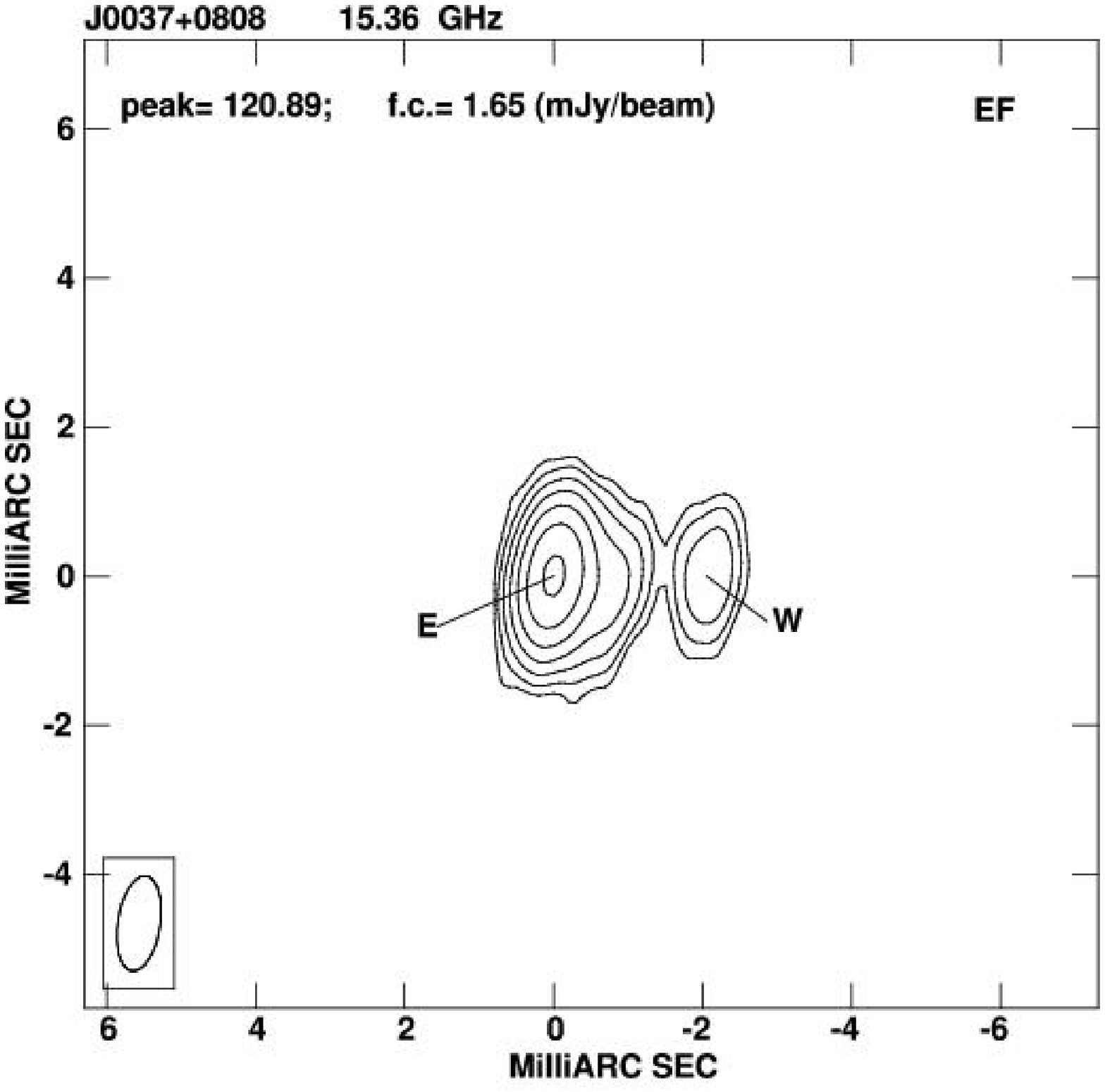}
\vspace{22.5cm}
\caption{An example of VLBA images at the two frequencies of a
  candidate HFP with a CSO morphology. 
  For each image we give the following information
  on the plot itself: a) Peak flux density in mJy/beam; b) First
  contour intensity ({\it f.c.}, in mJy/beam), which is generally 3 times
  the r.m.s. noise on the image plane; contour levels increase by a
  factor of 2; c) the optical identification; d) the restoring beam 
  is plotted on the bottom left corner of each image. All the images
  of Figure 1 are available in electronic form only at {\rm
  http://www.edpsciences.org/aa.}}
\label{CSO}
\end{center}
\end{figure*}

\addtocounter{figure}{-1}
\begin{figure*} 
\begin{center}
\includegraphics{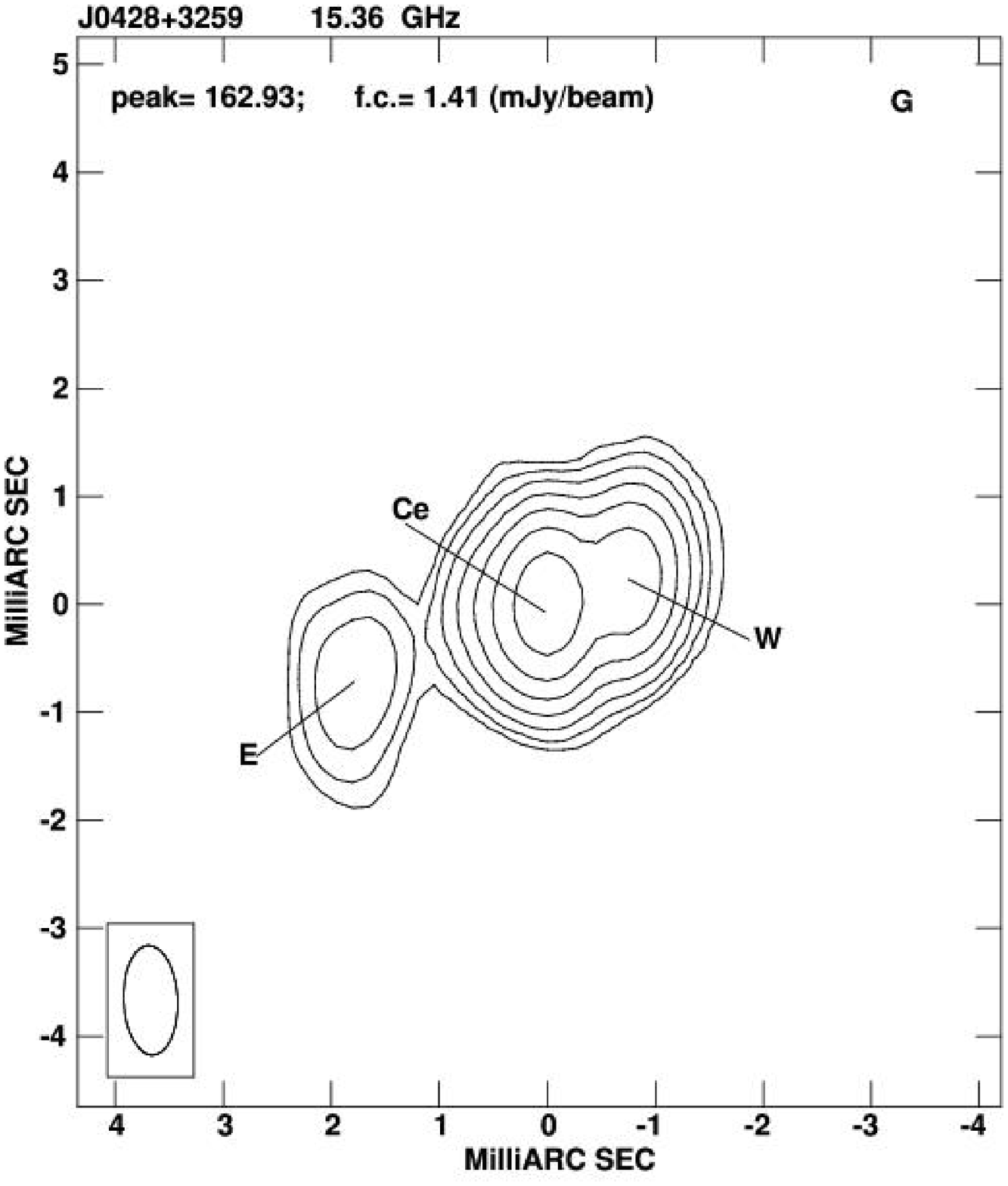}
\includegraphics{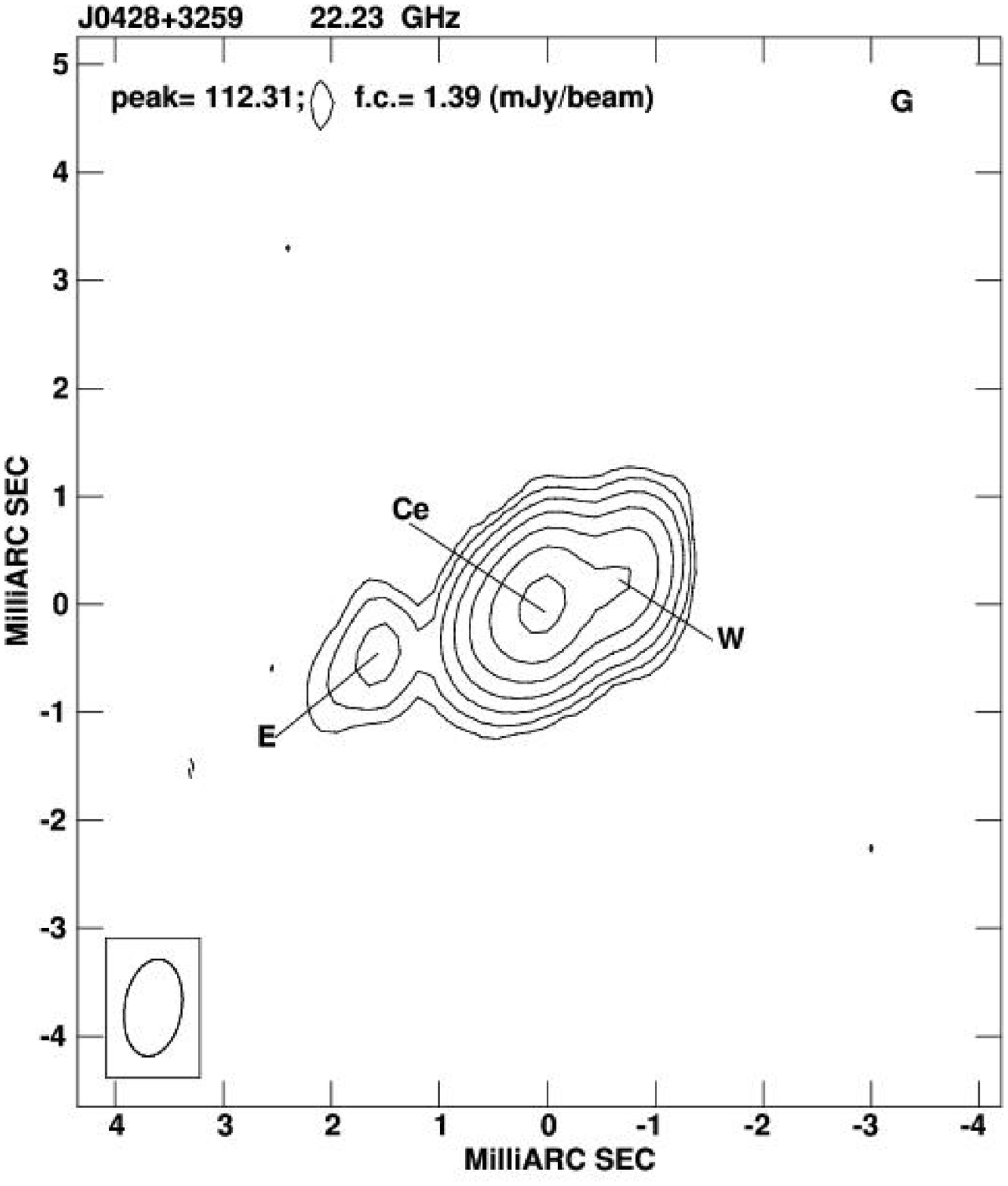}
\includegraphics{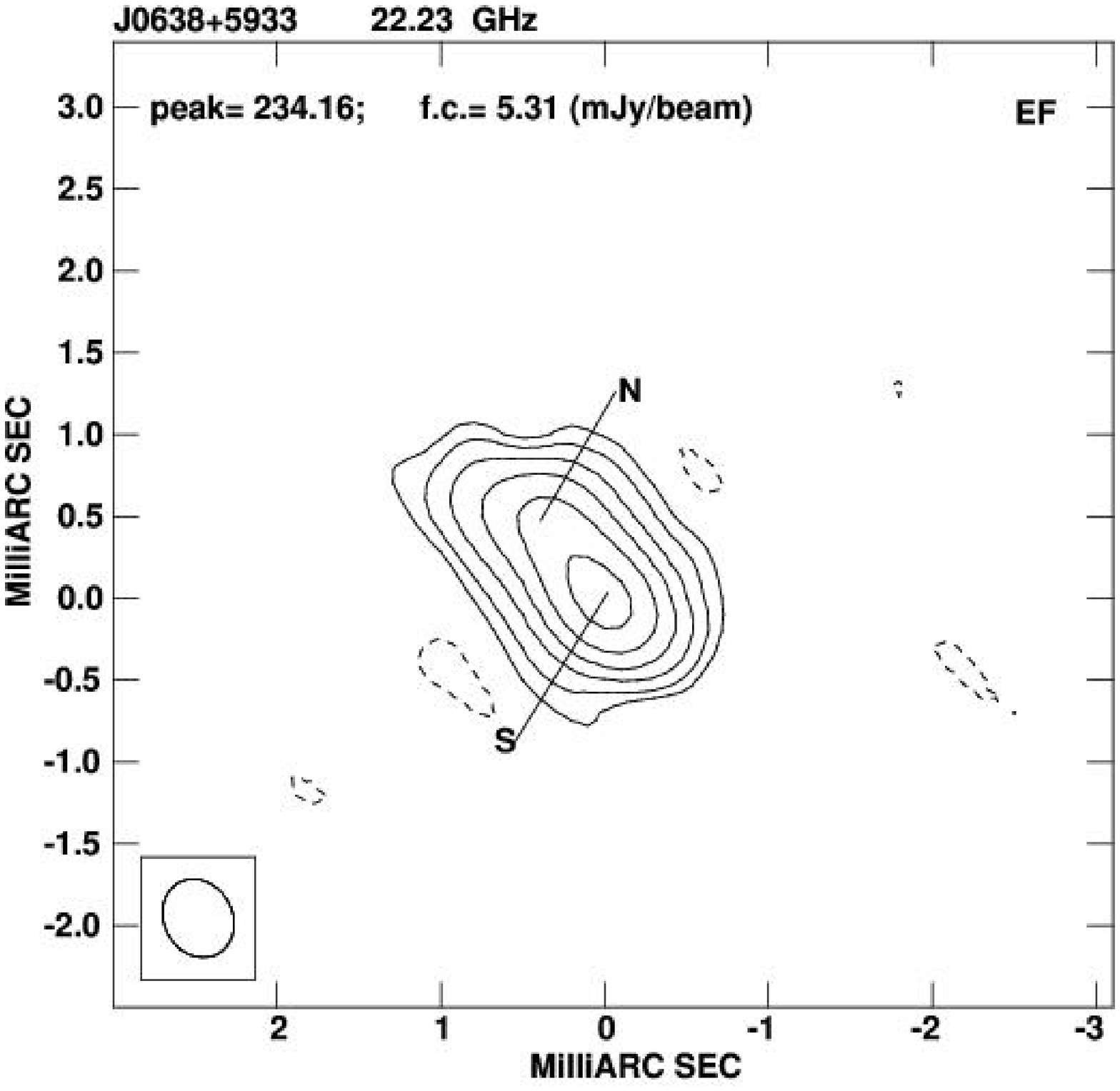}
\includegraphics{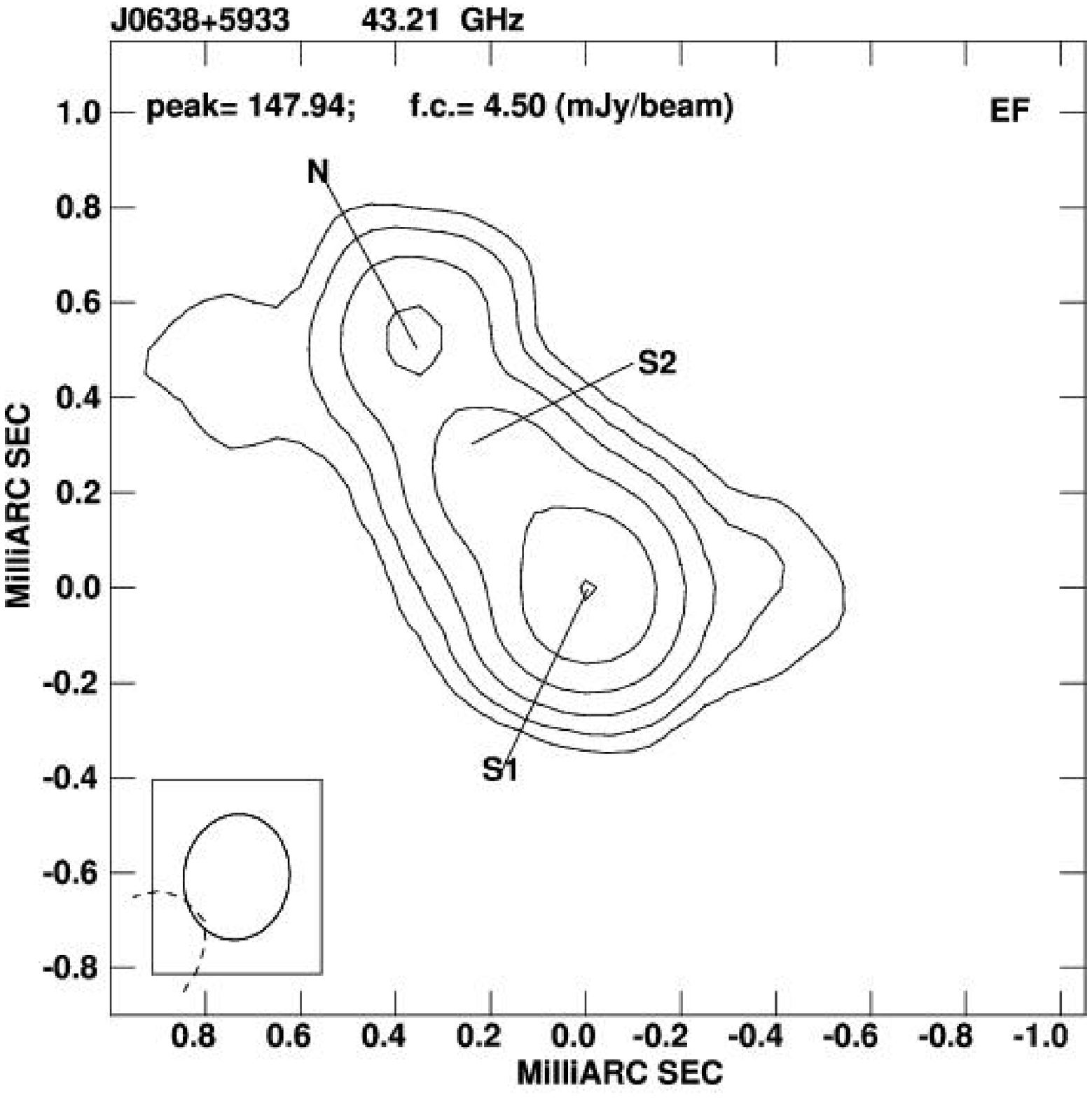}
\includegraphics{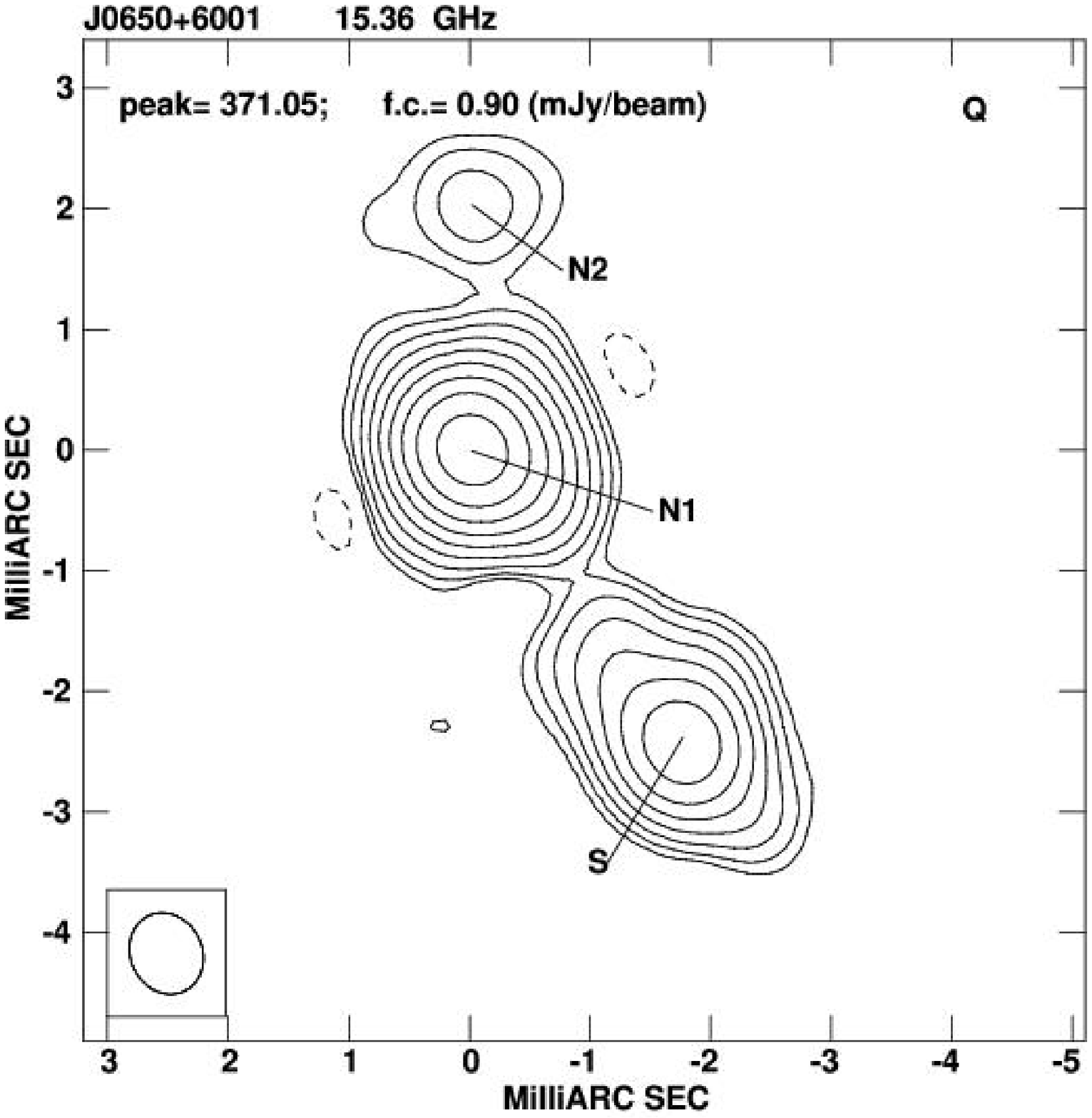}
\includegraphics{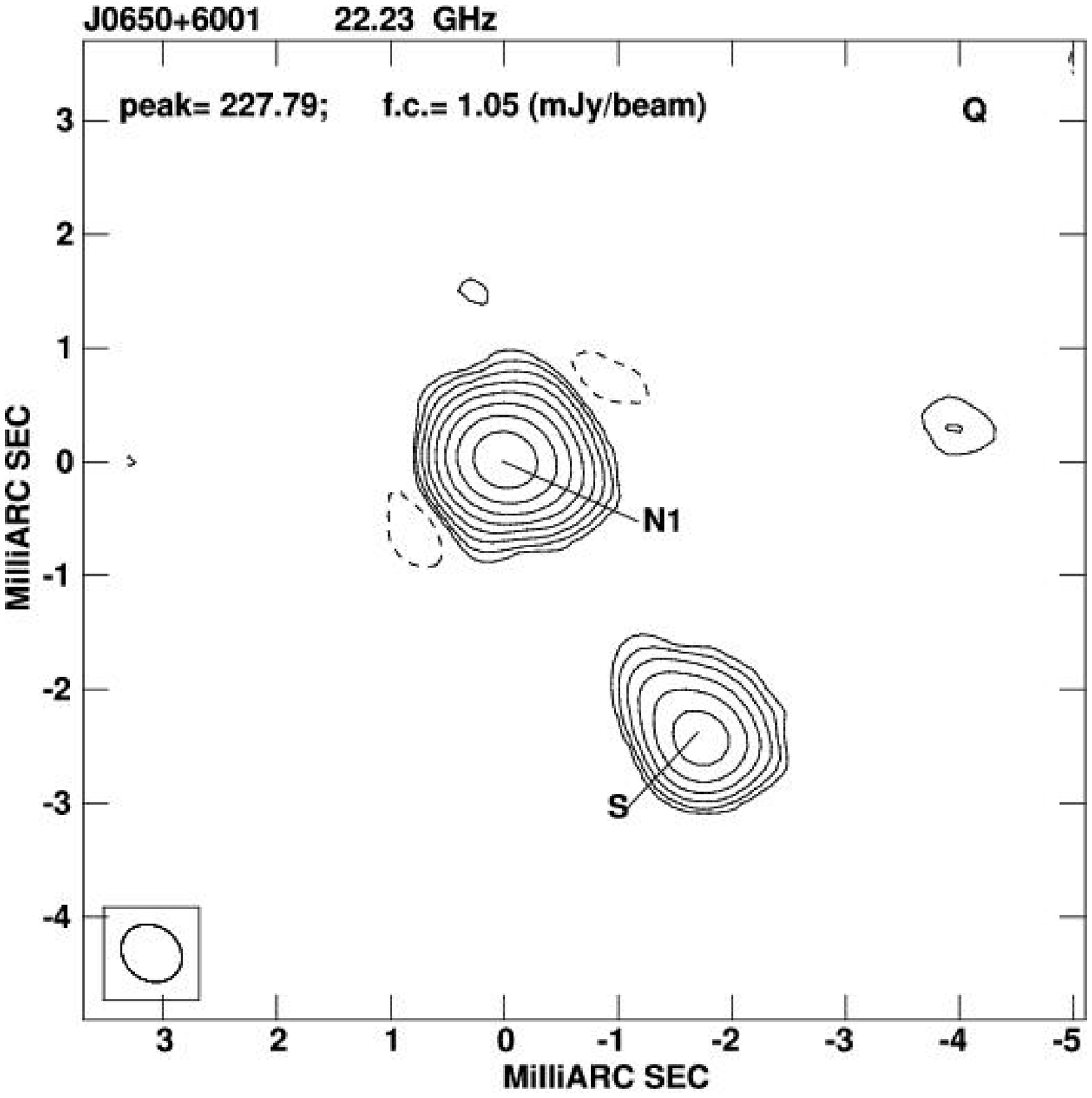}
\vspace{23.5cm}
\caption{Continued.}
\end{center}
\end{figure*}

\addtocounter{figure}{-1}
\begin{figure*} 
\begin{center}
\includegraphics{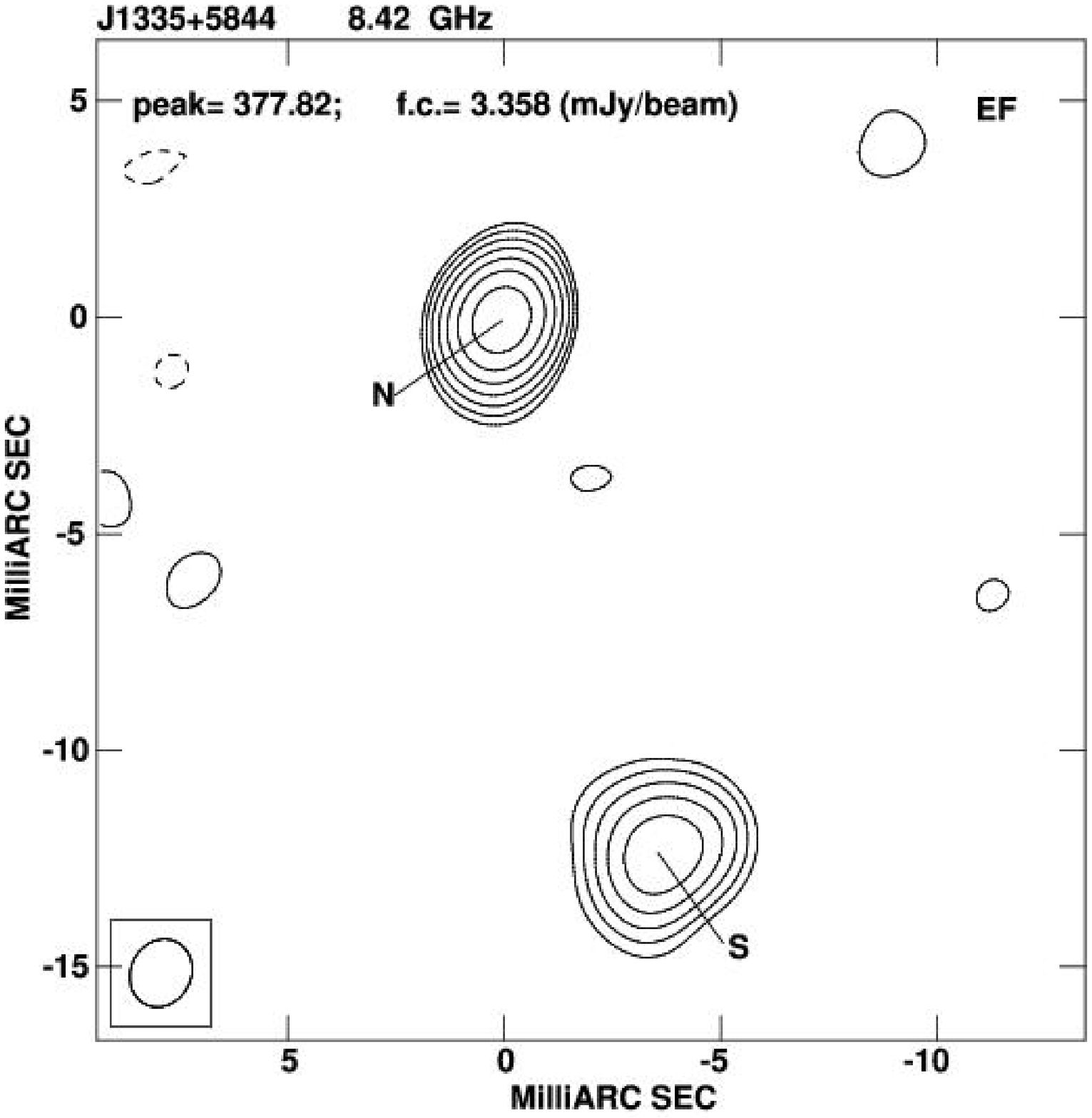}
\includegraphics{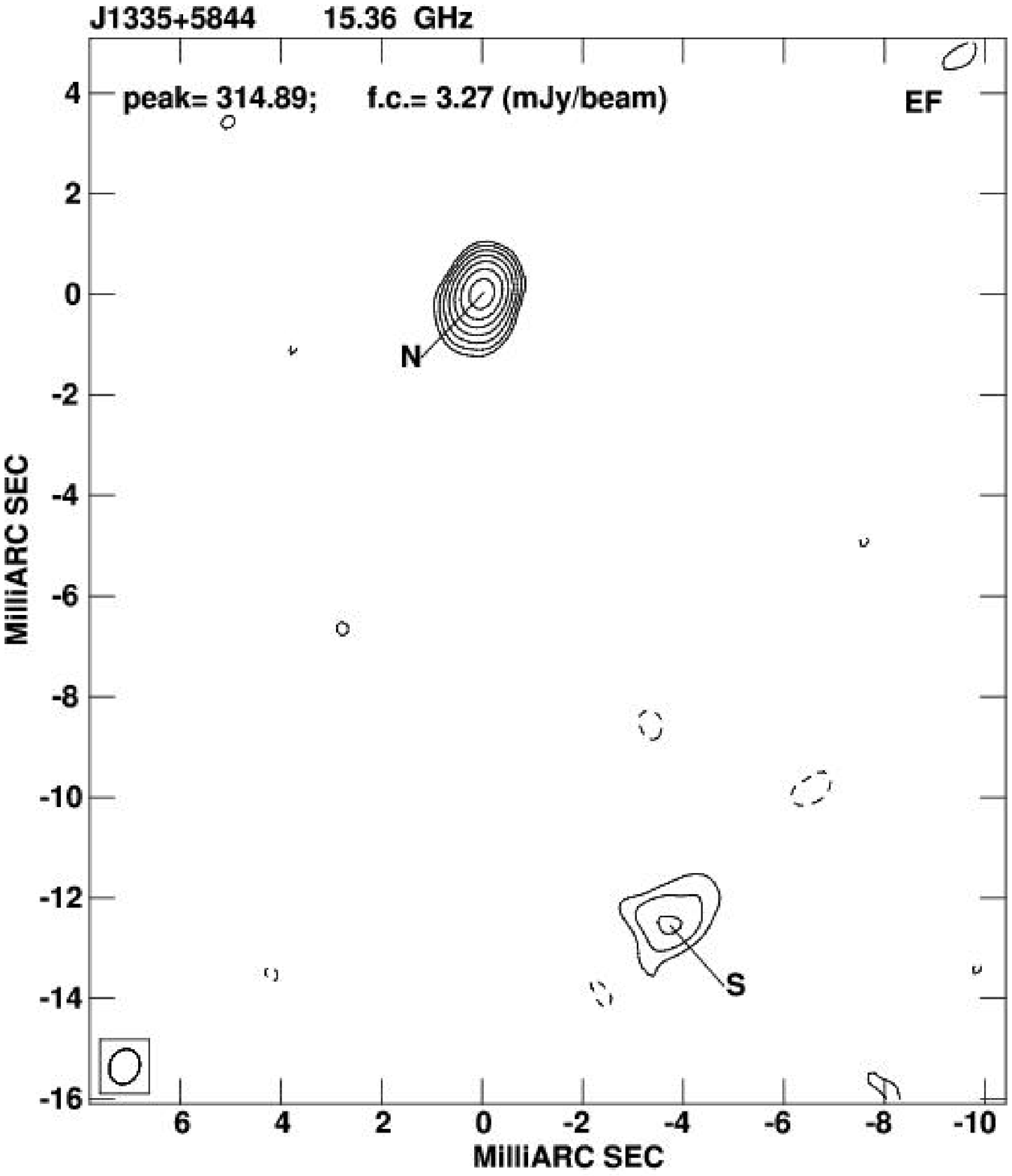}
\includegraphics{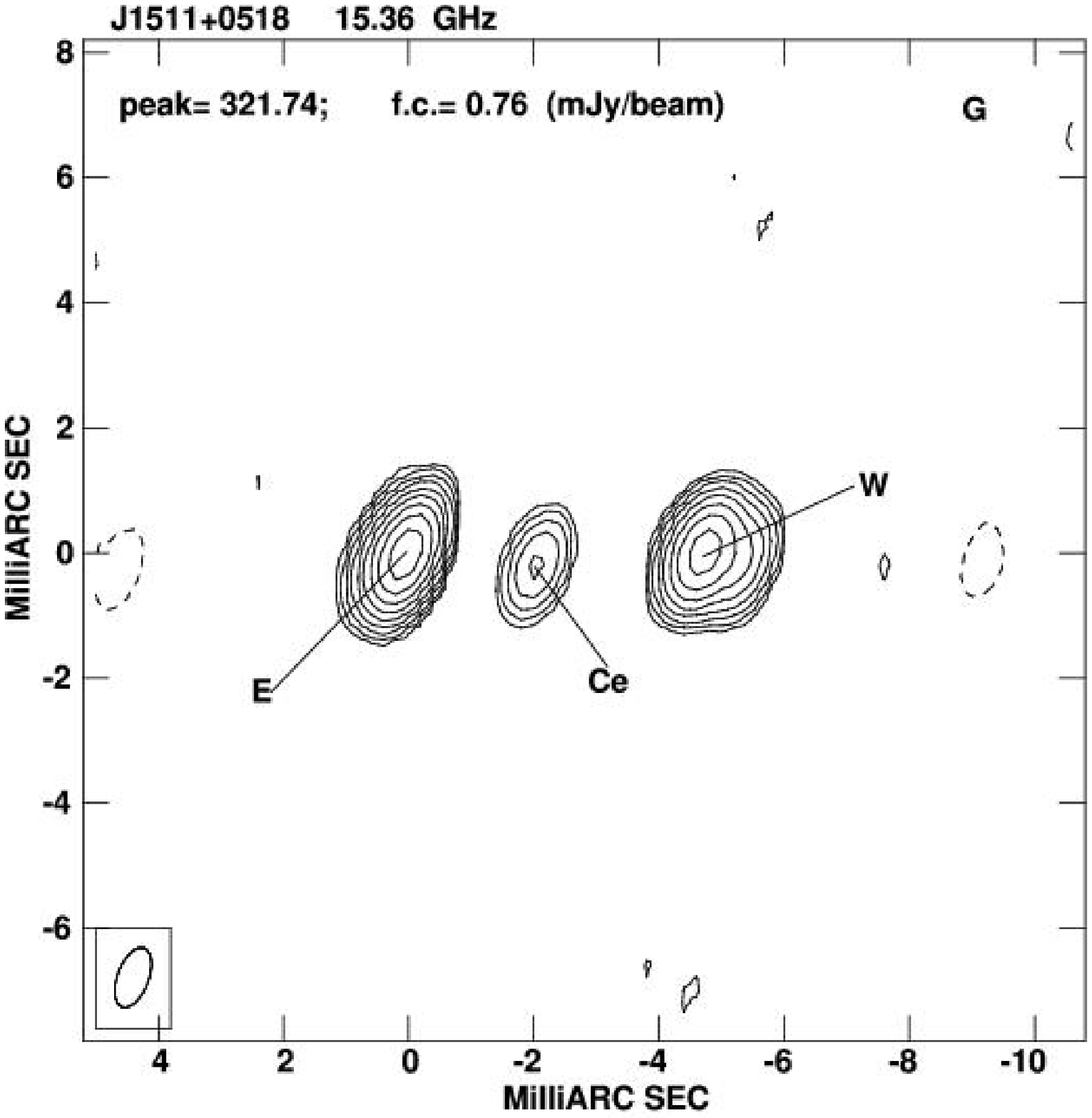}
\includegraphics{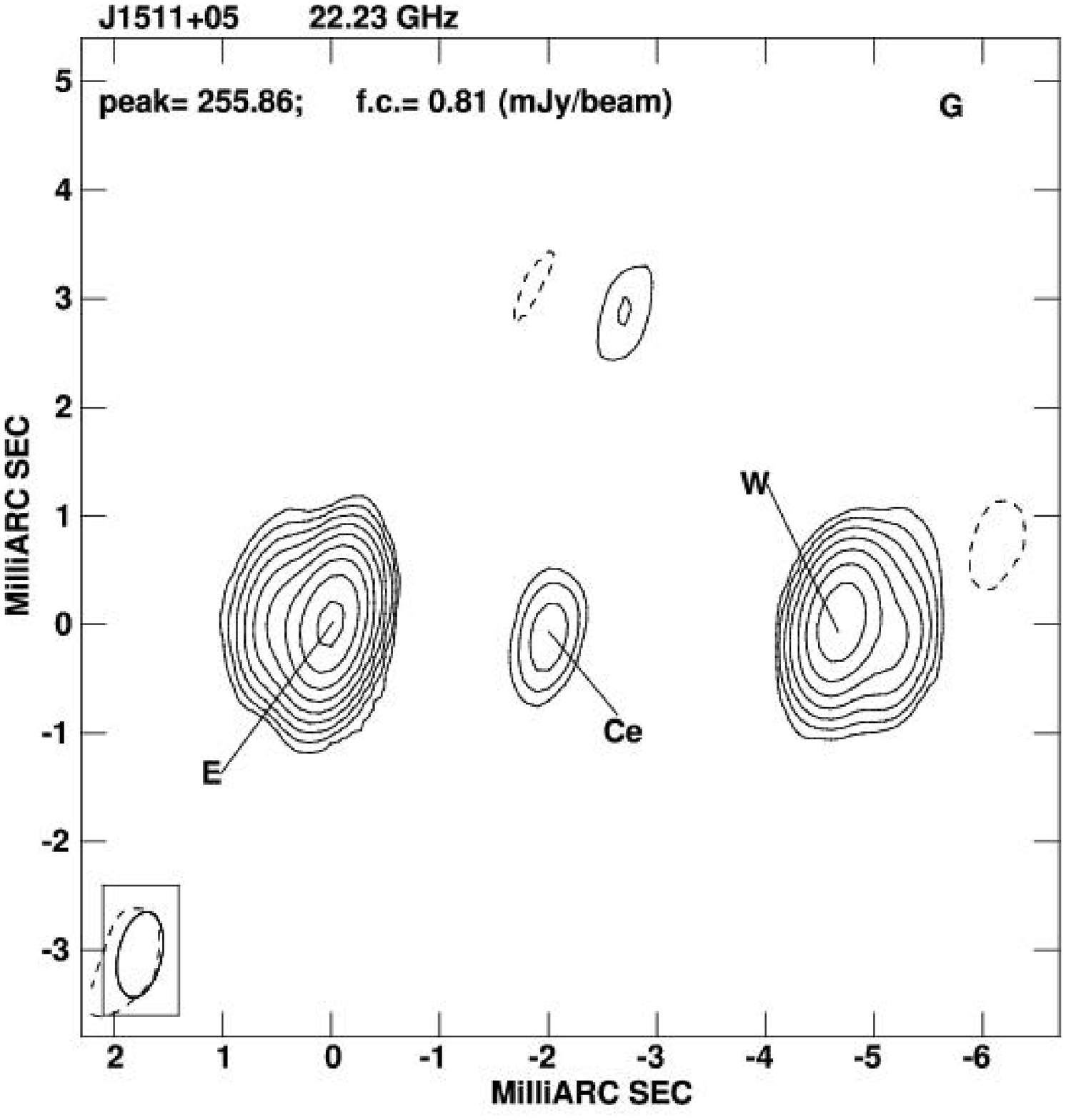}
\includegraphics{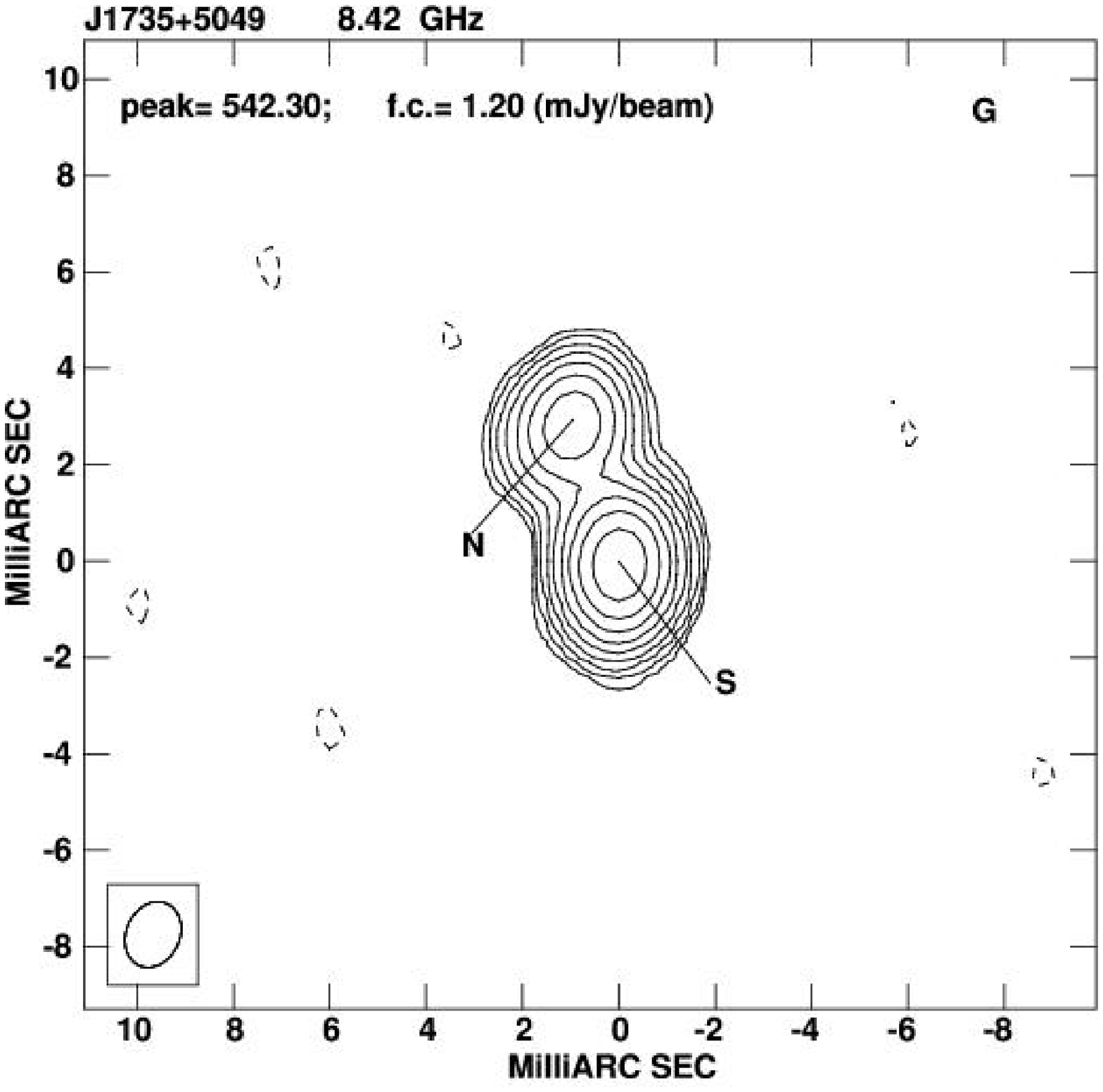}
\includegraphics{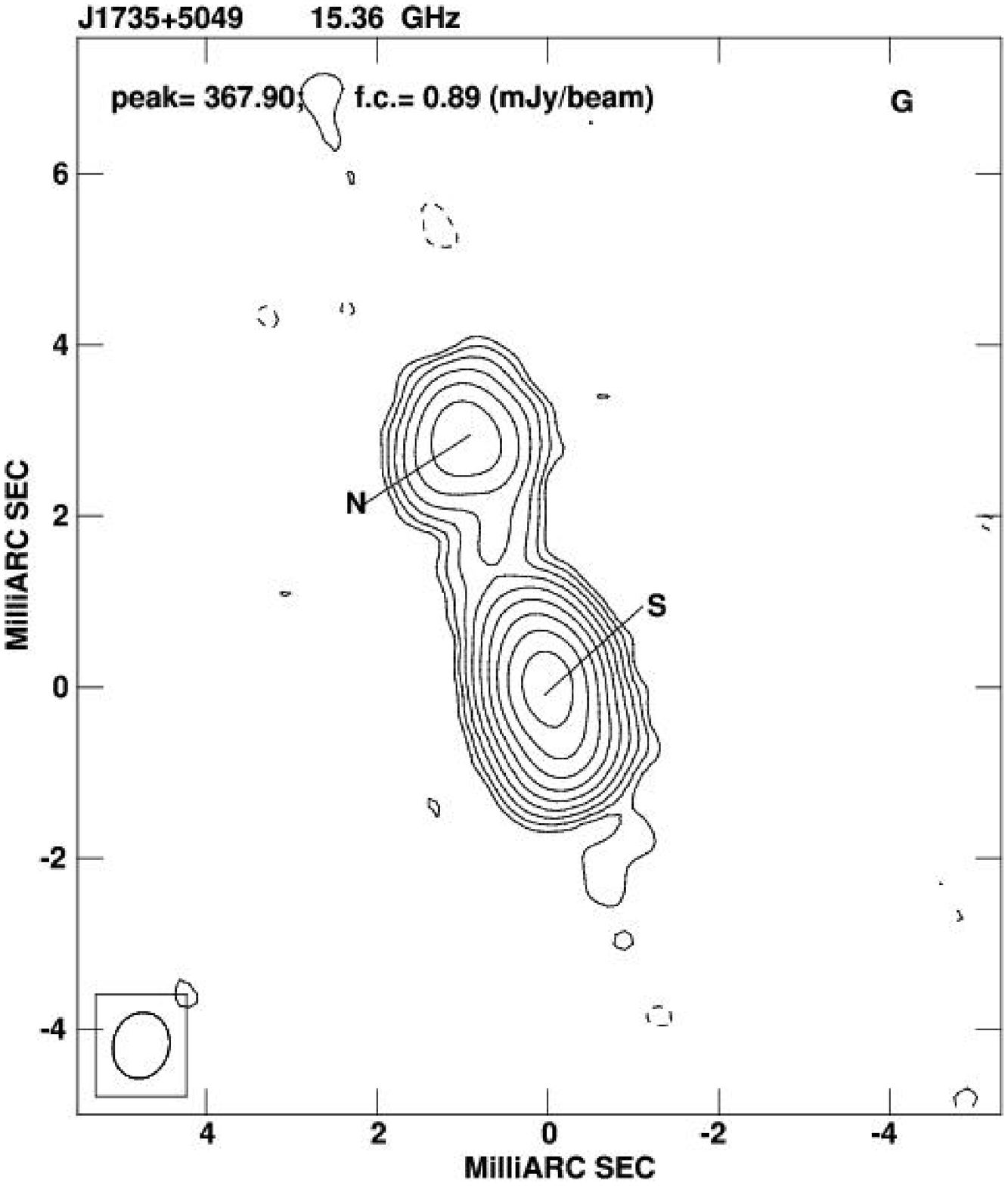}
\vspace{23.5cm}
\caption{Continued.}
\end{center}
\end{figure*}

\addtocounter{figure}{-1}
\begin{figure*} 
\begin{center}
\includegraphics{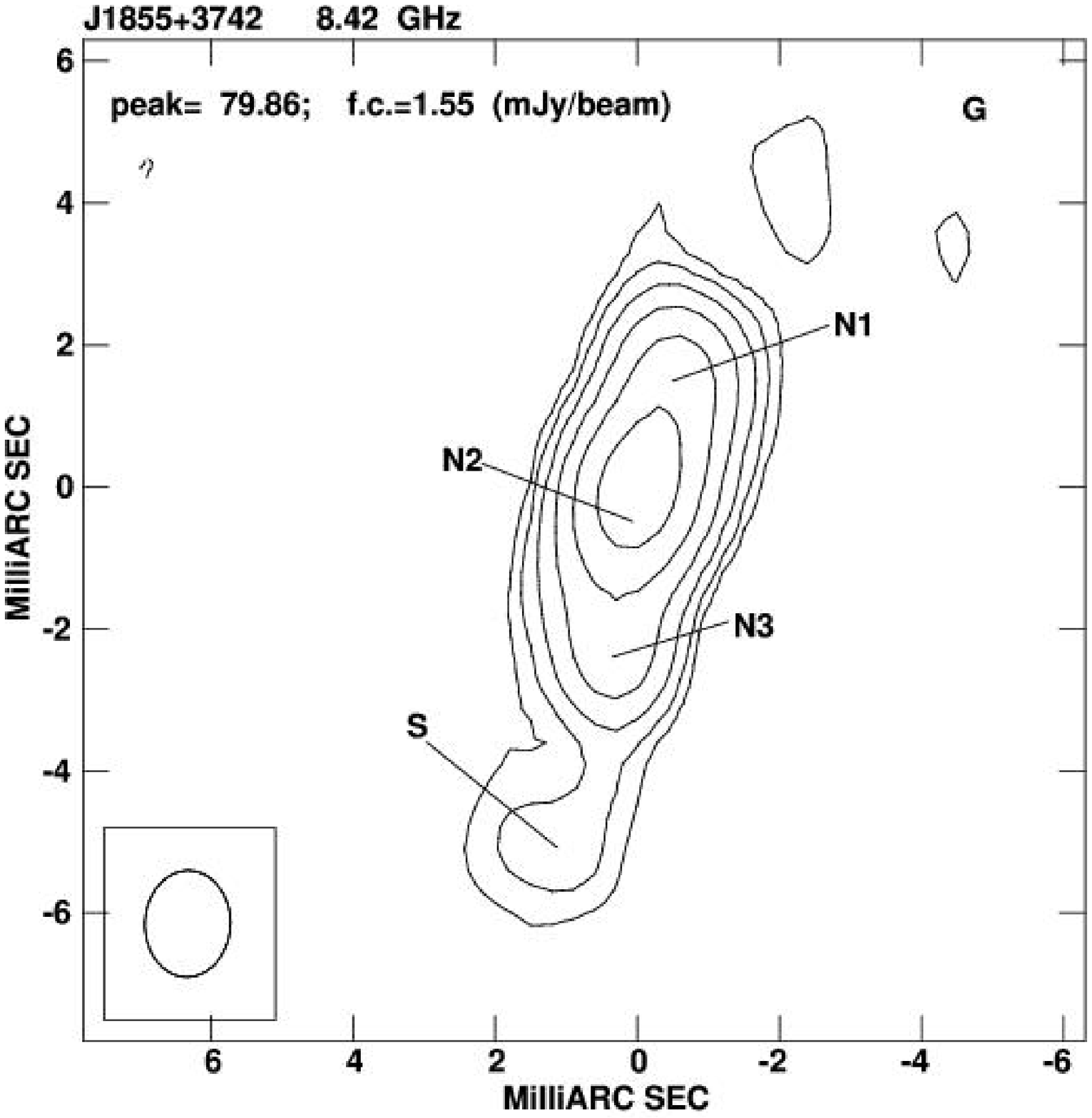}
\includegraphics{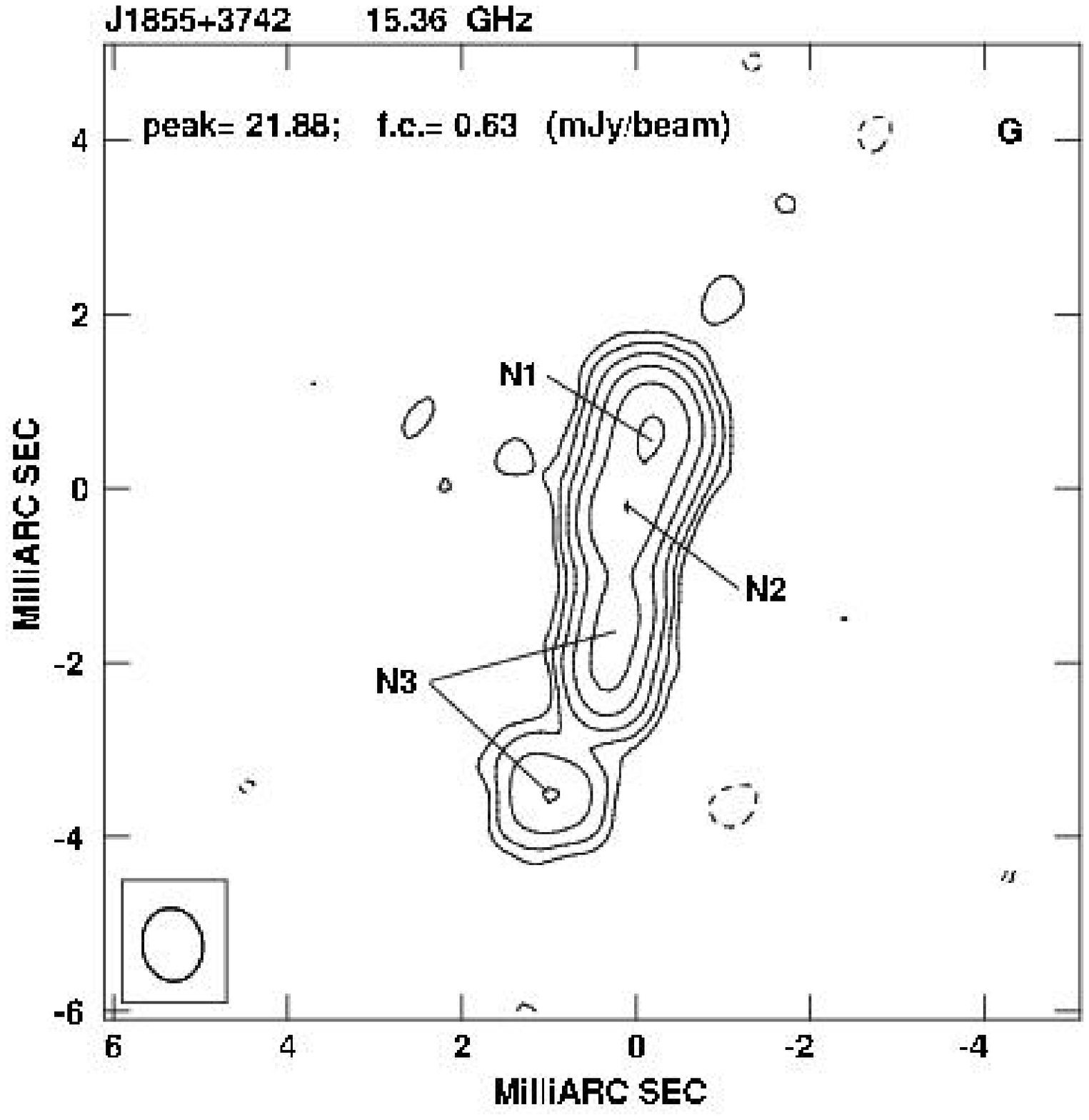}
\includegraphics{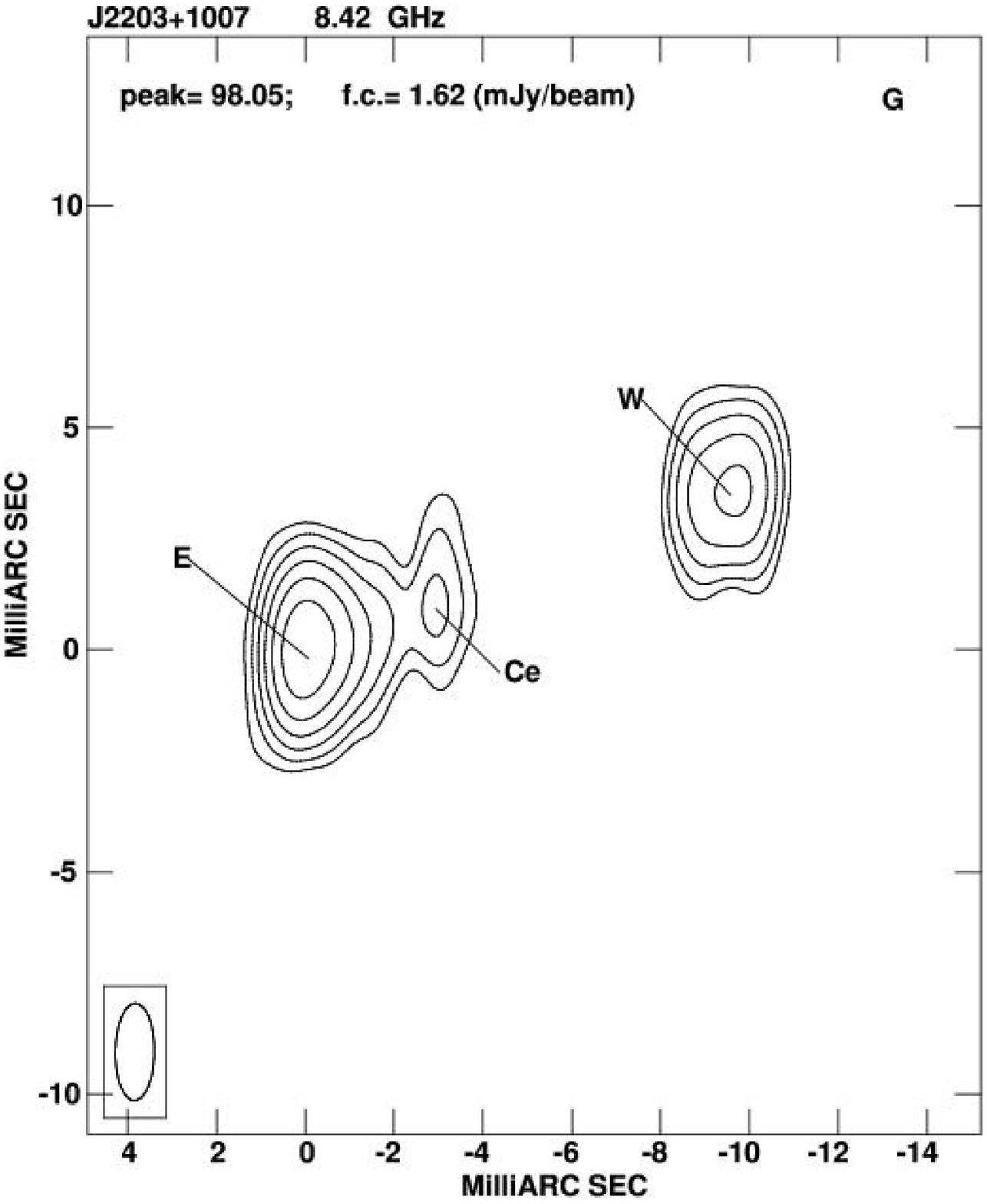}
\includegraphics{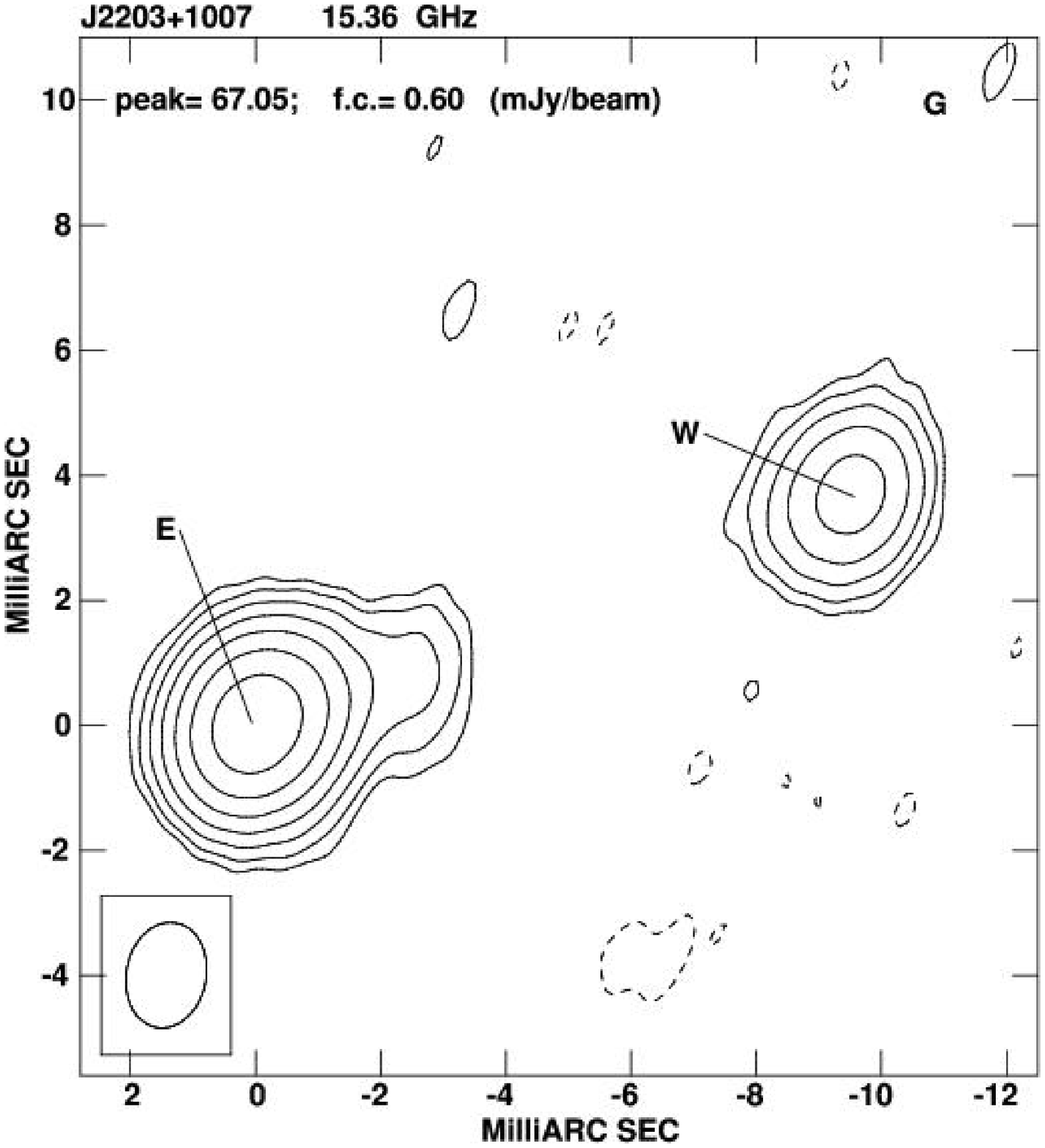}
\vspace{15.5cm}
\caption{Continued.}
\end{center}
\end{figure*}

\begin{figure} 
\begin{center}
\includegraphics{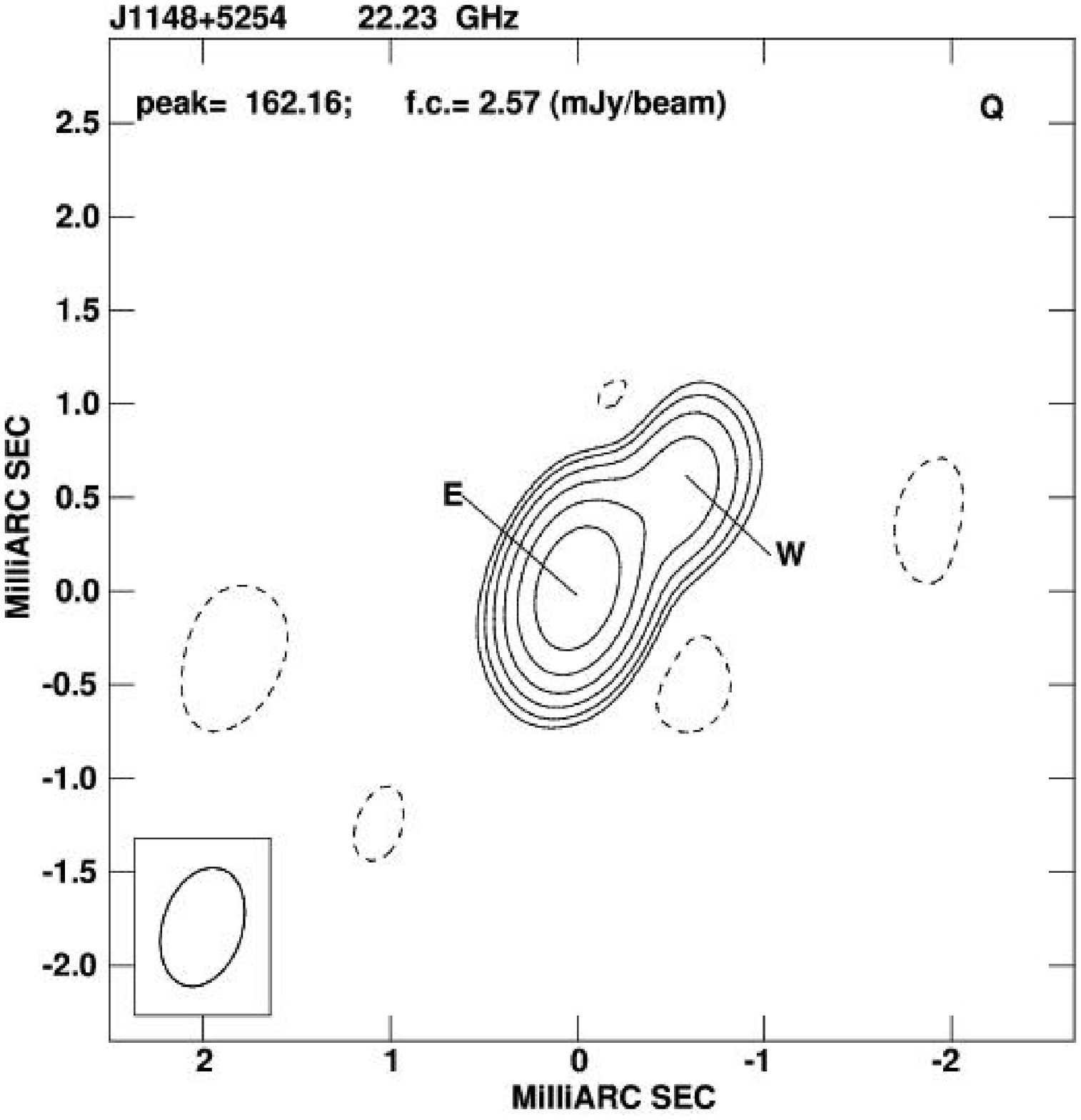}
\includegraphics{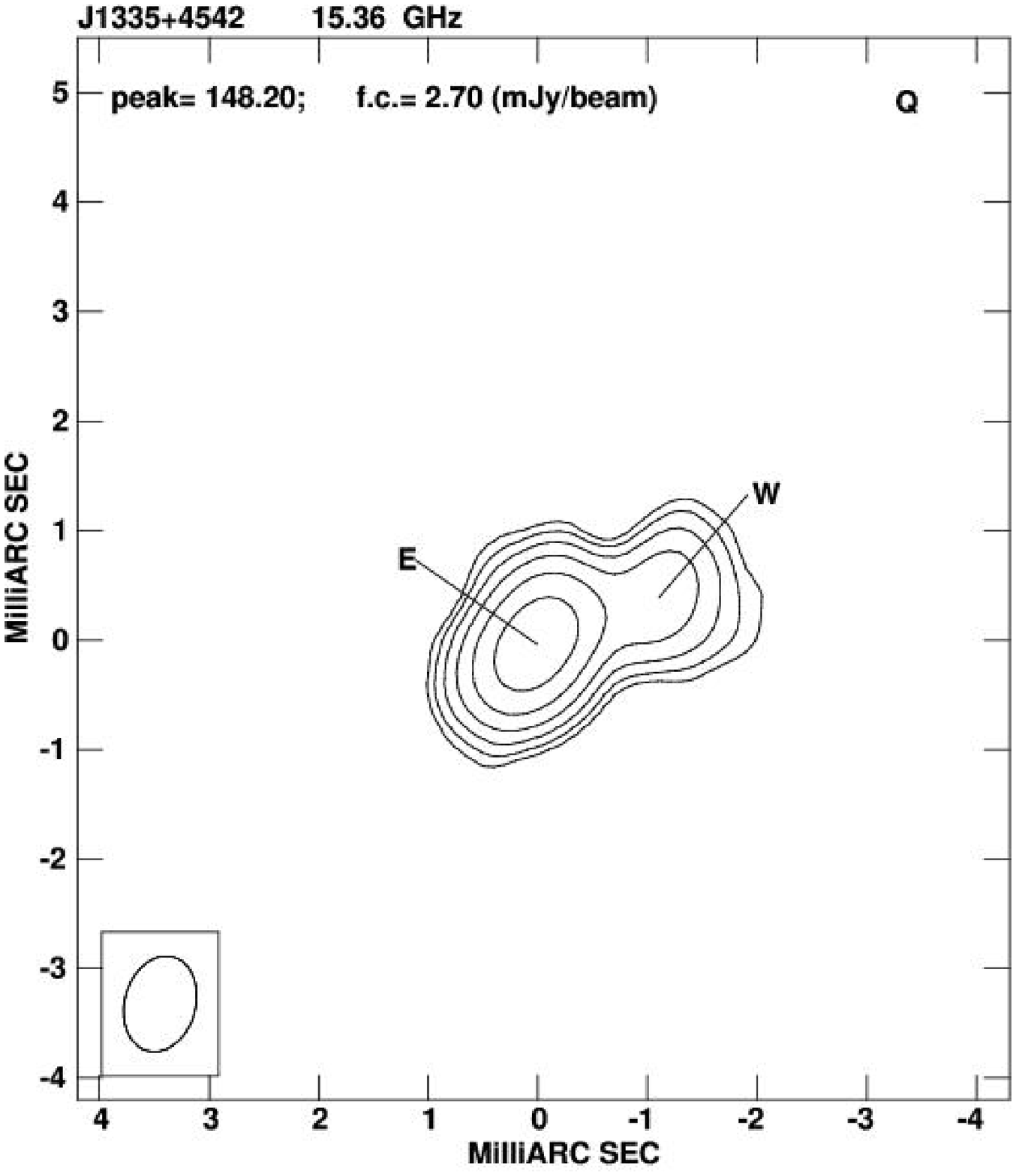}
\includegraphics{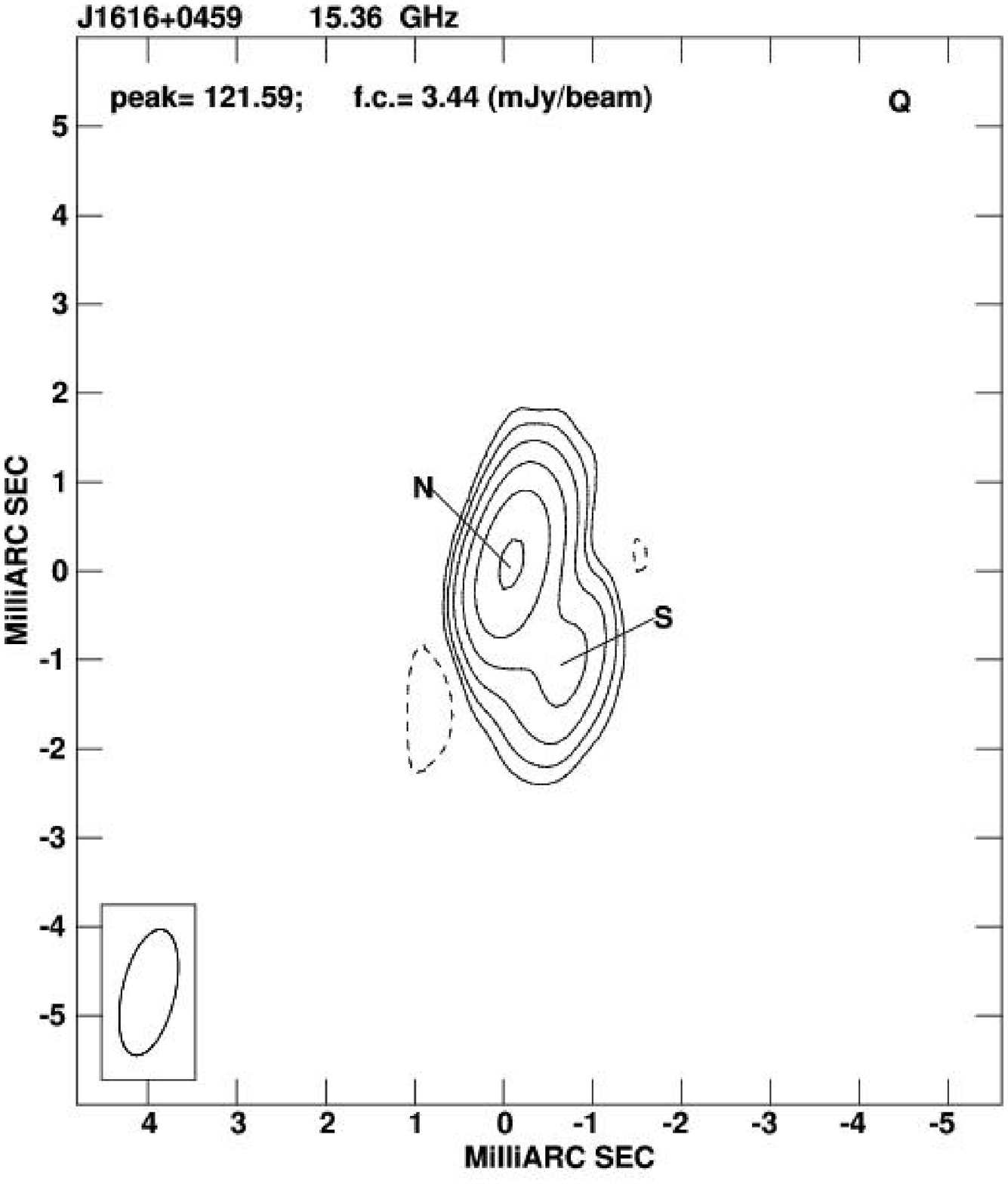}
\vspace{20.5cm}
\caption{VLBA images of the sources with a CSO-like morphology, 
which are resolved at highest frequency only. For each image we give the following information
  on the plot itself: a) Peak flux density in mJy/beam; b) First
  contour intensity ({\it f.c.}, in mJy/beam), which is generally 3 times
  the r.m.s. noise on the image plane; contour levels increase by a
  factor of 2; c) the optical identification; 
  d) the restoring beam is plotted on the bottom left
  corner of each image. }
\label{CSO-1f}
\end{center}
\end{figure}

\subsection{Comments on individual sources}

\begin{figure*} 
\begin{center}
\includegraphics{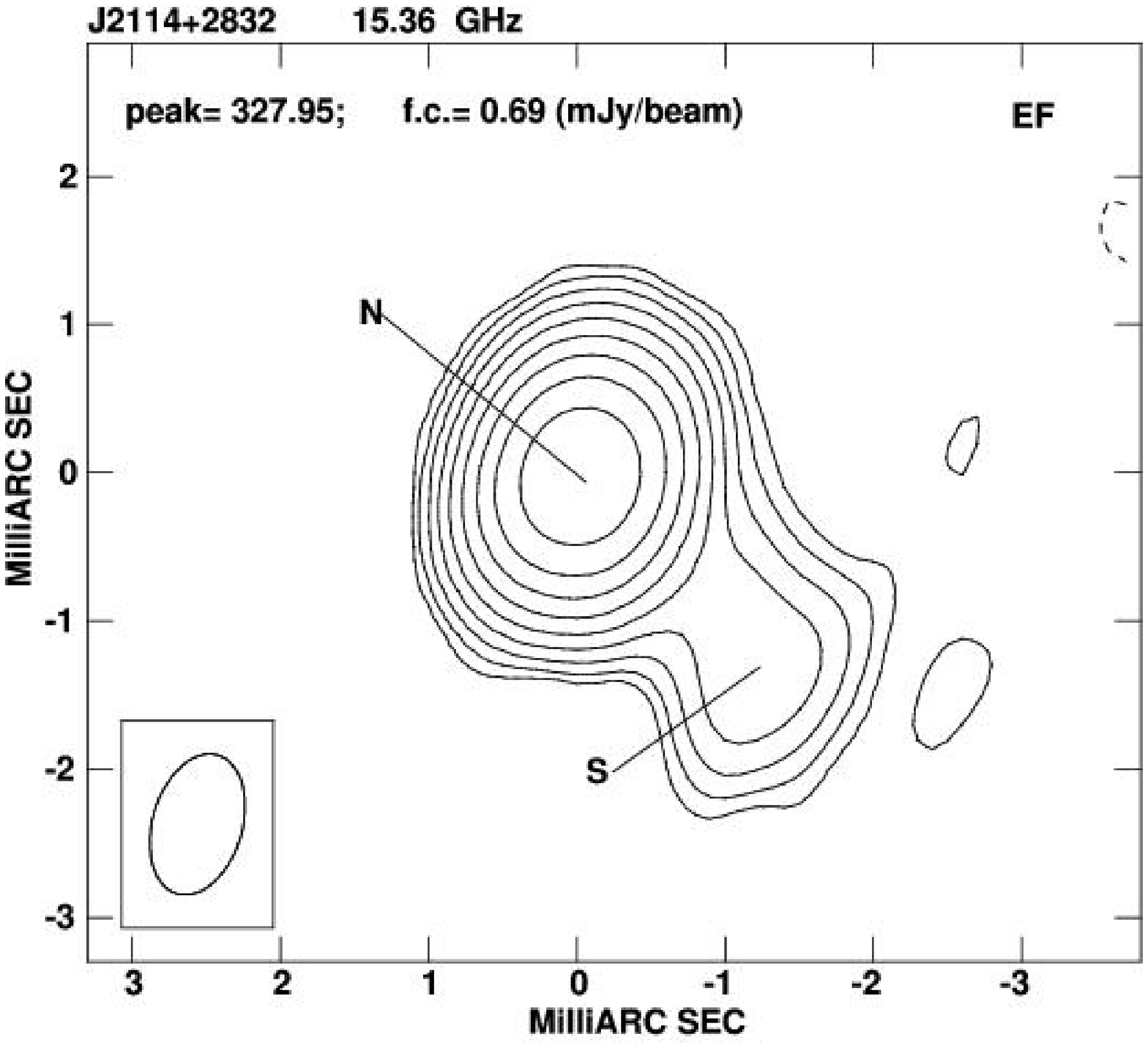}
\includegraphics{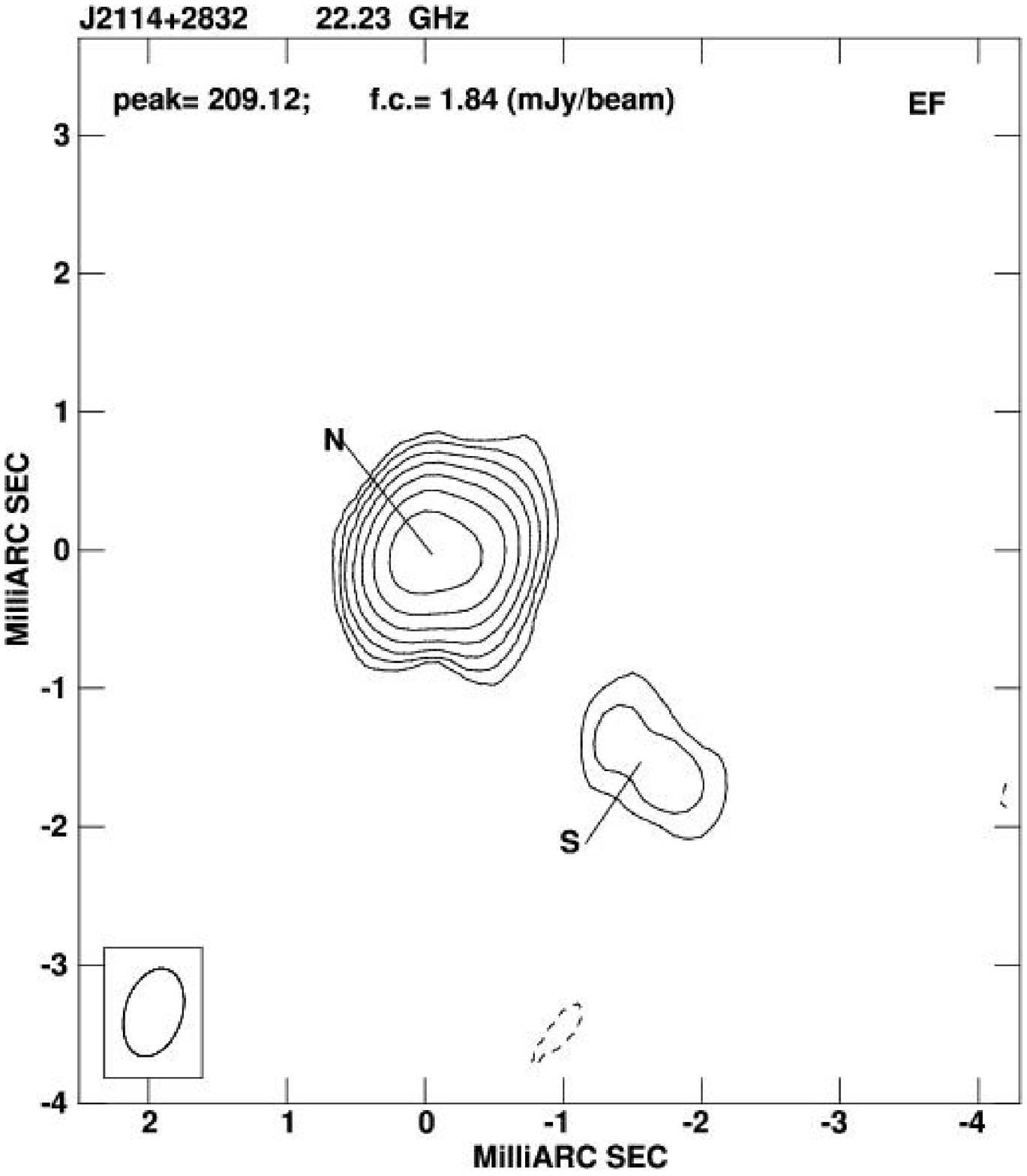}
\includegraphics{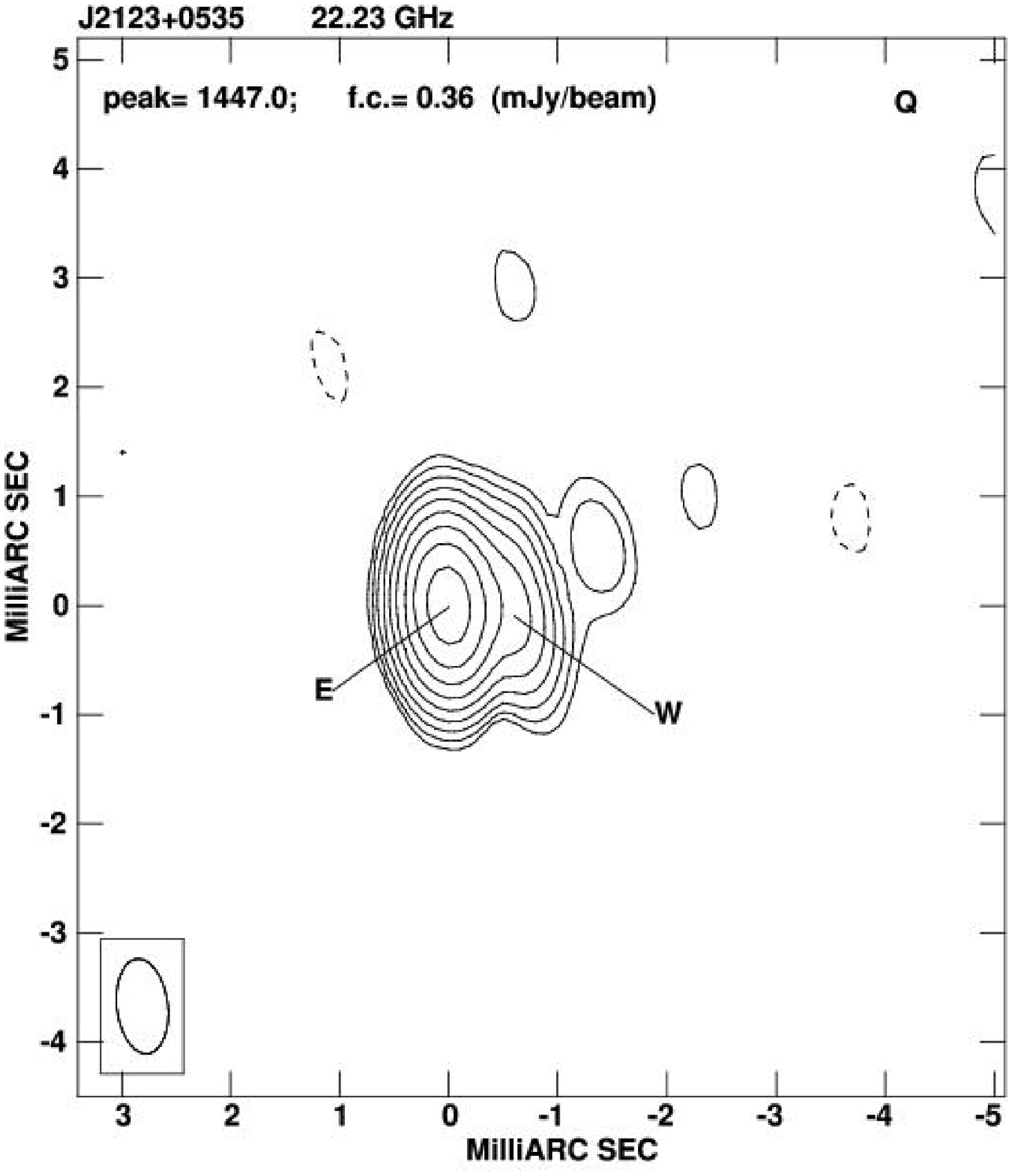}
\includegraphics{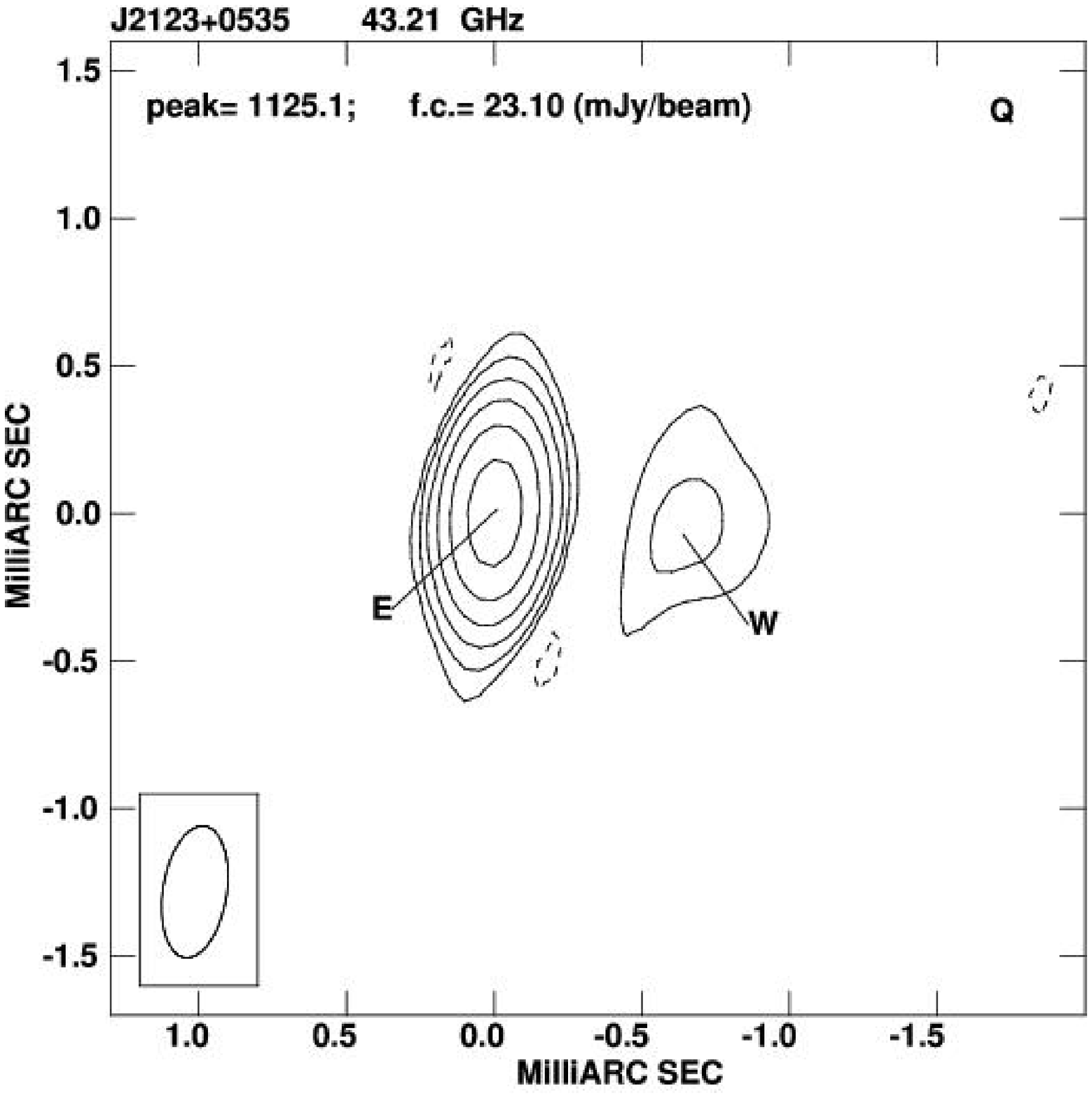}
\includegraphics{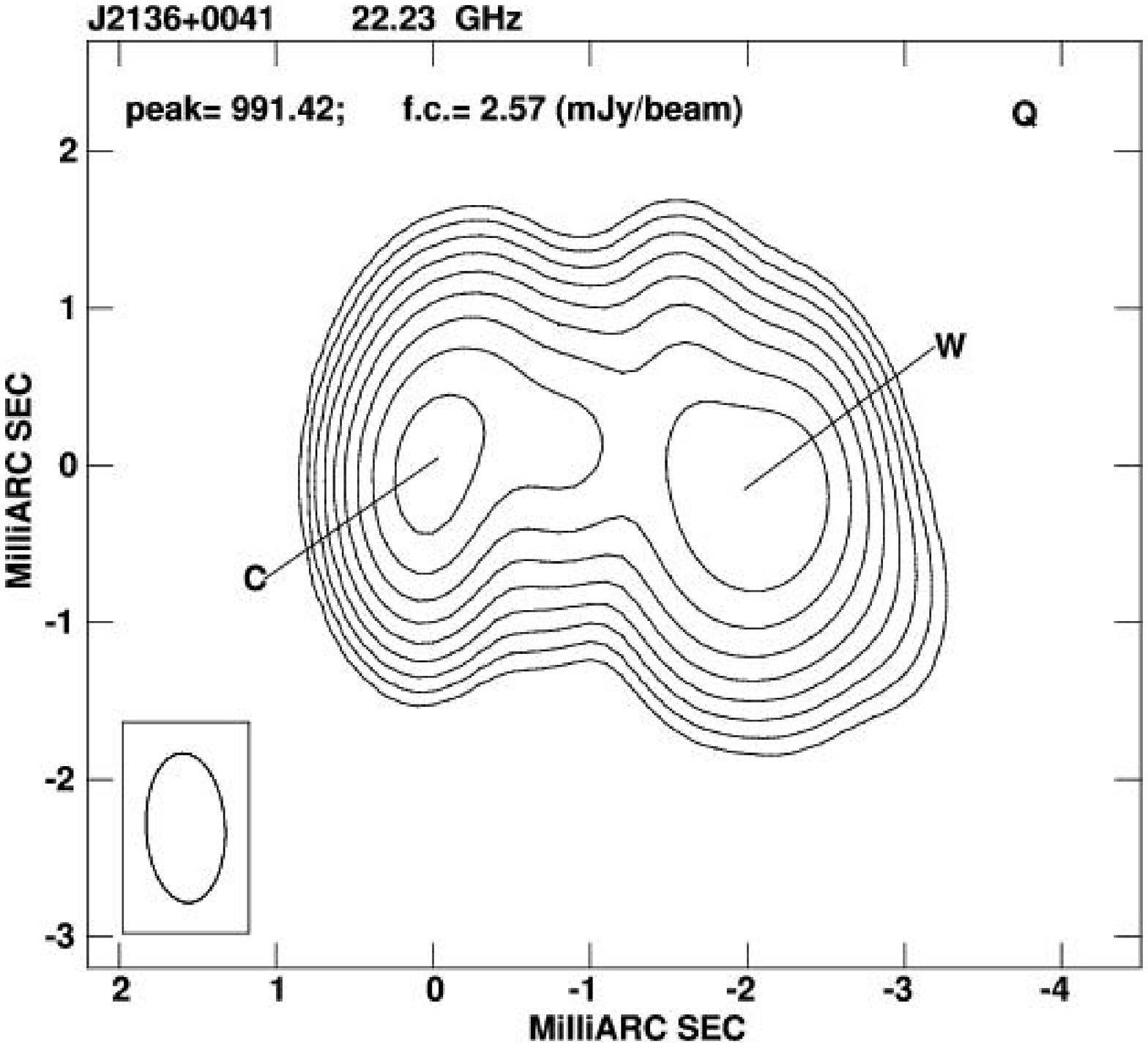}
\includegraphics{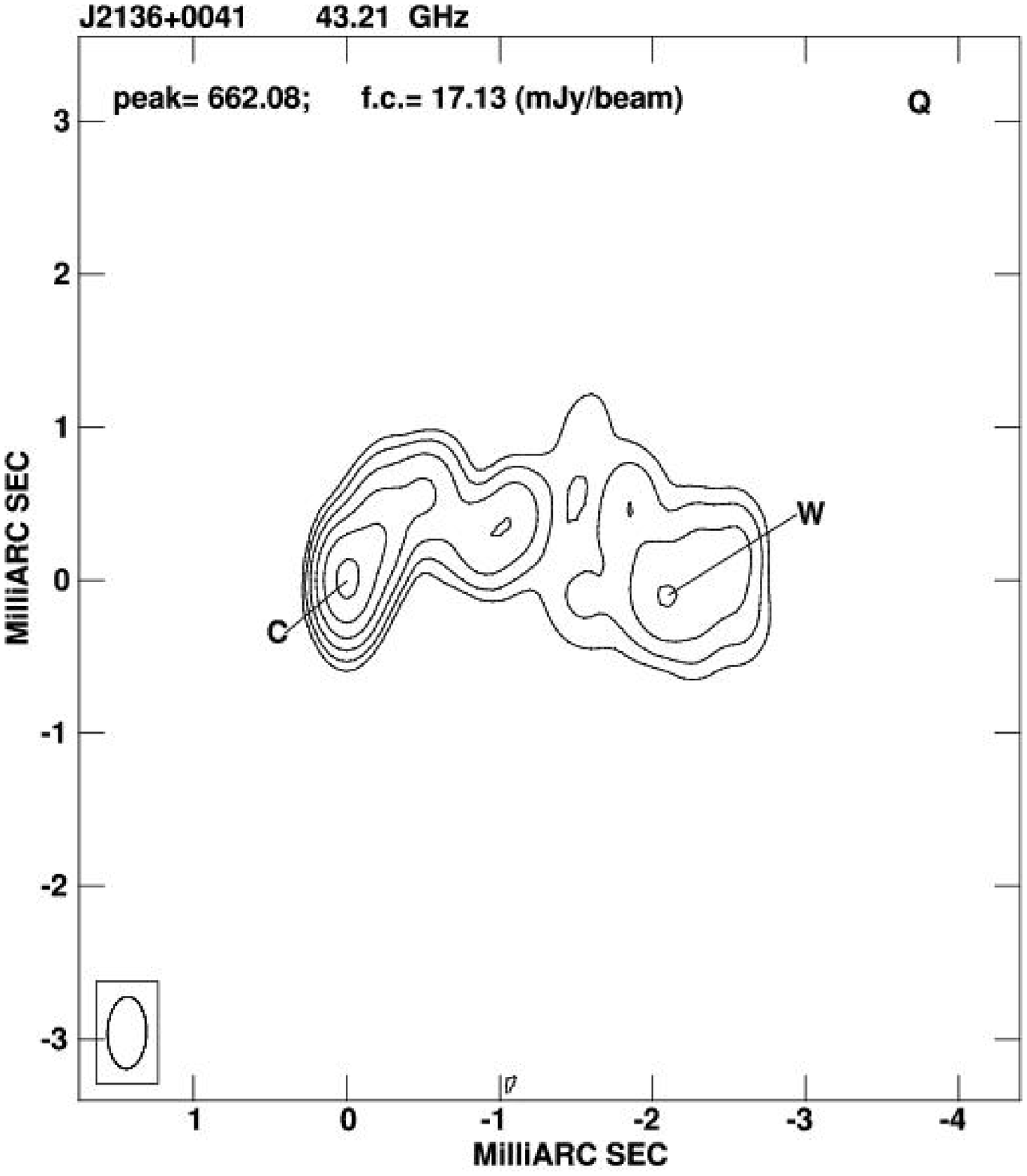}
\vspace{22.5cm}
\caption{VLBA images at the two frequencies of the sources with a 
Core-Jet morphology. For each image we give the following information
  on the plot itself: a) Peak flux density in mJy/beam; b) First
  contour intensity ({\it f.c.}, in mJy/beam), which is generally 3 times
  the r.m.s. noise on the image plane; contour levels increase by a
  factor of 2; c) the optical identification; 
  d) the restoring beam is plotted on the bottom left
  corner of each image.}
\label{jet}
\end{center}
\end{figure*}

\begin{figure} 
\begin{center}
\includegraphics{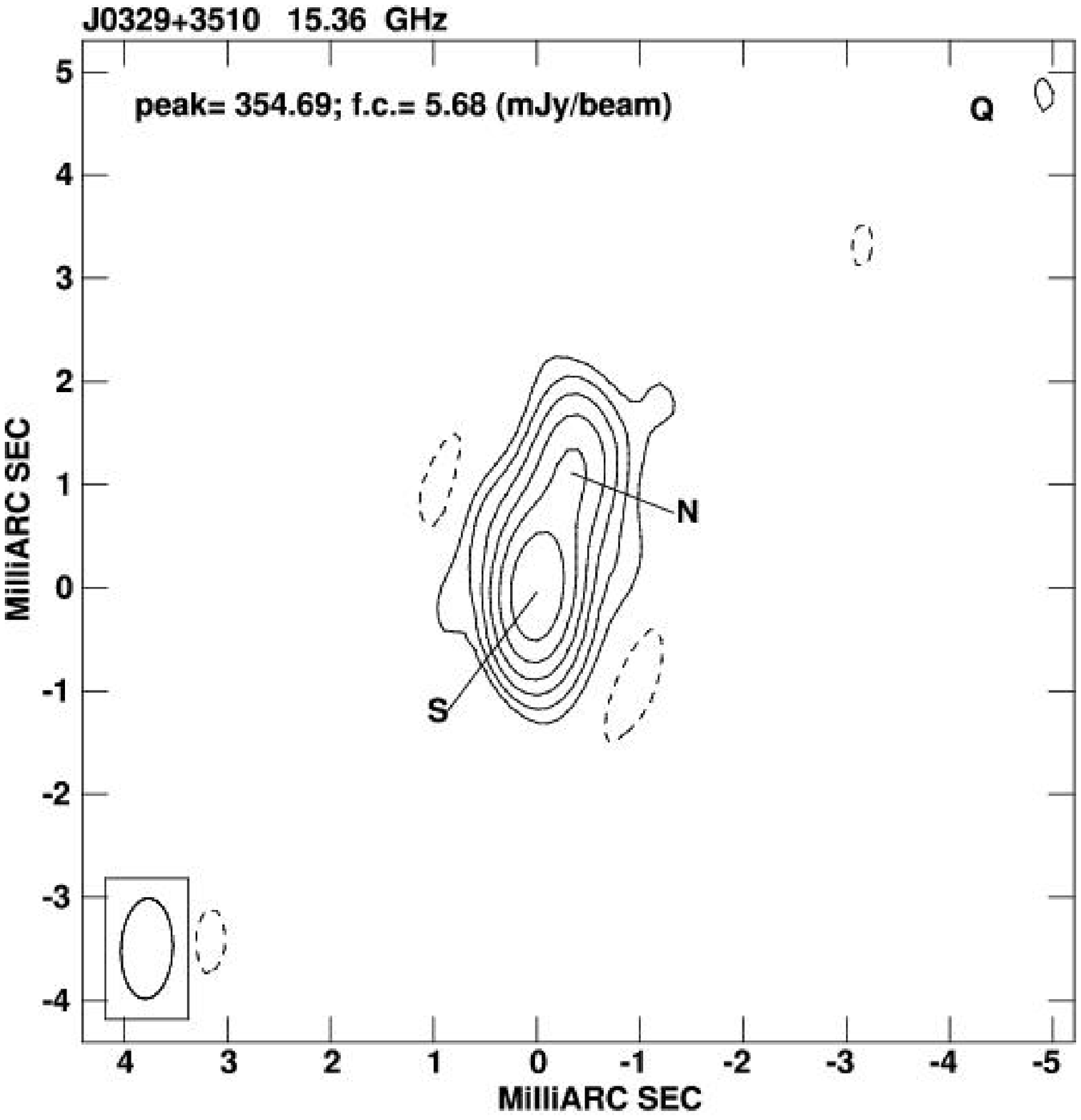}
\includegraphics{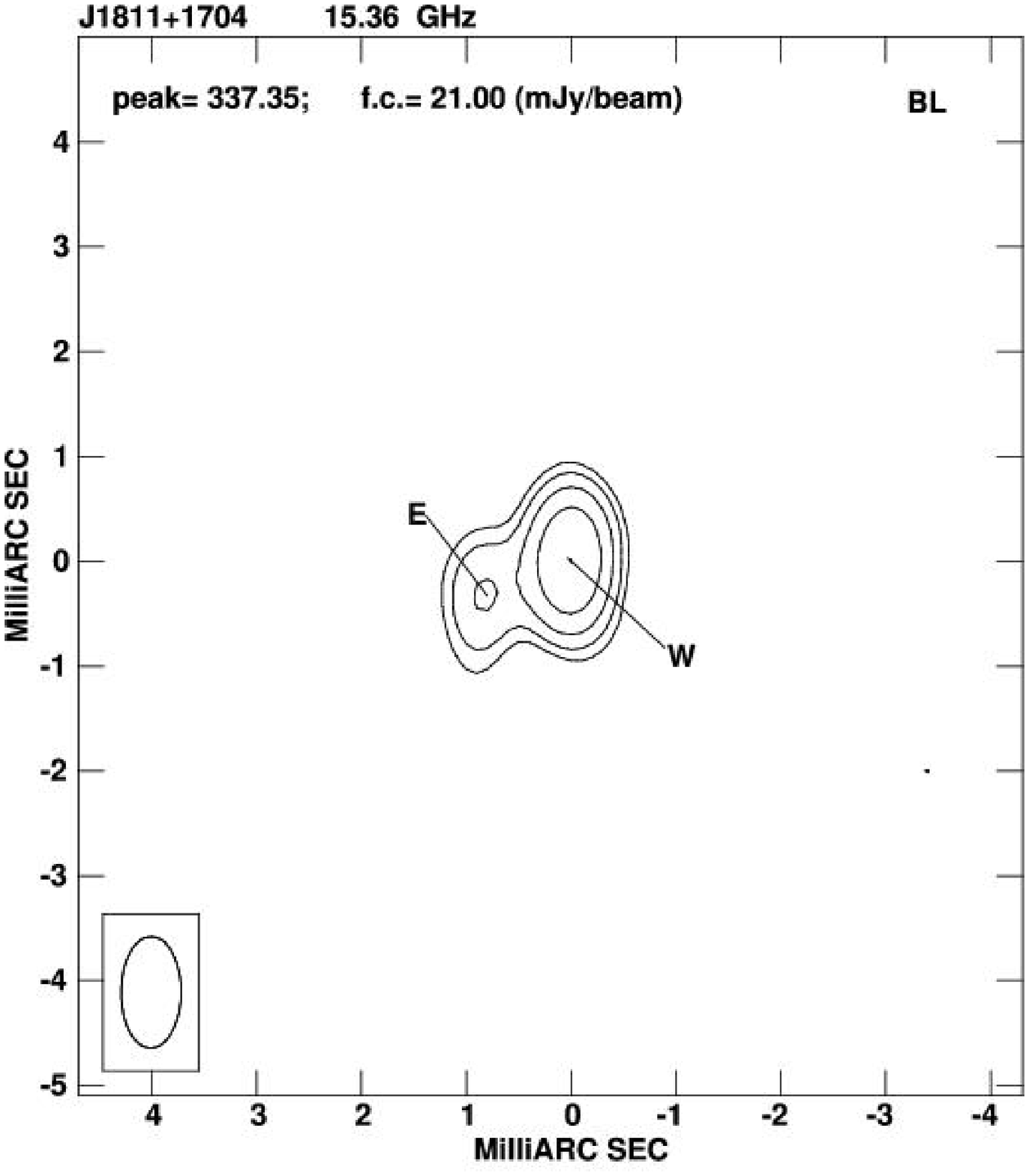}
\includegraphics{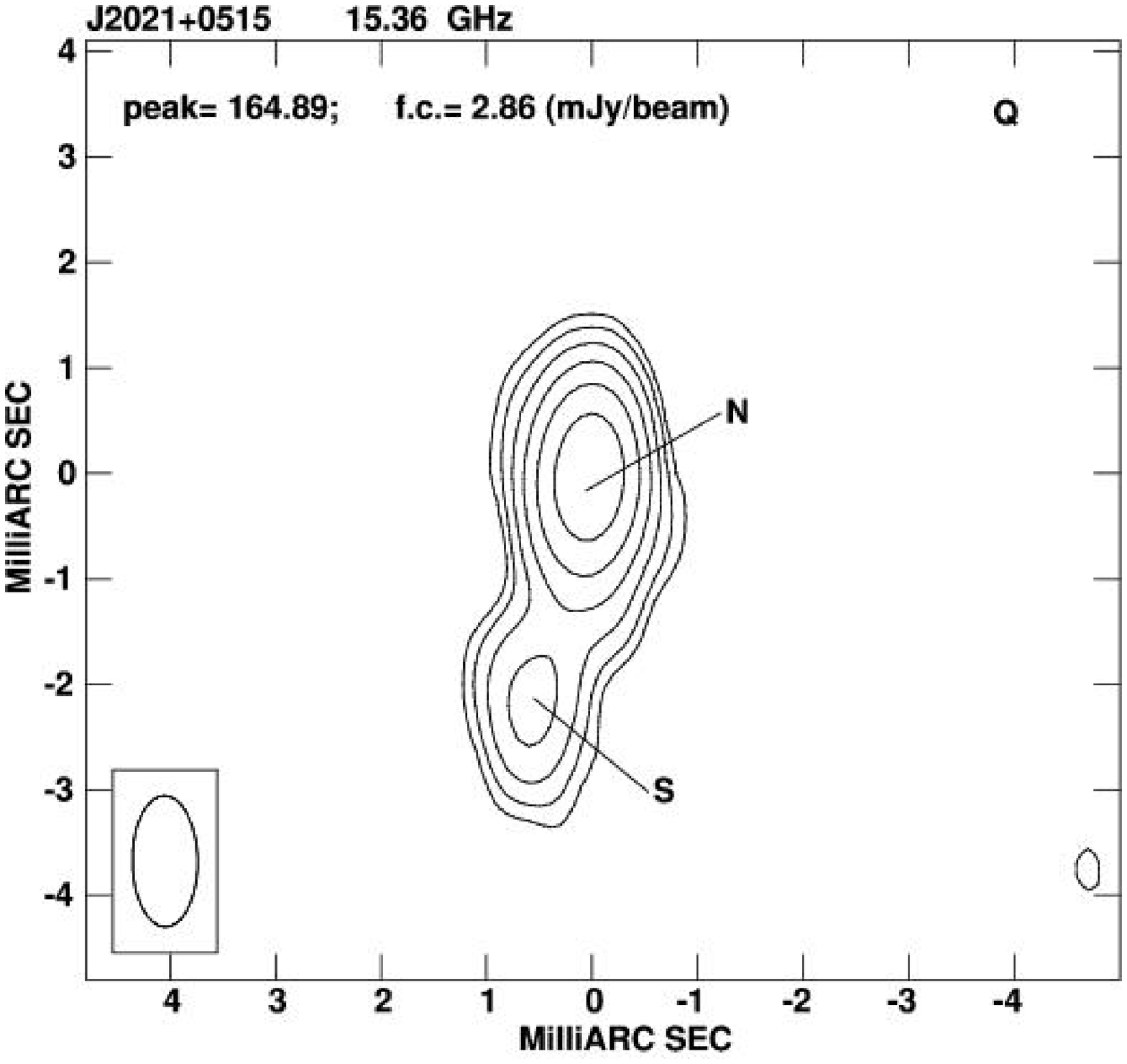}
\vspace{20.5cm}
\caption{VLBA images of the sources with a Core-Jet morphology, 
which are resolved at highest frequency only. For each image we give the following information
  on the plot itself: a) Peak flux density in mJy/beam; b) First
  contour intensity ({\it f.c.}, in mJy/beam), which is generally 3 times
  the r.m.s. noise on the image plane; contour levels increase by a
  factor of 2; c) the optical identification; 
  d) the restoring beam is plotted on the bottom left
  corner of each image. }
\label{jet-1f}
\end{center}
\end{figure}

In this section we discuss in detail the properties of the sources
which show a CSO-like or Core-Jet morphology, in the light of
Figs. \ref{CSO}, \ref{CSO-1f}, \ref{jet} and \ref{jet-1f}.\\
We note that the spectral index of the source components has been
computed considering the {\it full-resolution} images at both
frequencies, since they allowed the lowest r.m.s. noise levels in
the images. 
This may produce an artificial steepening of the spectral
index of the components, at least a few beam size wide. We note,
however, that the vast majority of the sources and/or components are
unresolved or marginally resolved, and the artificial steepening may
occur in a very few cases.\\  
Typical errors on the determination of the spectral index are of the
order of 0.1.\\

\subsubsection{CSO candidates}

We consider CSO candidates those sources which have a ``Double/Triple''
morphology in our VLBA observations (Figs. \ref{CSO}, \ref{CSO-1f}),
an optically thin total spectral index  $>$ 0.5 (S $\propto$ $\nu^{-
  \alpha}$; Table 2), and  
small value of the variability index, as described in Tinti et
al. (\cite{st05}). Moreover, each source component is
  characterised by steep spectral indices (Tab. 3), 
with the only exception of
  the sources \object{J0005+0524}, 
\object{J1335+5844} and \object{J1735+5049}, 
and our classification will
  be explained in more detail in the following discussion.\\

\noindent{--} \object{J0003+2129}: galaxy at z=0.45 (Dallacasa et al. \cite{dfs6}). 
This source appears as a very asymmetric Double, with flux density
ratio of $\sim$ 25:1 at both frequencies. 
Mostly of the flux density originates within the Eastern component,
while a weak feature, accounting for 8 and 5 mJy at 8.4 and 15 GHz
respectively, is present to the West, $\sim$ 4 mas apart. 
The total linear size (LS) is about
20 pc. \\  

\noindent{--} \object{J0005+0524}: quasar at z=1.887.
The radio emission originates within two well-resolved components,
separated by 2 mas. Their flux density is almost the same at 8.4 GHz,
while it increases significantly at 15 GHz where S$_{\rm W}$/S$_{\rm
E}$ $\sim$ 3.6. This source does not match one of our selection
criteria. Indeed, 
the spectral index of the Western 
component is quite flat ($\alpha^{15}_{8.4}$ $\sim$ 0.2),
while the other structure
has a steep spectrum ($\alpha^{15}_{8.4}$ $\sim$ 1.9; Table
3). Although these values, we still classify this source as a CSO
candidate, since the flattish spectrum of the component W could be due
either to a core component, or a very compact hot-spot. New
observations with higher dynamic range and resolution 
are necessary to unambiguously classify this source.\\    

\noindent{--} \object{J0037+0524}: this source has been found as an empty field
by Dallacasa et al. (\cite{dfs}). It shows a Double morphology and the
radio emission originates within two well-resolved components, $\sim$
2 mas apart. 
Their flux density ratio is S$_{\rm E}$/S$_{\rm W}\sim$ 5 and 10 at
8.4 and 15 GHz respectively.\\ 

\noindent{--} \object{J0428+3259}: galaxy at z=0.479 (Dallacasa et
al. \cite{dfs}; \cite{dfs6}). The images presented in Fig. \ref{CSO}
suggest that this source is an asymmetric Triple.  
The Eastern and the Western components are 2 mas and 1 mas
respectively, from the central component, which is also the brightest one.
Their flux density ratio is S$_{\rm E}$:S$_{\rm W}$:S$_{\rm Ce}$ =
1:4:10 and 1:3.3:14, at 15 and 22 GHz, respectively. 
On the arcsecond-scale, at 1.4, 1.7 and 5.0 GHz it shows an extended
emission accounting for 12, 8 and 3 mJy respectively, and with the
same position angle of the VLBA structure
(Tinti et al. \cite{st05}).\\ 

\noindent{--} \object{J0638+5933}: optically unidentified. 
This radio source can be classified as a
Triple. The main component is marginally resolved in two
different regions, with a flux density ratio of S$_{S1}$/S$_{S2}$
$\sim$ 2 and 4 at 22 and 43 GHz respectively. 
Since the two components labelled as N and N1 are a single unresolved
component at 22
GHz, when we compute the spectral index we must take into account
the sum of their flux density at 43 GHz, obtaining $\alpha^{43}_{22}$
0.4, similar to the spectral index of the Southern component ($\sim$
0.4). The similarity of the spectral indices, together with the fact
that values of 0.4 are usually found in hot-spot region, lead us to
still classify this source as a CSO candidate, although it does not
match our selection criteria. 
New high-resolution observations with a higher dynamic range are
necessary to unambiguously classify this source.\\
About 11\% of the total flux density is missing in our VLBA
observations at 22 GHz.\\   

\noindent{--} \object{J0650+6001}: quasar at z=0.455.
This source has been confirmed as a CSO by Polatidis et
al. (\cite{p99}). The radio emission originates within two components,
$\sim$ 3 mas ($\sim$ 17 pc) apart. A weak feature, visible only at 15
GHz (Fig. \ref{CSO}) and accounting for 8 mJy, is present to the North
of the Northern component, in agreement with other works at lower
frequencies (Stanghellini et al. \cite{cstan99}). The Northern
component (labelled as N) is compact and the brightest one, while the
Southern is clearly resolved N-S. Their flux density ratio S$_{\rm
  N}$/S$_{\rm S}$ is $\sim$ 3.5 at both frequencies. 
About 18\% of
the total flux density is missing in our VLBA image at 22 GHz.
We note that in Tinti et al. (\cite{st05}) the source shows 
some variability in flux density at high frequencies, although the
spectral shape remains convex.\\ 

\noindent{--} \object{J1148+5254}: quasar at z=1.632.
Although at 15 GHz, this source appears only marginally resolved in
the NW direction, at 22 GHz (Fig. \ref{CSO-1f}) it is clearly separated
in two different asymmetric components. Their flux density ratio
accounts for S$_{\rm E}$/S$_{\rm W}$ $\sim$ 5.8 and 4.7 at 15 and 22
GHz respectively.   
We note that, although both components have steep spectra, the
faintest is the flattest as well, with $\alpha^{22}_{15}$ $\sim$ 0.6, instead
of 1.2 shown by the brightest.
About 29\% and 48\% of the total flux density is missing in our VLBA 
images at 15 and 22 GHz respectively, and this may complicate our
interpretation further.\\ 
No information on the flux density variability is currently available
given that this source was not observed by Tinti et al. (\cite{st05}). 
The VLA
observations considered here as a comparison with the VLBA total
flux densities were carried out by Dallacasa et al. (\cite{hfp0}),
therefore it is possible that some amount of missing flux density
could arise from intrinsic variability.\\ 

\noindent{--} \object{J1335+4542}: quasar at z=2.449. 
At 15 GHz (Fig. \ref{CSO-1f}), the radio emission clearly originates
within two different components, while at the lower frequency the
source appears marginally resolved in the NW direction only. 
Their flux density ratio is S$_{\rm E}$/S$_{\rm W}$ $\sim$ 4
at 15 GHz. The Eastern component has a spectral index
$\alpha^{15}_{8.4}$ $\sim$ 0.6, while the other component has a 
very steep spectrum ($\alpha^{15}_{8.4}$ $\sim$ 1.9; Table 3), 
although the fit to the images, from which the flux densities
were derived, proved to be rather problematic. \\
About 11\% and 14\% of the total flux density are missing in our VLBA 
images at 8.4 and 15 GHz, respectively.\\

\noindent{--} \object{J1335+5844}: there is an optical identification with a
very weak object (likely a galaxy) reported by Dallacasa, Falomo \&
Stanghellini (\cite{dfs6}). 
The optically thin spectral index of this source is 0.5. 
At 8.4 GHz the source is clearly resolved into two compact components
in agreement with what found by Xiang et al. (\cite{x02}),
while at 15 GHz the Southernmost component becomes weak and extended.  
As shown in Fig. \ref{CSO}, the two components are separated by $\sim$
15 mas. 
Their flux density ratio is S$_{\rm N}$/S$_{\rm S} \sim$ 3 and 8 at
8.4 and 15 GHz respectively. 
This source does not match one of our selection criteria. In fact, 
the Northern has a flattish spectrum
($\alpha^{15}_{8.4}$ $\sim$ 0.1),  
while the spectral index of the Southern component is
very steep $>$ 2 (Table 3). 
For these reason, Peak \& Taylor (\cite{pt00}) did not consider this
  source as a CSO, and it was not included in the COINS sample.
However, we still classify
this source as a CSO candidate, since the flattish spectrum of the
Northern component could be due to either an embedded core component,
or a very compact hot-spot where self-absorption is relevant (Xiang et
al., in preparation). Indeed,
in VLBI images with a dynamic range higher than those presented
here, it has been possible to identify the core component of this
source, which is located about midway from the outer and much brighter 
components (Dallacasa et al. \cite{dd05}).     
About 10\% and 15\% of the total flux density is missing in
our VLBA images at 8.4 and 15 GHz respectively.\\ 

\noindent{--} \object{J1511+0518}: it is optically identified with a Seyfert
galaxy at redshift 0.081 (Chavushyan et al. 2001).
The optically thin spectral index of the source is very steep with
$\alpha^{22}_{15}$ $\sim$ 1.2. \\
At 15 GHz it is clearly resolved in three well-separated, aligned
  components. The flux density of the outer components is S$_{\rm
    E}$/S$_{\rm W}$ $\sim$ 2 at both frequencies. On the other hand, 
the inner component
  accounts for 14 and 4 mJy at 15 and 22 GHz respectively, 
indicating a very steep spectral index
  $\alpha^{22}_{15}$ $>$ 3.
About 30\% of the total flux density is
missing in our VLBA images at both frequencies.\\   
   
\noindent{--} \object{J1616+0459}: quasar at z=3.197. At 8.4 GHz this source
consists of a bright component, labelled as N, and a slightly resolved
structure in the SW direction (S component), while at 15 GHz the
radio emission clearly originates within two different components,
$\sim$ 1.4 mas (11 pc) apart (Fig. \ref{CSO-1f}). 
Their flux density ratio is S$_{\rm
  N}$/S$_{\rm S} \sim$ 
3.5 at both frequencies. 
About 11\% and 21\% of the total flux density is
missing in our VLBA images at 8.4 and 15 GHz respectively.\\

\noindent{--} \object{J1735+5049}: optically identified with a very faint
object, tentatively classified as a galaxy by Stickel \& K\"{u}hr
(\cite{sk96}).  
The optically thin spectral index of this source is $<$ 0.5. 
This source shows a Double morphology, in agreement with what found by
Xiang et al. (\cite{x02}). 
As for \object{J1335+5844}, it does not match our selection criteria, since 
the Northern component is characterised by a steep spectrum
$\alpha^{15}_{8.4}$ $\sim$ 1.2, while the Southern component has a
flattish spectrum of about 0.3. For this reason, 
Peck \& Taylor (\cite{pt00}) did not consider this source as a CSO,
  and it was not included in the COINS sample.
However, we still classify
this source as a CSO candidate, since the flattish spectrum of the
Southern component could be due to either a core component, or a very
compact hot-spot.
New high-resolution observations are necessary to
unambiguously classify this source.
About 9\% of the total flux density is missing in our VLBA image at 15
GHz.\\ 

\noindent{--} \object{1855+3742}: optically identified with a galaxy 
(Dallacasa et al. \cite{dfs6}).
The pc-scale structure is quite different from all the other sources
discussed so far, presenting an elongated structure which is well
resolved in the 15 GHz image. At 8.4 GHz the flux density is clearly
peaked on the source centre, while at 15 GHz is more uniformly
distributed. 
About 14\% and 33\% of the total flux density is missing in our VLBA
observations at 8.4 and 15 GHz, respectively.\\ 

\noindent{--} \object{J2203+1007}: weak galaxy with unknown redshift 
(Dallacasa et al. \cite{dfs}).
This source has been confirmed as a CSO by Gugliucci et
al. (\cite{gu05}). It shows a Triple morphology, but the radio
emission mainly originates in the outer components, located about 10
mas apart. Their flux density ratio is S$_{\rm   E}$/S$_{\rm W}$
$\sim$ 2.2 and 4.8 at 8.4 and 15 GHz respectively. 
As in J1511+0518, the central component, which accounts for 11 mJy at
8.4 GHz, is completely resolved out at 15 GHz, indicating a very steep
spectral index. 
About 19\% of the total flux density is missing in our
VLBA image at 15 GHz.\\    

\begin{figure} 
\begin{center}
\includegraphics{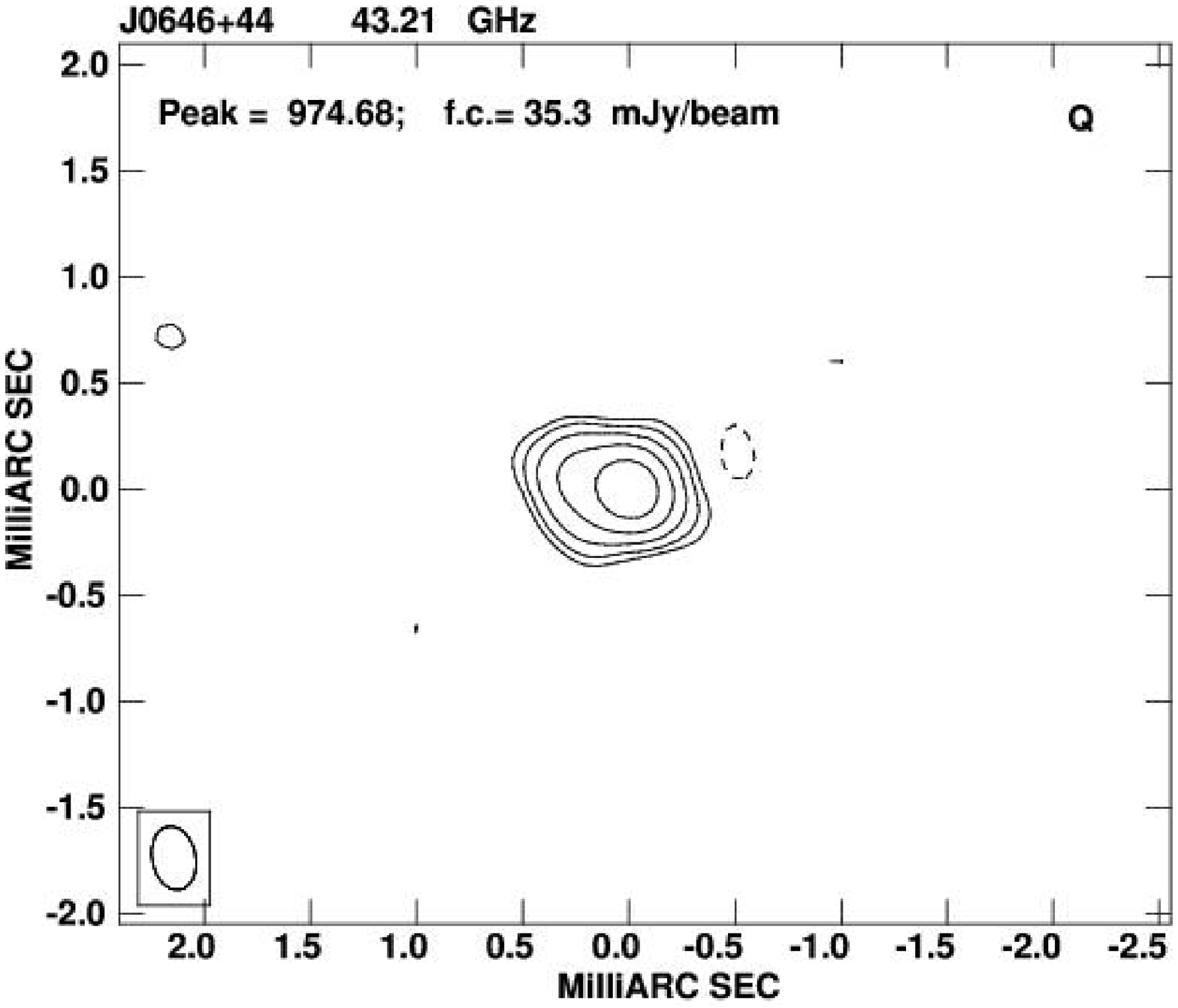}
\vspace{5.5cm}
\caption{VLBA image at 43 GHz of the Marginally Resolved source
  \object{J0646+4451}.
We give the following information
  on the plot itself: a) Peak flux density in mJy/beam; b) First
  contour intensity ({\it f.c.}, in mJy/beam), which is 3 times
  the r.m.s. noise on the image plane; contour levels increase by a
  factor of 2; c) the optical identification; 
  d) the restoring beam is plotted on the bottom left
  corner. }
\label{MR}
\end{center}
\end{figure}

\subsubsection{Core-Jet sources}

Here we describe the properties of the six Core-Jet sources.
We consider Core-Jet those sources which have a 
compact component (containing the source core) 
with a flat spectrum, and the
other structures with a steeper spectrum (Table 4).\\

\noindent{--} \object{J0329+3510}: quasar at z=0.50 (Dallacasa et al. \cite{dfs6}).
The pc-scale radio structure of this
source is characterised by a compact and bright component, and a second
one located at $\sim$ 1 mas (7 pc) 
apart in the NW direction (Fig. \ref{jet-1f}).    
Their flux density ratio is S$_{S}$/S$_{N}$ $\sim$ 2 at both
frequencies and both the radio components have an inverted spectral
index ($\alpha^{15}_{8.4}$ $\sim$ -0.2; Table 4). 
Extended emission have been detected on the arcsecond-scale with the
VLA at 1.4, 1.7 and 4.9 GHz (Tinti et al. \cite{st05}).
In the VLA second epoch observations it no longer shows the convex
spectrum, leading us to the conclusion that this is a blazar object,
and it will be rejected from the sample of genuine young and small
radio sources.\\  

\noindent{--} \object{J1811+1704}: optically identified with a stellar object 
by Dallacasa et al. (\cite{dfs}). Its optical spectrum has been found
featureless (Dallacasa et al. \cite{dfs6}), and then
we can consider it as a BL Lac object. \\
This source is characterised by a Double morphology at the highest
frequency, while it is only marginally resolved in the E direction at
8.4 GHz. The brightest component has an 
inverted spectrum ($\alpha^{15}_{8.4}$ $\sim$ --0.1), indicating the 
presence of the source core. It shows an arcsecond-scale structure in
the VLA images at 1.4 and 1.7 GHz (Tinti et al. \cite{st05}). 
In the
VLA second epoch observations it no longer shows the convex spectrum,
leading us to confirm it as a blazar object, and then it will be
rejected from the sample of young and small radio sources.\\  

\noindent{--} \object{J2021+0515}: optically identified with a galaxy by
Dallacasa et al. (\cite{dfs}). 
At 15 GHz, the radio emission originates within two well-resolved
components, $\sim$ 2 mas apart, while at 8.4 GHz it appears
Marginally Resolved. Super-resolving the lower-frequency image in the
N-S direction, indicates that the overall structure might be an
asymmetric double. Although the radio emission mainly originates
within the brightest component, labelled as N in Fig. \ref{CSO-1f}, 
its
spectrum is steeper than the Southern component ($\alpha^{15}_{8.4}$
$\sim$ 0.8 and -0.4 for N and S component, respectively). 
About 10\% of the total flux density is missing in our VLBA image at
15 GHz.\\

\noindent{--} \object{J2114+2832}: optically identified with an unresolved
object with R=18.35 (Dallacasa et al. \cite{dfs}). 
This source is characterised by a compact and bright component, which
accounts for 93\% of the total flux density and with a spectral index
of $\alpha^{22}_{15}$ $\sim$ 0.6.  
At 22 GHz this component is marginally resolved in the E-W direction,
suggesting that we are looking at the jet-base.  
Another extended low-brightness structure with a steep spectral index
($\alpha^{22}_{15}$ $\sim$ 1) is located $\sim$ 2 mas apart. About
39\% and 47\% of the total flux density is missing in our VLBA images
at 15 and 22 GHz, respectively. It should be noted that this 
source has not been observed by Tinti et al. (\cite{st05}). The VLA 
observations considered as a comparison between the total flux
densities were carried out by Dallacasa et al. (\cite{hfp0}), 
therefore it is likely that intrinsic variability may account for some
amount of missing flux density.\\    

\noindent{--} \object{J2123+0535}: quasar at z=1.878. 
The radio emission of this source mainly
originates within a compact region (E component) 
which accounts for $\sim$ 90\% of the total
flux density. A weak feature, visible in the VLBA image at 43 GHz and
only marginally resolved at the lower frequency, is present to the
Western part of the main component, so the radio morphology is that of a
Double. Both spectral indices are quite flat ($\alpha^{43}_{22}$
$\sim$ 0.3 and 0.5 for the E and W component, respectively).
About 33\% of the total flux density is missing in our VLBA
image at 22 GHz. 
This source was a GPS candidate, observed by the ATCA monitoring
program. Although its convex spectrum, it has been rejected as new GPS 
candidate for the evidence of significant variability (Edwards et
al. \cite{e04}). In the second-epoch VLA observations it no longer
shows the convex spectrum, leading us the conclusion that this is a
blazar object, and it will be dropped from the sample.\\  

\noindent{--} \object{J2136+0041}: quasar at z=1.932. 
This object was recognised early as a GPS source (Shimmins et
al. \cite{sh68}). 
It has a Core-Jet morphology, although at low frequency it may
appears as a Double with components of similar flux density. The
source core is located in the Easternmost component (labelled as C),
while the jet is initially directed to the North and then turns to the
West. The spectral index of the jet is rather steep
$\alpha^{43}_{22}$ $\sim$ 1.3, while the core has a flatter
spectrum ($\alpha^{43}_{22}$ $\sim$ 0.7), although this value may be
slightly increased by some emission from the steep-spectrum jet-base.
It has been noted that, although this source presents some amount of
flux density variability, its GPS spectral shape does not change with
time (Tornikoski et al. \cite{t01}). About 57\% of the total flux 
density is missing in our VLBA image at 22 GHz. The VLA
observations considered as a comparison between the total flux densities
were carried out by Dallacasa et al. (\cite{hfp0}), 
therefore it is possible that some amount of
missing flux density could be related to an intrinsic variability.\\

\begin{figure*} 
\begin{center}
\includegraphics{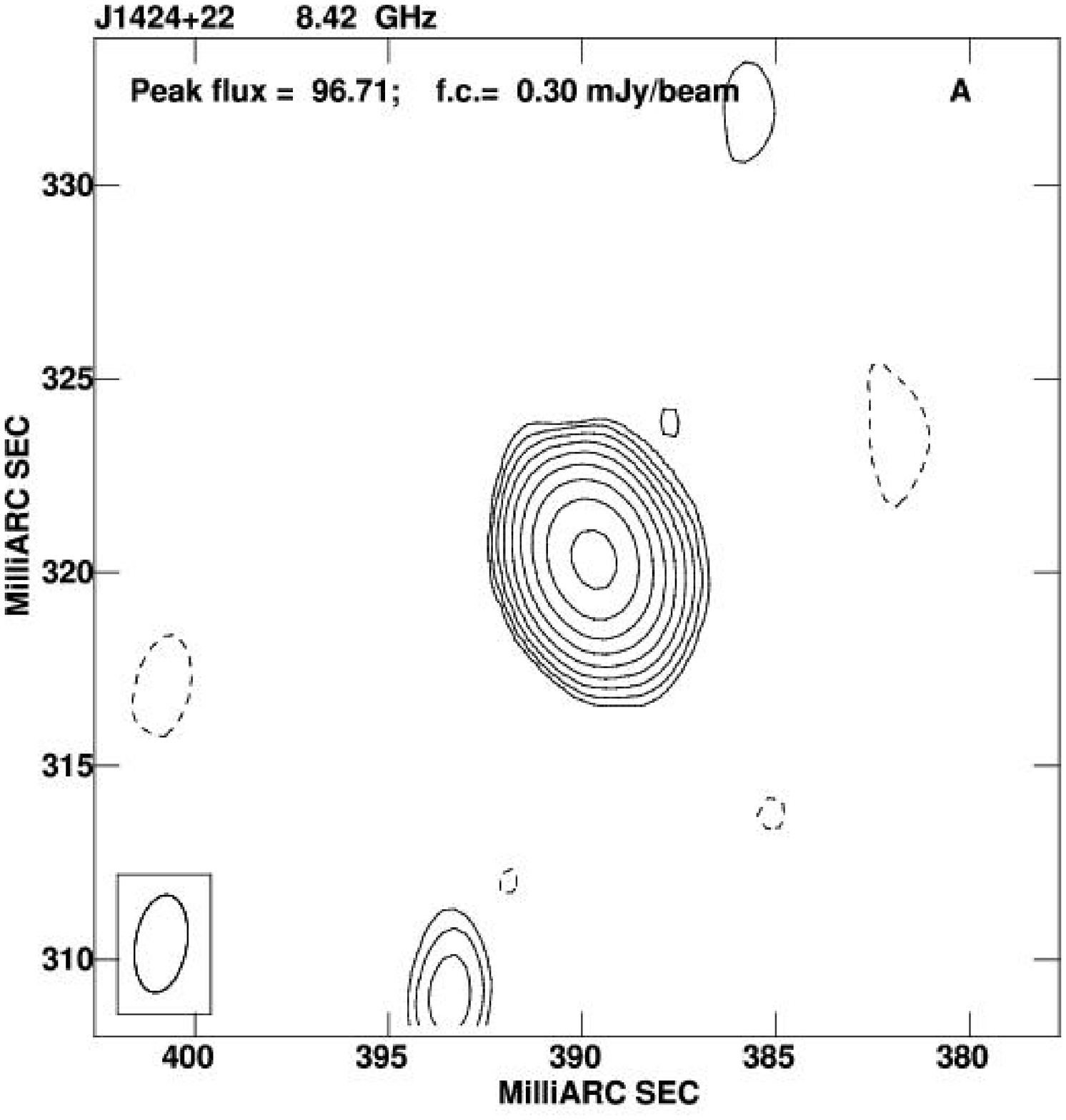}
\includegraphics{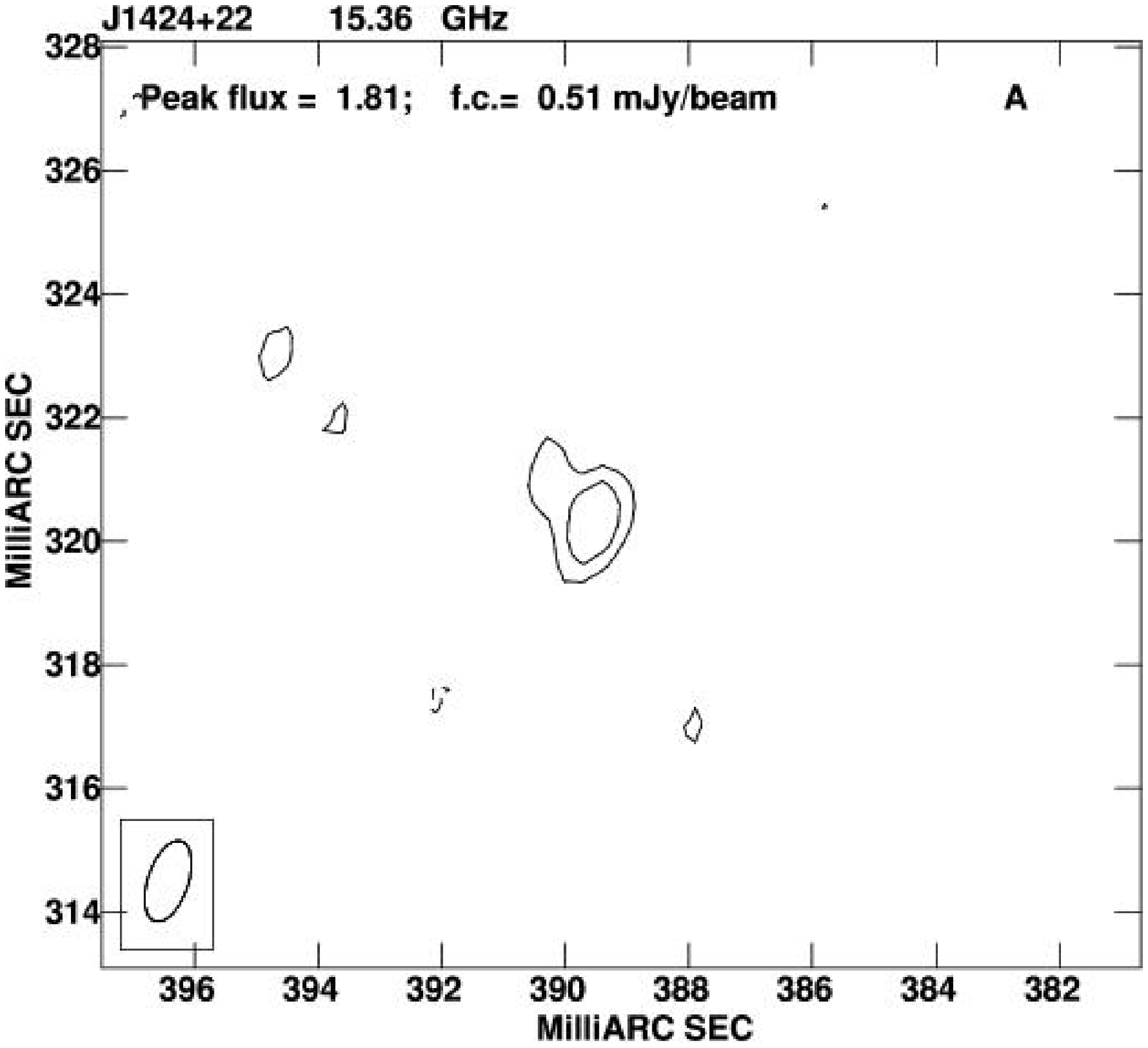}
\includegraphics{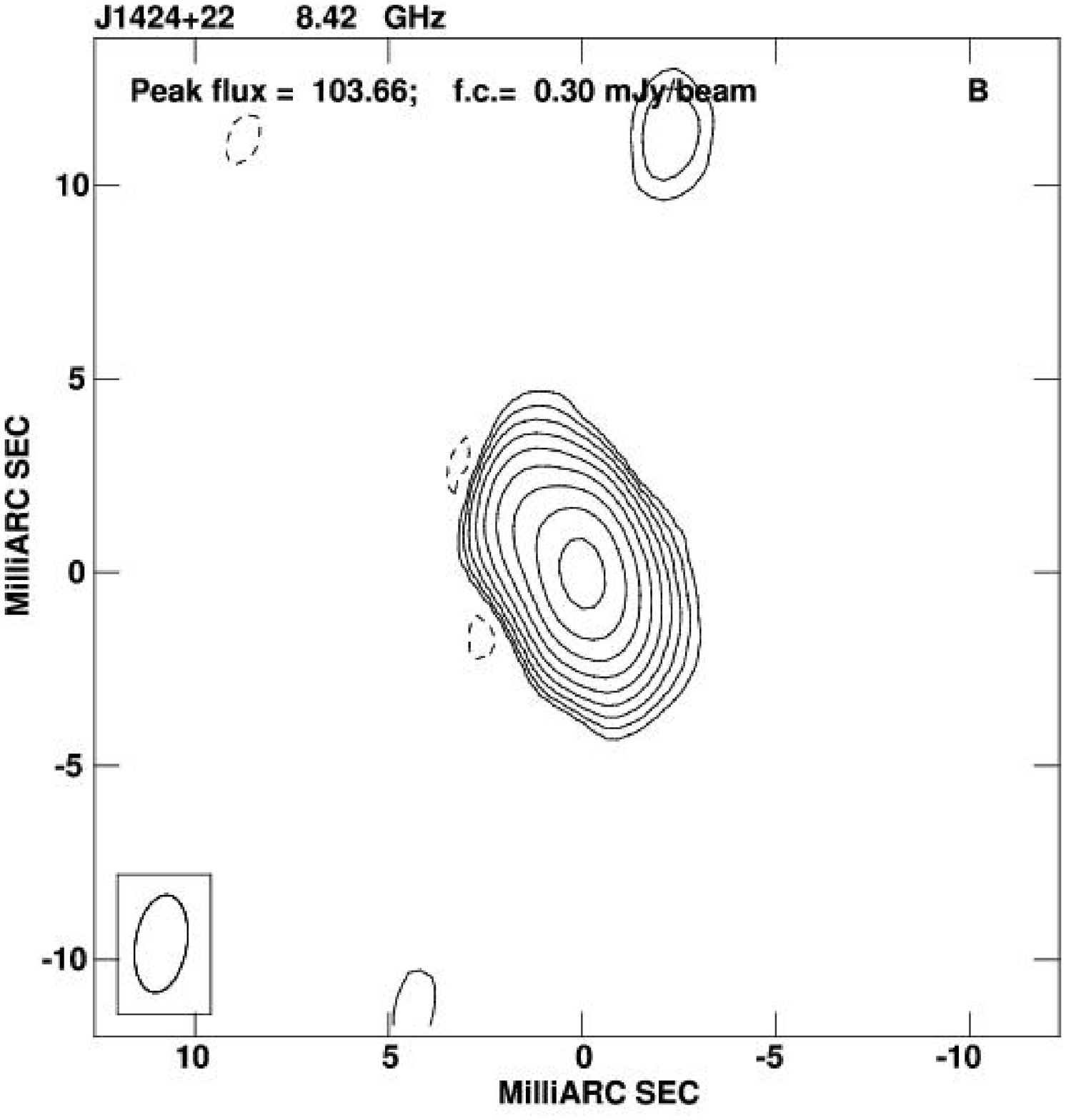}
\includegraphics{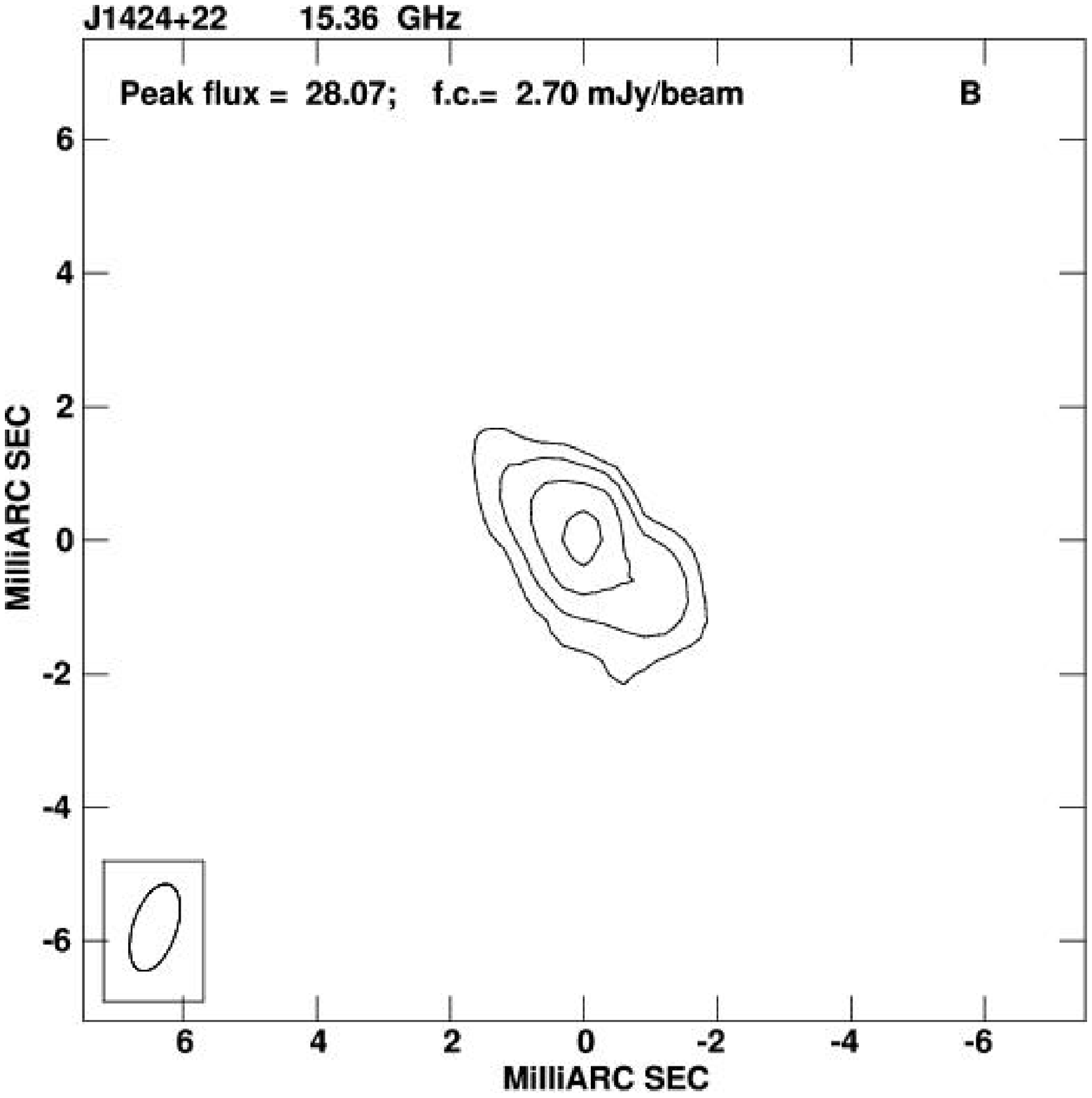}
\includegraphics{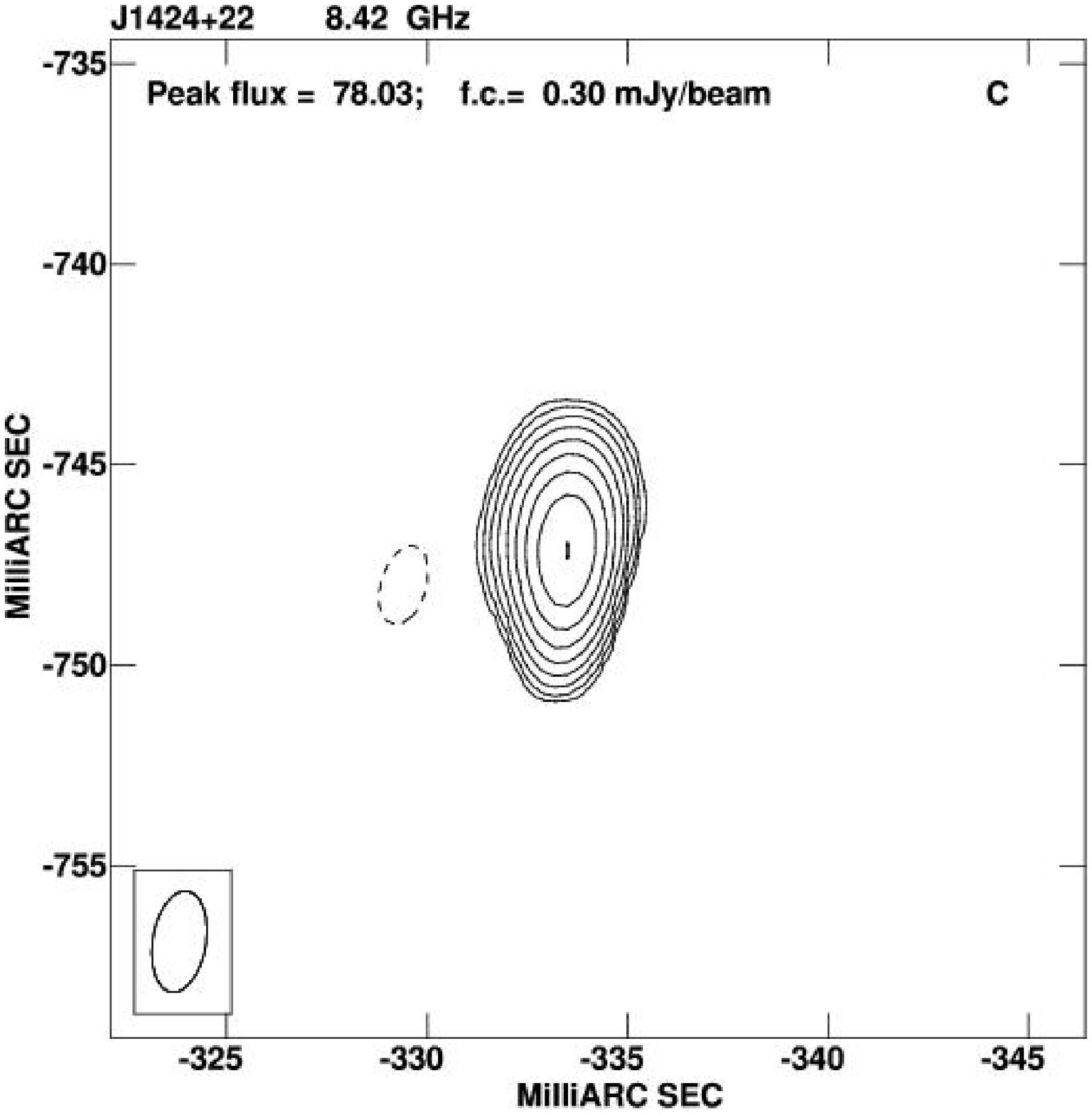}
\includegraphics{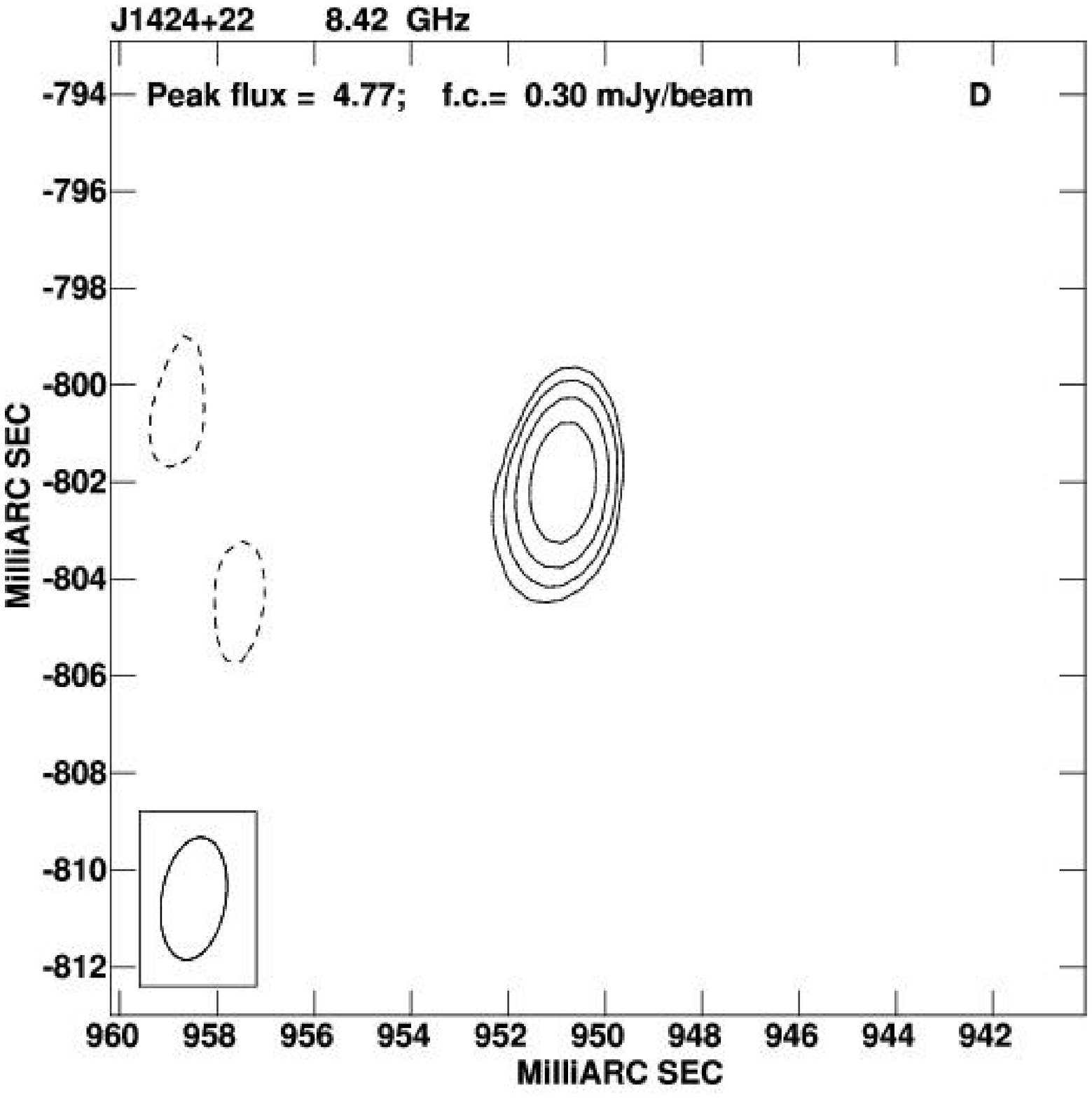}
\vspace{22.0cm}
\caption{VLBA images of the gravitational lens system \object{J1424+2256}. {\it
    top}: Images at 8.4 and 15 GHz of Component A. {\it middle}:
    Images at 8.4 and 15 GHz of Component B. {\it bottom}: Images at
    8.4 of Component C ({\it left}) and D ({\it right}). For each image we give the following information
  on the plot itself: a) Peak flux density in mJy/beam; b) First
  contour intensity ({\it f.c.}, in mJy/beam), which is generally 3 times
  the r.m.s. noise on the image plane; contour levels increase by a
  factor of 2; c) the restoring beam is plotted on the bottom left
  corner of each image. }
\label{lente}
\end{center}
\end{figure*}

\subsubsection{The gravitational lens \object{J1424+2256}}
The gravitational lens \object{J1424+2256} is associated with a quasar of 16.5
magnitude at z=3.62. For this reason is one of the highest apparent
luminosity lens system (Patnaik et al. \cite{pat92}).
The lensing object belong to a group of galaxies at z=0.338 
(Kudnic et al. \cite{kud97}).\\
In our VLBA observations the lens system consists of four components
(Fig. \ref{lente}). Considering the position angle of the three
brightest components, PA$_{\rm A}$ 56$^{\cdot}$, PA$_{\rm B}$
43$^{\cdot}$ and PA$_{\rm C}$ 14$^{\cdot}$, they are
tangentially elongated as expected from lens models. Component D is
unresolved in our observations. Their flux
densities at 8.4 GHz, accounting for S$_{\rm A}$ 156 mJy, S$_{\rm B}$ 169
mJy, S$_{\rm C}$ 89 mJy and S$_{\rm D}$ 4.6 mJy, as well as their
position angles, are in agreement with
what found by Patnaik et al. (\cite{pat99}). 
At 15 GHz, only the two brightest components have been
detected. Component B shows the same elongation as at lower frequency,
while component A is almost resolved out.\\  
From VLA and VLBI observations some amount of polarised emission have
been found in the three brightest components, accounting for 2.5\%,
1.8\% and 1.2\% in A, B and C respectively. Component D was too faint
to detect any polarisation (Patnaik et al. \cite{pat99},
\cite{pat92}).\\

\section{Discussion}

Here we present an overview of the properties derived from the
pc-scale structure as seen in our observations. A detailed discussion
on the physical properties will be given in a forth-coming paper.\\ 

\subsection{Morphology}
The search for young radio sources based only on their spectral shape
has proven successful, although there is some amount of contamination
by flat-spectrum, variable objects caught when their radio emission is
likely dominated by a self-absorbed jet component (see Tinti et
al. \cite{st05}). Flux density and spectral shape variability is an
important tool to make a proper classification, but it needs to be
complemented by an accurate pc-scale radio morphological
classification. 
Intrinsically small and young radio sources are expected to 
have a ``Double/Triple'' structure like in CSOs, while a Core-Jet
morphology is more typical of blazar objects.\\
These two pieces of evidence need to be considered
together.  \\
Given the short time spent on each source,
our VLBA observations have little or no sensitivity to complex structures
covering a large number of pixels, but this should not play a major
role for most of the sources discussed here.
In the morphological classification given in Table 2, also the
spectral information derived by comparing the flux densities of the
components detected at two frequencies is very relevant. 
In general all the source components visible in
Figs. \ref{CSO} and \ref{CSO-1f} have
steep spectra, while it is not the case for those in
Figs. \ref{jet} and \ref{jet-1f}.\\

In Table 3 and Table 4 for each component of
our CSO and Core-Jet candidate sources, we 
present the physical parameters. 
The total flux density
and the Full Width Half Maximum (FWHM) have been obtained by fitting
the components with a Gaussian model (task JMFIT).\\
Steep spectrum components are labelled as North (N), South (S),
East (E), West (W), Central (Ce) depending on the source orientation
and the component location,
while a flat spectrum core is labelled C, when detected. Beyond this,
our core candidates must be unresolved.\\

\subsubsection{CSO morphology}
The sources discussed here have been observed in the optically thin
part of their radio spectrum, but at frequencies at a factor of a few
within their peak. In case of different component size and/or flux
density within the same object, it is possible that the individual
spectra of the two (or more) components, peak at different
frequencies. Therefore, the observed spectral index in our data may be
affected showing different values.\\   
Fourteen sources (7 galaxies \object{J0003+2129}, \object{J0037+0808}, \object{J0428+3259},
\object{J1511+0518}, \object{J1735+5049}, \object{J1855+3742} and \object{J2203+1007}, 5 quasars \object{J0005+0524},
\object{J0650+6001}, \object{J1148+5254}, \object{J1335+4542} and \object{J1616+0459}, and 2
empty field \object{J0638+5933}, \object{J1335+5844}) 
show a
``Double/Triple'' morphology, similar to those found in 
CSO sources (Figs. \ref{CSO}, \ref{CSO-1f}). 
Ten sources have a
Double structure characterised by two separated components
where the flux density asymmetry is rather large, like in \object{J0003+2129}.
Four sources are characterised by a Triple morphology (\object{J0428+3259},
\object{J0650+6001}, \object{J1511+0518} and \object{J2203+1007}), where all the components have a steep
spectrum. Since the presented data do not have either
the dynamic range or the resolution suitable to detect weak flat
components, there is no secure core identification so far.
The source \object{J1855+3742} is the only one which shows a quite complex
structure with steep spectrum.\\
Their optically thin spectral index is much steeper than 0.5. 
Only the sources \object{J0638+5933}, \object{J1335+5844}
and \object{J1735+5049} have spectral indices of $\sim$ 0.5 or little
flatter. Despite Peck \& Taylor (\cite{pt00}) did not consider
  \object{J1335+5844} and \object{J1735+5049} CSO candidates, 
  not including them in the COINS sample,
we still classify
them as young object candidates. For \object{J1335+5844},
the source core has been identified, using high dynamic range VLBI
images (Dallacasa et al. \cite{dd05}), and this object can be
considered a case study.\\  
All these sources were not found to be variable in multi-epoch
VLA observations, and they are characterised by low value of
the variability index V, as defined by Tinti et al. (\cite{st05}). Two
exceptions are \object{J0650+6001}, which shows a little variability in its flux
density at high frequencies, and \object{J1616+0459} where the flux density has
decreased at all frequencies, but keeping the spectral shape unchanged.
For the source \object{J1148+5254}, only a single-epoch VLA observation
is available, and then it is not possible to assess its variability.\\
The source \object{J0428+3259} is the only object which shows some extended
emission on the arcsecond-scale (Tinti et al. \cite{st05}), although
such extension could not be properly imaged in a relatively deep L
band image.\\

\subsubsection{Core-Jet morphology}
In this sub-section will be discussed objects not considered 
genuine young radio sources anymore.
Six objects (five quasars \object{J0329+3510}, \object{J2021+0515}, \object{J2114+2832}, 
\object{J2123+0535}, \object{J2136+0041} and one BL Lac \object{J1811+1704}), 
show a
Core-Jet morphology, where one compact and bright component with a
quite flat or inverted spectrum dominates the radio emission 
(from the 70\% up to
more than 90\% of the total flux density of the source).
On the other hand, the jet component accounts only for a little
percentage of the total flux density and generally has a steeper spectrum.\\ 
Three sources (\object{J0329+3510}, \object{J1811+1704} and \object{2123+0535}) did not show the
convex form in the second epoch spectrum (Tinti et al. \cite{st05}), which
actually turns out to be flat. Therefore, these sources are classified
as blazar and they will not be considered HFP candidates anymore. 
All these three blazar sources also show some extended
emission on the arcsecond-scale (Tinti et al. \cite{st05}).\\
The Core and Jet components are also characterised by very different
physical parameters: high magnetic field and small emitting area for
the Core component, while lower magnetic fields, more similar to those
found in CSO components, and a larger emitting area for the Jet
structure. These properties will be discussed in more detail in
section 5.3.\\  

\subsection{Radio structure and optical identification}
There is a clear segregation in radio morphology between quasars
and galaxies. The majority of galaxies ($\sim$78\%) show a ``Double/Triple''
morphology while quasars
are generally either Core-Jet ($\sim$ 16\%) or Unresolved ($\sim$
71\%).\\
This is consistent with the idea that the HFP spectrum in galaxies and
quasars appears originated from intrinsically different emitting regions:
mini-lobes and/or hot-spots in galaxies, compact regions related to
the core and to the jet base in quasars. This is in agreement with
other results obtained by comparing the properties of galaxies and
quasars in GPS and bright CSS samples (Stanghellini et
al. \cite{cstan05}; Fanti et al. \cite{rf90}). 
It is likely that HFP quasars (as well as GPS quasars) are
intrinsically similar to the flat spectrum radio sources, and their
convex radio spectrum is due to a single homogeneous component, like a
knot in a jet, which dominates the radio emission, although some HFP
quasars can still be genuine young radio sources.\\
Strong support to the idea that a significant fraction of HFP quasars
represent a different population of beamed radio sources like blazars,
comes from the detection of substantial variability of their radio
spectra (Tinti et al. \cite{st05}). 
Indeed, during a second epoch of simultaneous multi-frequency VLA
observations, carried out for monitoring the flux density, seven
objects no longer exhibit the convex spectrum, which actually turned
out to be flat.
All these sources are
associated with quasars with ``Core-Jet'', ``Unresolved'' or
``Marginally Resolved'' radio structure
in our VLBA images, suggesting that they are blazars
previously selected during a flare of a self-absorbed component.\\        

\begin{figure} 
\begin{center}
\includegraphics{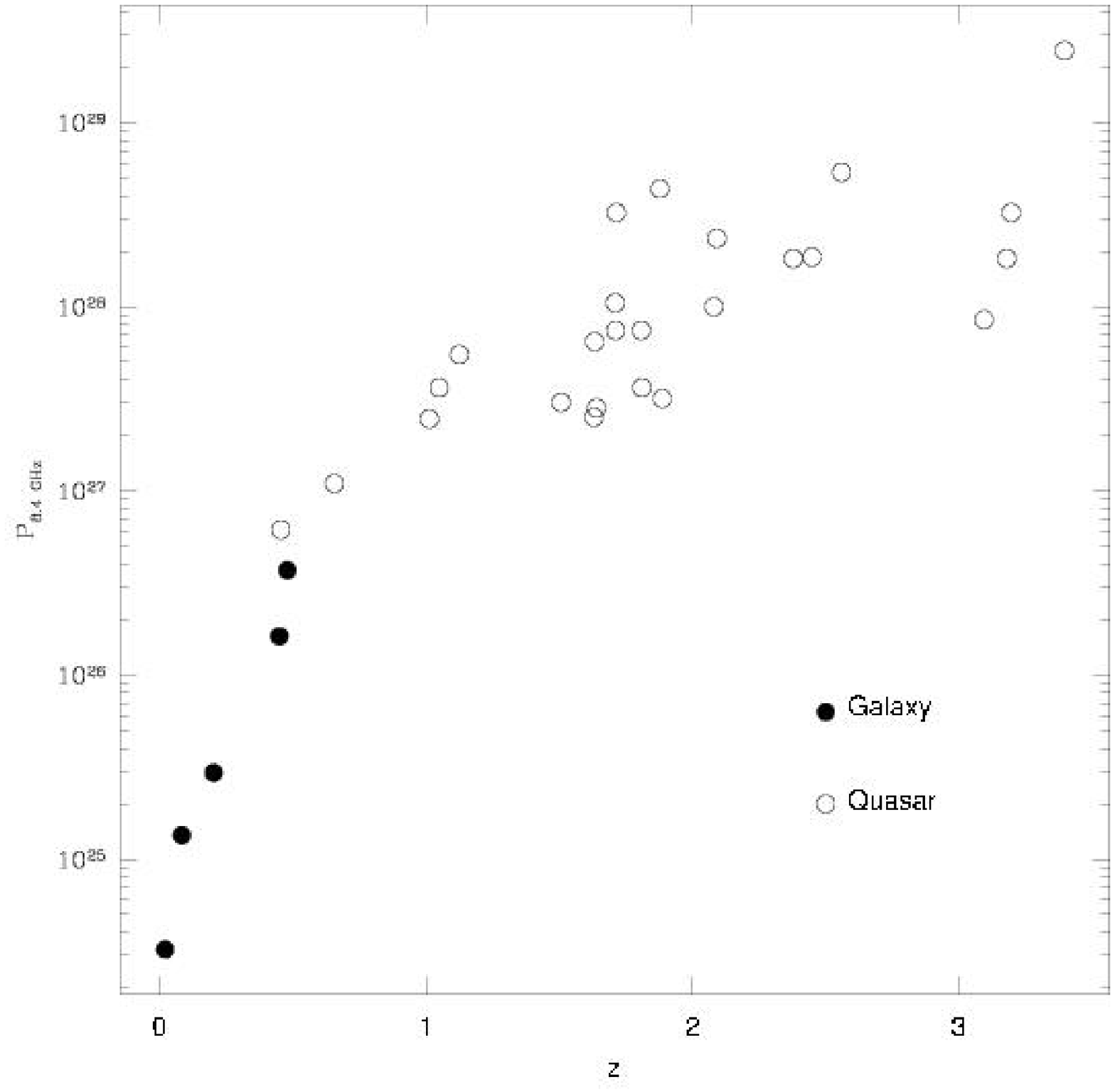}
\vspace{7.5cm}
\caption{Distribution of the monochromatic luminosity at 8.4 GHz (W/Hz)
versus the redshift, for the HFP objects with known z. 
The apparent luminosity has been computed on the basis 
of the VLA flux density. There is a clear influence of the redshift:
galaxies, characterised by low redshift, are located in the bottom
right part of the plot, while quasars dominate the upper left region.}
\label{pot_z}
\end{center}
\end{figure}
\begin{figure} 
\begin{center}
\includegraphics{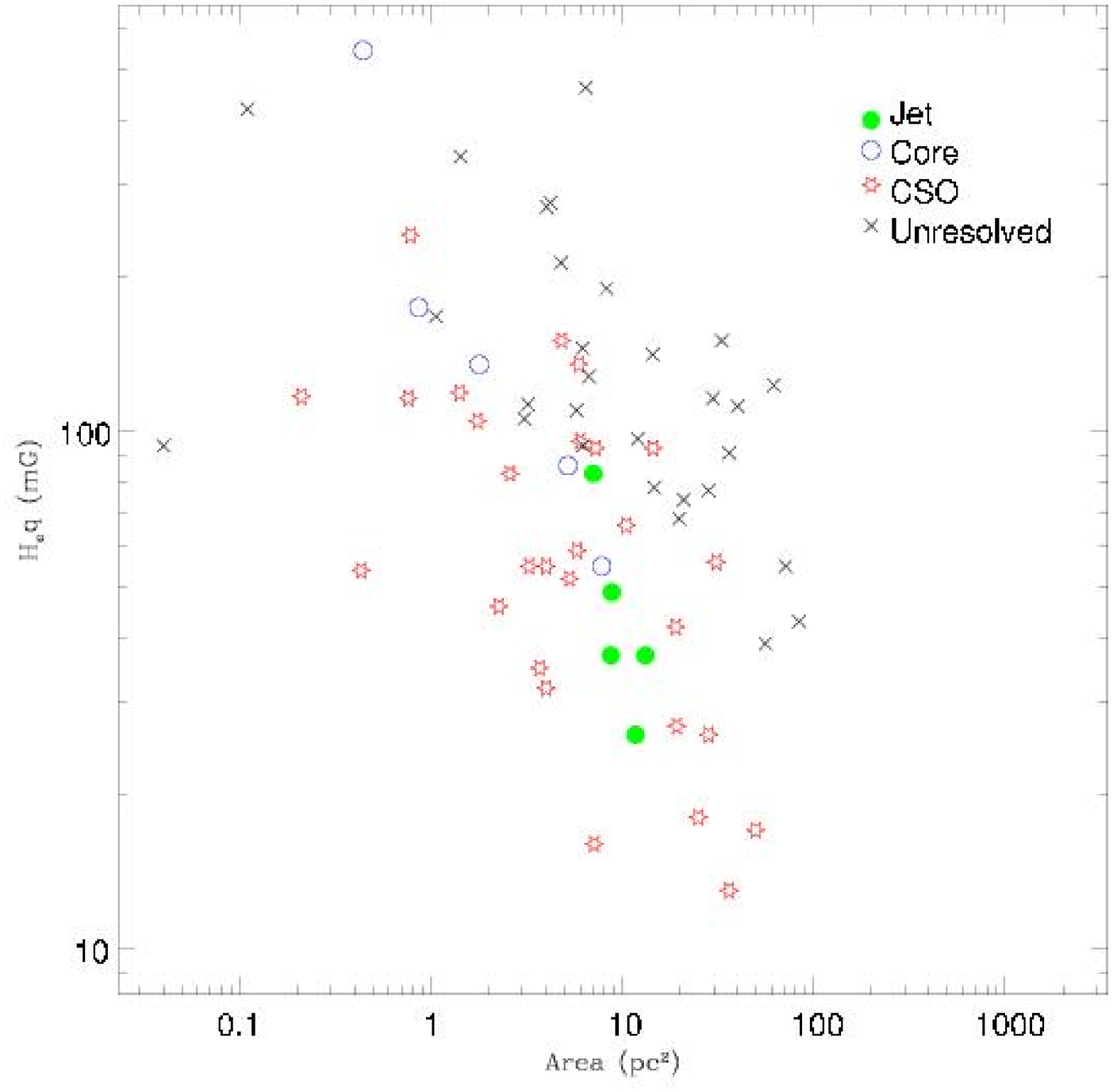}
\vspace{7.5cm}
\caption{Distribution of the emitting area versus the equipartition
  magnetic field.}
\label{magn}
\end{center}
\end{figure}

\subsection{Physical parameters}

Physical parameters for the radio sources have been computed 
assuming equipartition and using standard formulae (Pacholczyk,
\cite{pach70}). Proton and electron energies have been assumed equal,
with a filling factor of unity; ellipsoidal geometry and an average
optically thin spectral index of 0.7 have been adopted. The average
luminosity of the sources is Log P $\sim$ 10$^{27.5}$ W/Hz with a clear
influence of redshift, being the most distant sources (quasars) the
most luminous as well (Fig. \ref{pot_z}). 
In the components of candidate CSOs, typical
values obtained for minimum total energy (U$_{\rm min}$), minimum
energy density (u$_{\rm min}$), equipartition magnetic field 
(H$_{\rm eq}$), and brightness temperature (T$_{\rm b}$) are:\\

\noindent
(U$_{\rm min}$) $\sim$ 10$^{53}$--10$^{54}$ erg;\\
(u$_{\rm min}$) $\sim$  10$^{-4}$--10$^{-5}$ erg cm$^{-3}$;\\
(H$_{\rm eq}$)  $\sim$  10$^{-2}$ G;\\
(T$_{\rm b}$)   $\sim$  10$^{8}$--10$^{11}$ K.\\

\noindent  
The same values are found also in the jets of the HFPs with 
a Core-Jet morphology,
while the core components are generally characterised by higher energy
density and brightness temperatures, 
as they have larger luminosities and smaller sizes.\\
For comparison, the minimum energies stored in CSS radio 
sources (Fanti et al. \cite{rf90}) 
of similar radio power are 2 - 3 order of
magnitude larger and the corresponding energy densities are 2 - 3
order of magnitude lower.\\
In Fig. \ref{magn} we plot the magnetic field versus the emitting
area. There is a clear segregation among the different morphologies.
The bottom left part of
the plot is dominated by CSO and jet components, while the core
components and the unresolved sources are located in the upper part.  
This relation may be very useful to better classify CSO
candidate. Indeed, as Core and Jet structures have
deep differences in their equipartition magnetic field and emitting
area, if we find such a trend also in a CSO candidate, it may suggest
a possible wrong classification.\\ 
Following the definition by Readhead (\cite{rh94}), 
we computed the {\it ``equipartition brightness
  temperature''}. Since it depends mainly on the redshift and only
very weakly on observed parameters,
like the observed peak frequency and the flux density, it can be
determined to high precision, becoming a good observable quantity
against which to test theoretical models.
The values we
obtain range from 7$\times$10$^{10}$ K,
up to 2.5$\times$10$^{11}$ K, in
agreement with what
found by Readhead (\cite{rh94}).\\
Moreover, assuming the turnover frequency from Dallacasa et
al. (\cite{hfp0}) and the
equipartition magnetic field,
we compute the {\it  maximum brightness temperature}
T$_{\rm  b,max}$.  
The
comparison between the T$_{eq}$ and T$_{\rm  b,max}$
is an important tool to investigate a possible
departure from the minimum energy and equipartition condition in
these object, since T$_{\rm  b,max}$/T$_{eq}$ is an independent way to
derive the equipartition Doppler factor (Readhead
\cite{rh94}). Therefore, a T$_{\rm  b,max}$ $>>$ T$_{eq}$ is a strong
indication that that source is a blazar.\\ 
For all the unresolved sources, the core components and few of
the brightest CSO components, we find that T$_{\rm
  b,max}$/T$_{eq}$ $\sim$ 2-3, while for a good fraction of CSO and jet
components, it is $\leq$ 1, indicating that in these structures the total
energy is not far from the minimum energy condition.\\

\section{Summary}

We have presented the results of new VLBA observations at two
different frequencies in the optically thin part of  the spectrum for
51 HFPs from the ``bright HFP sample'' (Dallacasa et al. \cite{hfp0}). 
The conclusion from this investigation can be summarised as follows:\\

a) Morphology:\\

\begin{itemize}

\item Fourteen sources (27\%) show a CSO-like morphology in our VLBA
  images;
\item Six sources (12\%) have a Core-Jet morphology; 

\item Thirty-one sources (61\%) are Unresolved or Marginally resolved
  even at the VLBA resolution;

\end{itemize}

b) Optical identification related evidence:\\

\begin{itemize}

\item Seven sources all identified with quasar (3 with a Core-Jet
  structure and 4 Unresolved) have been rejected, since their radio
  spectrum turned out to be flat;

\item HFP galaxies and quasars have different morphological
  properties. The 78\% of the galaxies show a CSO-like morphology,
  while quasars have Core-Jet (16\%) or Unresolved (71\%) morphology;
\end{itemize}

c) Relationship with flux density variability:\\

\begin{itemize}

\item Three Unresolved and Marginally Resolved sources (2 quasars and
  1 BL Lac), show substantial flux density
  variability, and they are not considered HFP candidates anymore;

\item Generally galaxies do not show any evidence of variability;

\end{itemize}

In general, the pc-scale information available from these short VLBA
observations and low-dynamic range images, becomes more effective if
complemented with observations of the flux density and spectral shape
variability.\\
Milli-arcsecond images with high dynamic range, polarisation
measurements, together with a multi-epoch flux density monitoring can
help to find other contaminating objects of the original
sample. The analysis of a 
third epoch of simultaneous multi-frequency VLA observations and
polarisation data is in progress. At the end, we will be able to
construct a sample of confirmed genuine and young HFP radio sources.\\    

\begin{table*}[h]
\small{ 
\begin{center}
\begin{tabular}{|c|c|c|c|c|c|c|c|c|c|c|c|c|c|}
\hline
Source&Comp&S$_{\rm 8.4 GHz}$&S$_{\rm 15.3 GHz}$&S$_{\rm 22.2
  GHz}$&S$_{\rm 43.2 GHz}$&$\alpha$&$\theta(1)$&$\theta(2)$&P.A.&H$_{\rm
  eq}$&$\nu_{\rm max}$&LAS&LLS\\
 & &mJy&mJy&mJy&mJy& &mas&mas& &mG&GHz&mas&pc\\
\hline
&&&&&&&&&&&&&\\
J0003+2129&E&214&126& & &0.9&0.32&0.28&150&46&4.8& & \\
 &W&8&5& & &0.8&0.95&0.22&120&52&1.3&3.8&22\\
J0005+0524&W&90&82& & &0.2&0.97&0.27&95&56&4.0& & \\
 &E&73&23& & &1.9&0.64&0.53&92&42&3.0&1.7&14\\
J0037+0808&E&215&150& & &0.6&0.44&0.26&80&59&5.3& &\\
 &W&41&15& & &1.7&0.37&0.21&75&32&2.9&2.1&17$^{1}$\\
J0428+3259&E& &23&12 & &1.7&1.01&0.26&100&16&1.8&1.9&11\\
 &Ce& &233&168& &0.9&0.63&0.19&109&55&5.3& &\\
 &W& &93&40& &2.2&0.20&0.08&147&54&4.9&0.8&5\\
J0638+5933&S& & &313&236&0.4&0.22&0.16&91&105&11.4& & \\
 &N& & &179&60&1.6&0.19&0.08&104&116&10.0& &\\
 &N1& & & &73& &0.26&0.20&-&83&7.03&0.76&-\\
J0650+6001&N& &479&327& &1.0&0.40&0.39&95&55&6.5& &\\
 &S& &144&78& &1.6&0.45&0.32&37&35&3.9&3.2&18\\
J1148+5254&E& &306&199& &1.2&0.23&0.06&3&240&13.9& &\\
 &W& &53&42& &0.6&0.25&0.10&9&119&7.2&0.8&7\\
J1335+4542&E&345&242& & &0.6&0.67&0.17&125&135&7.7& &\\
 &W&188&60& & &1.9&0.67&0.30&130&66&4.2&1.3&11\\
J1335+5844&N&422&388& & &0.1&0.44&0.22&161&150&7.5& &\\
 &S&165&46& & &2.1&1.19&0.84&121&17&1.9&14.8&118$^{1}$\\
J1511+0518&E& &415&285& &1.0&0.24&0.06&-&114&7.7&2.1&3\\
 &Ce& &15&4& &3.5&0.23&0.11&177&35&2.3& & \\
 &W& &256&150& &1.4&0.55&0.21&130&117&4.6&2.8&4\\
J1616+0459&N&390&208& & &1.1&0.74&0.43&157&93&5.8& &\\
 &S&106&60& & &1.0&0.46&0.29&123&96&5.2&1.4&11\\
J1735+5049&N&182&87& & &1.2&0.69&0.56&33&27&3.1&3.4&27$^{1}$\\
 &S&692&573& & &0.3&0.91&0.16&18&93&7.6& &\\
&&&&&&&&&&&&&\\
\hline
\end{tabular}
\end{center}  }
\caption{The VLBA flux density of each component, for the sources with
  a CSO-like morphology. Columns 1 and 2:
  source name and sub-component label; Columns 3, 4, 5 and 6: VLBA
  flux density at 8.4, 15.3, 22.2 and 43.2 GHz respectively; Column 7:
  spectral index between the two frequencies where VLBA images
  are available; Columns 8, 9
  and 10: Deconvolved angular sizes of major and minor axis of the
  best-fitting Gaussian component and the position angle of major axis
  as estimated on the most suitable images at the different
  frequency, using JMFIT.Column 11: Equipartition magnetic field; we
  assume $\alpha$ = 0.7. Column 12: Turnover frequency; Columns 13 and
  14: the
  angular and linear distance between the components.
$^{1}$For the sources
  with redshift unknown, we adopt z=1.00.}
\label{cso}
\end{table*}

\addtocounter{table}{-1}
\begin{table*}[h]
\small{ 
\begin{center}
\begin{tabular}{|c|c|c|c|c|c|c|c|c|c|c|c|c|c|}
\hline
Source&Comp&S$_{\rm 8.4 GHz}$&S$_{\rm 15.3 GHz}$&S$_{\rm 22.2
  GHz}$&S$_{\rm 43.2 GHz}$&$\alpha$&$\theta(1)$&$\theta(2)$&P.A.&H$_{\rm
  eq}$&$\nu_{\rm max}$&LAS&LLS\\
 & &mJy&mJy&mJy&mJy& &mas&mas& &mG&GHz&mas&pc\\
\hline
&&&&&&&&&&&&&\\
J1855+3742&N1&43&29& & &0.7&0.58&0.41&179&26&2.0&1.9&15$^{1}$\\
 &N2&106&25& & &2.1&0.67&0.15&157&43&3.2& &\\
 &N3&29&33& & &-0.2&1.27&0.64&137&16&1.5&2.4&19$^{1}$\\
 &S&10& & & & &1.91 &1.19&131&8&0.8&5.3&42$^{1}$\\
J2203+1007&E&129&92& & &0.6&0.98&0.58&112&26&2.7&3.2&25$^{1}$\\
 &W&58&18& & &2.0&1.09&0.46&110&18&1.7&7.3&58$^{1}$\\
 &Ce&11& & & & &1.30&0.56&170&13& & &\\
&&&&&&&&&&&&&\\
\hline
\end{tabular}
\end{center}  }
\caption{Continued.}
\label{cso}
\end{table*}

\begin{table*}[b]
\small{ 
\begin{center}
\begin{tabular}{|c|c|c|c|c|c|c|c|c|c|c|c|c|c|}
\hline
Source&Comp&S$_{\rm 8.4 GHz}$&S$_{\rm 15.3 GHz}$&S$_{\rm 22.2
  GHz}$&S$_{\rm 43.2 GHz}$&$\alpha$&$\theta(1)$&$\theta(2)$&P.A.&H$_{\rm
  eq}$&$\nu_{\rm max}$&LAS&LLS\\
 & &mJy&mJy&mJy&mJy& &mas&mas& &mG&GHz&mas&pc\\
\hline
&&&&&&&&&&&&&\\
J0329+3510&S&340&374& & &-0.2&0.30&0.12&173&107&9.0& & \\
 &N&126&143& & &-0.2&0.57&0.31&163&39&4.1&1.2&7\\
J1811+1704&W&342&367& & &-0.1&0.178&0.097&168&174&12.1& &\\
 &E&176&119& & &0.7&0.54&0.49&173&37&3.17&0.84&7$^{1}$\\
J2021+0515&N&327&202& & &0.8&0.49&0.32&156&55&5.3& &\\
 &S&32&40& & &-0.4&0.76&0.23&147&37&3.1&2.4&27$^{1}$\\
J2114+2832&N& &447&360& &0.6&0.55&0.19&81&86&7.9& &\\
 &S& &19&13& &1.0&0.84&0.28&47&26&2.3&1.9&15$^{1}$\\
J2123+0535&E& & &1528&1269&0.3&0.16&0.05&18&545&36.1& &\\
 &W& & &191&139&0.5&0.37&0.34 &50&83&7.6&0.68&6\\
J2136+0041&C& & &1614&1038&0.7&0.54&0.15&146&230&16.5& &\\
 &W& & &1695&718&1.3&1.24&0.71&65&76&8.0&2.2&19\\
&&&&&&&&&&&&&\\
\hline
\end{tabular}
\end{center}  }
\caption{The VLBA flux density of Core-Jet source components. Columns 1 and 2:
  source name and sub-component label; Columns 3, 4, 5 and 6: VLBA
  flux density at 8.4, 15.3, 22.2 and 43.2 GHz respectively; Column 7:
  spectral index between the two frequencies where VLBA images
  are available; Columns 8, 9
  and 10: Deconvolved angular sizes of major and minor axis of the
  best-fitting Gaussian component and the position angle of major axis
  as estimated on the image at the highest
  frequency, using JMFIT. Column 11: Equipartition magnetic field; we
  assume $\alpha$ = 0.7. Column 12: Turnover frequency; Columns 13 and
  14: the
  angular and linear distance between the components. 
$^{1}$For the sources
  with redshift unknown, we adopt z=1.00.}
\label{corejet}
\end{table*}

\begin{acknowledgements}
The VLBA is operated by the US National Radio Astronomy Observatory
which is a facility of the National Science Foundation operated under
a cooperative agreement by Associated Universities, Inc. This work has
made use of the NASA/IPAC Extragalactic Database NED which is operated
by the JPL, California Institute of Technology, under contract with
the National Aeronautics and Space Administration. 
\end{acknowledgements}

\end{document}